\documentclass[12pt]{article}

\usepackage{graphicx}
\usepackage{multirow}
\usepackage{booktabs}
\usepackage{array}
\usepackage{booktabs,subcaption,amsfonts,dcolumn}
\usepackage{soul}
\usepackage{authblk} 
\usepackage{natbib} 
\usepackage{url} 
\usepackage{amsmath}
\newcommand{\blind}{0}
\usepackage{enumitem}
\usepackage[doublespacing]{setspace}
\usepackage{bm}
\usepackage{amssymb}
\usepackage[font=scriptsize]{caption} 

\DeclareMathOperator{\Cov}{Cov}
\DeclareMathOperator{\E}{E}

\DeclareMathOperator*{\argmin}{argmin}

\newcolumntype{M}[1]{>{\centering\arraybackslash}m{#1}}

\usepackage{booktabs}
\newcommand{\bigcell}[2]{\begin{tabular}{@{}#1@{}}#2\end{tabular}}

\begin{document}

\def\spacingset#1{\renewcommand{\baselinestretch}%
{#1}\small\normalsize} \spacingset{1}


\if0\blind
{
  \title{\bf Robust Multivariate Functional Control Chart}
   
\author[]{Christian Capezza}
\author[]{Fabio Centofanti \thanks{Corresponding author. e-mail: \texttt{fabio.centofanti@unina.it}}}
\author[]{Antonio Lepore}
\author[]{Biagio Palumbo}

\affil[]{Department of Industrial Engineering, University of Naples Federico II, Piazzale Tecchio 80, 80125, Naples, Italy}

\setcounter{Maxaffil}{0}
\renewcommand\Affilfont{\itshape\small}
\date{}
\maketitle

} \fi

\if1\blind
{
  \begin{center}
    {\LARGE\bf Robust Multivariate Functional Control Chart}
\end{center}
} \fi
\large{\noindent \textbf{This is an original manuscript of an article published by Taylor \& Francis in \textit{Technometrics} on 12 April 2024, available at: \\\url{https://doi.org/10.1080/00401706.2024.2327346}}}
\normalsize
\begin{abstract}
In modern Industry 4.0 applications, a huge amount of data is acquired during manufacturing processes that are often contaminated with anomalous observations  in the form of both casewise and cellwise outliers. 
These can seriously reduce the performance of control charting procedures, especially in complex and high-dimensional settings.
To mitigate this issue in the context of profile monitoring, we propose a new framework, referred to as robust multivariate functional control chart (RoMFCC), that is able to monitor multivariate functional data while being robust to both functional casewise and cellwise  outliers. 
The RoMFCC relies on four main elements: (I) a functional univariate filter to identify functional cellwise outliers to be replaced by missing components; (II) a robust multivariate functional data imputation method of missing values; (III) a casewise robust dimensionality reduction; (IV) a monitoring strategy for the multivariate functional quality characteristic.
An extensive Monte Carlo simulation study is performed to compare the RoMFCC with competing monitoring schemes already appeared in the literature.
Finally, a motivating real-case study is presented where the proposed framework is used to  monitor a resistance spot welding process in the automotive industry.
\end{abstract}

\noindent%
{\it Keywords:}  Functional Data Analysis, Profile Monitoring, Statistical Process Control, Robust Estimation, Casewise and Cellwise Outliers


\spacingset{1.45} 

\section{Introduction}
\label{sec_intro}

Control charts are known as the main tools for statistical process monitoring (SPM), whose aim is to detect when the process is out-of-control (OC), i.e., when special causes of variation act on it. 
On the contrary, when only common causes are present, the process is said to be in control (IC).
The application of control charts in modern industrial applications must however turn into value the massive amounts of data collected at high frequency by modern data acquisition systems. 
Several examples may be found in the current Industry 4.0 framework, which is reshaping the format of measurements that can be gathered in manufacturing processes. 

In many cases, the experimental measurements of the quality characteristic of interest are in fact characterized by complex and high dimensional formats that are well represented as one or multiple functional data \citep{ramsay2005functional,kokoszka2017introduction} also referred to as profiles.
The simplest approach for SPM through one or multiple functional quality characteristics is based on the  extraction of scalar features from each profile observation and the application of classical multivariate SPM techniques \citep{montgomery2012statistical}.
However, the feature extraction step is known to be problem specific, arbitrary, and possibly masking useful information.
Thus, this issue stimulated a growing interest in \textit{profile monitoring} \citep{noorossana2011statistical} that is the monitoring of a process through quality characteristic observations in the form of one or multiple profiles.
Some recent examples of profile monitoring applications can be found in  \cite{menafoglio2018profile,capezza2020control,capezza2021functional_qrei,capezza2021functional_clustering,capezza2022funchartspaper,centofanti2020functional}.

Control charts are currently implemented in two phases. 
The first is referred to as Phase I and is concerned to identify a clean data set to be assumed as representative of the IC state of the process, named Phase I sample or Phase I observations. 
This data set is then used to quantify the expected variation of a future observation to be used for the prospective process monitoring, referred to as Phase II.
Unfortunately, the identification of a Phase I sample in high-dimensional contexts is not an easy task due to the large presence of outlying observations in at least one component of a multivariate functional quality   characteristic.
On the other hand, control charts are very sensitive to the presence of outlying observations in the Phase I sample that inflate control limits and, therefore, reduce the detection  power in Phase II.

\begin{figure}
\begin{center}
\includegraphics[width=\textwidth]{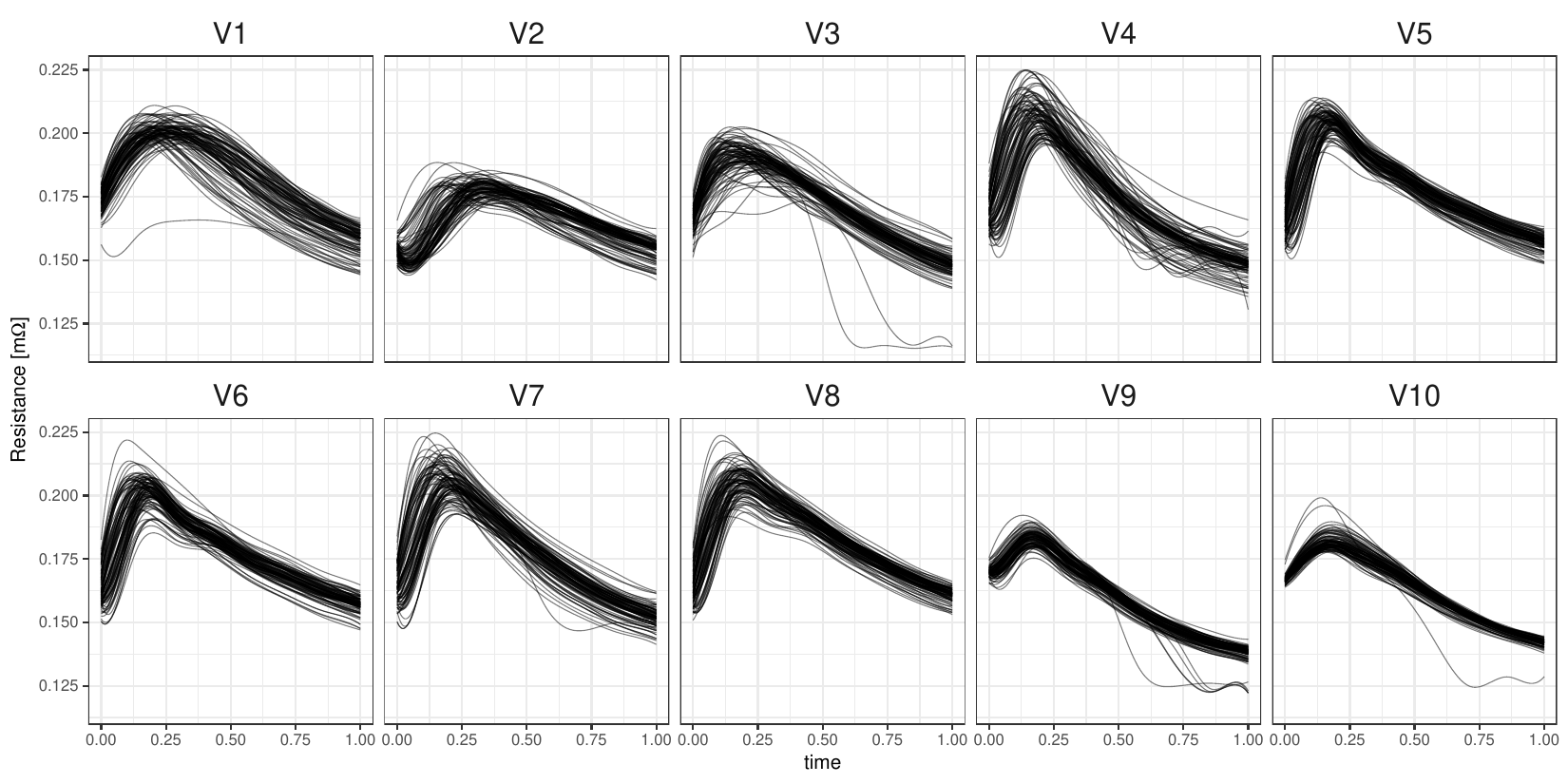}
\caption{Sample of 100 DRCs, measured in $m\Omega$, that are acquired during the RSW process from the real-case study in Section \ref{sec_real}. The different panels refer to the corresponding different spot welds, denoted with names from V1 to V10.}
\label{fig_drc}
\end{center}
\vspace{-1cm}
\end{figure}

In this view, let us consider the real-case study on the SPM of a resistance spot welding (RSW) process in the automotive body-in-white manufacturing that motivated this research and will be thoroughly presented in Section \ref{sec_real}.
RSW is the most common technique employed in joining metal sheets, mainly because it guarantees the structural integrity and solidity of welded items while being adaptable for mass production \citep{martin}.
Among on-line measurements of RSW process parameters, the so-called dynamic resistance curve (DRC) is recognized as the full technological signature of the metallurgical development of a spot weld \citep{dickinson,capezza2021functional_clustering} and, thus, it can be regarded as an in-line low-cost proxy of the RSW process quality with respect to off-line costly destructive tests.
For illustrative purposes, each panel of Figure \ref{fig_drc} displays 100 multivariate DRCs randomly sampled from the real-case study data set, measured in $m\Omega$, acquired during RSW lab tests at Centro Ricerche Fiat (Italy) and refers to the same spot weld locations on 100 different items.
Figure \ref{fig_drc} clearly highlights the motivating challenge of handling outliers possibly occurring in a few components (i.e., panels) and different items. 

SPM methods use two common alternatives to deal with outliers, namely the \textit{diagnostic} and the \textit{robust} approaches \citep{kruger2012statistical,hubert2015multivariate}.
The diagnostic approach is based on standard statistics  after the removal of sample units identified as outliers 
that translates into SPM methods where iterative re-estimation procedures are considered in Phase I. This approach could  be often safely applied to eliminate the eﬀect of a small number of very extreme observations In the Phase I sample.
However, this approach may fail to detect  moderate outliers that are not always as easy to label correctly.
On the contrary, the robust approach  accepts all the data points by means of a robust estimator that reduces the impact of outliers on the final results \citep{maronna2019robust}.
Throughout this paper, the term \textit{robust} means ``outlier resistant" and not ``robust to model misspecification" as it is sometimes used in the literature. Under the robust approach, the terms Phase I sample or Phase I observations will be  used as before to refer to the data set representative of the IC state even though it may be contaminated by outliers.

In the SPM literature, several robust approaches have been proposed for monitoring a  multivariate scalar quality characteristic. 
\cite{alfaro2009comparison} show a comparison of robust alternatives to the classical Hotelling's control chart.
They include two alternative Hotelling's $T^2$-type control charts for individual observations, proposed by \cite{vargas2003robust} and \cite{jensen2007high}, based on the minimum volume ellipsoid and the minimum covariance determinant estimators \citep{rousseeuw1984least}, respectively.
Moreover, the comparison includes the control chart based on the reweighted minimum covariance determinant estimators, proposed by \cite{chenouri2009multivariate}.
More recently, \cite{cabana2021robust} propose a robust Hotelling's $T^2$ procedure using the robust shrinkage reweighted estimator.
To the best of the authors' knowledge, \cite{kordestani2020monitoring} and \cite{moheghi2020phase} are the only ones to propose robust estimators to monitor simple linear profiles, which are not able to capture the functional nature of a multivariate functional quality characteristic.

Beyond the SPM literature, several works have been proposed to deal with outlying functional observations.
The classical L-estimator, which is the linear combination type estimator \citep{maronna2019robust}, is extended to the functional data setting to robustly estimate  the center of a functional distribution through trimming \citep{fraiman2001trimmed,cuesta2006impartial} and  functional data depths \citep{cuevas2009depth,lopez2011half}.
\cite{sinova2018m} introduce in the same setting the notion of maximum-likelihood type estimator, referred to as M-estimator.
 More recently, \cite{centofanti2021rofanova} propose a robust functional analysis of variance (ANOVA) that reduces the weights of outlying functional data on the results of the analysis.

Robust approaches appeared also in functional principal component analysis (FPCA) and are classified by \cite{boente2021robust} in three groups, based on the specific property enjoyed by the resulting principal components.
Methods in the first group perform the eigenanalysis of a robust estimator of the scatter operator, as the spherical principal components method of \cite{locantore1999robust} and the indirect approach of \cite{sawant2012functional}. 
The latter  performs a robust PCA method (e.g., the ROBPCA proposed by \cite{hubert2005robpca}) on the matrix of the basis coefficients corresponding to a basis expansion representation of the  functional data.
The second group contains projection-pursuit approaches as the one proposed by \cite{hyndman2007robust}, that sequentially search for the directions that maximize a robust estimator of the spread of the data projections.
The third group is composed of methods that estimate the principal component space by minimizing a robust reconstruction error measure \citep{lee2013m}.
Additionally, it is worth mentioning diagnostic approaches for functional outlier detection for both  univariate \citep{hyndman2010rainbow,arribas2014shape} and multivariate functional data \citep{hubert2015multivariate,aleman2022depthgram}.

In presence of many functional variables, the curse of dimensionality exacerbates the development of scalable robust approaches.
In fact, traditional multivariate robust estimators assume only a so-called \textit{casewise} contamination model for the data, which consists of a mixture of two distributions, one representing  the majority of the cases that are free of contamination and the second describing the minority of the cases assumed as generated by an unspecified outlier  distribution.
These traditional estimators work well when a small number of cases are contaminated.
However, they are affected by the outlier propagation problem
\citep{alqallaf2009propagation} and may fail when the fraction of perfectly observed cases is small.
Unfortunately, this is very common when the data are high dimensional and outliers are \textit{cellwise}, that is contamination in each variable is independent of the other ones.

Moreover, as pointed out by \cite{agostinelli2015robust}, in the multivariate scalar setting, casewise and cellwise data contamination may occur together.
To overcome this problem, they propose a two steps method. 
In the first step, they use a univariate filter to detect large cellwise outliers and replace them with missing values.
Then, in the second step, a robust estimation specifically designed to deal with missing data is applied to the incomplete data to overcome the casewise contamination.
\cite{leung2016robust} notice however that the univariate filter does not handle moderate-size cellwise outliers well and introduce in the first step of \cite{agostinelli2015robust} a consistent bivariate filter in combination with the univariate filter.
\cite{rousseeuw2018detecting} propose a new method for detecting deviating data cells that takes into account variables correlation, whereas, \cite{tarr2016robust} devise a method for robust estimation of precision matrices under cellwise contamination.

In this paper, we propose a new framework, referred to as robust multivariate functional control chart (RoMFCC), for the SPM of multivariate functional data that is robust to both functional  casewise and cellwise outliers. 
As detailed in Section \ref{sec_method}, to deal with the latter, the proposed framework considers an extension of the filtering approach proposed by \cite{agostinelli2015robust} for univariate functional data and an imputation method  inspired by the robust imputation technique of  \cite{branden2009robust}.
Moreover, it also considers a robust multivariate  functional principal component analysis (RoMFPCA) based on the ROBPCA method \citep{hubert2005robpca}, and a profile monitoring strategy  built on the Hotelling's $T^2$ and the squared prediction error ($SPE$) control charts  \citep{noorossana2011statistical,grasso2016using,centofanti2020functional,capezza2020control,capezza2021functional_qrei,capezza2022funchartspaper}.

A Monte Carlo simulation study is performed in Section \ref{sec_sim} to compare the RoMFCC with competing monitoring schemes that already appeared in the literature based on the ability in Phase II of detecting a process mean shift, where the Phase I sample is contaminated by casewise and cellwise outliers in different scenarios and severity levels. 
The practical applicability of the proposed control chart is illustrated in Section \ref{sec_real} by means of the motivating real-case study introduced above.
Section \ref{sec_conclusions} concludes the article. 
Supplementary Materials are available online and  contain details on data generation in the simulation study and additional simulation results. 
All computations and plots have been obtained using the programming language R \citep{r2021}.

\section{The Robust Multivariate Functional Control Chart Framework}
\label{sec_method}
\begin{figure}[h]
	\caption{Scheme of the  RoMFCC framework.}
	\label{fi_diag}
	
	\centering
	\includegraphics[width=\textwidth]{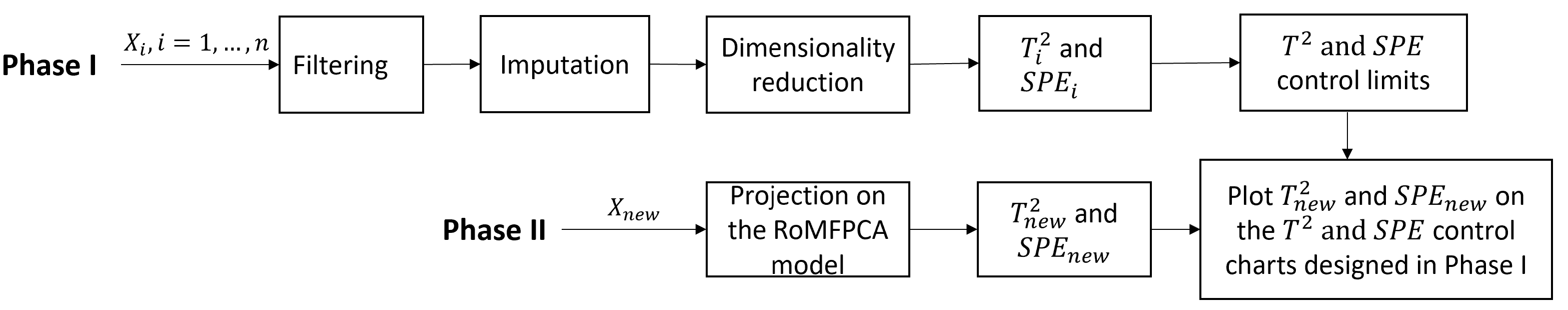}
\vspace{-1cm}
\end{figure}

The proposed RoMFCC is a new general framework for the SPM of multivariate functional data and is able to deal with both Phase I functional casewise and cellwise  outliers. It relies on the following four main elements:
\begin{enumerate}[label=(\Roman*)]
\itemsep-0.2em 
    \item \textit{functional univariate filter} is used to identify functional cellwise outliers and replace them by missing component values;
    \item \textit{robust multivariate functional data imputation} to replace missing values;
     \item \textit{casewise robust dimensionality reduction} to reduce the infinite dimensionality of the  multivariate functional data that is robust to casewise outliers;
      \item \textit{monitoring strategy} of the process through the multivariate functional data  obtained as the output of the previous steps.
\end{enumerate}
These elements are combined in the proposed RoMFCC framework in a Phase II monitoring strategy described in the scheme of Figure \ref{fi_diag}, where a set of Phase I observations $ \bm{X}_i $, $ i=1,\dots,n $ of the multivariate functional quality characteristic $\bm{X}$, which can be contaminated with outliers, is used for the design of the control chart.

\noindent In what follows, we describe a specific implementation of the RoMFCC framework.
In the \textit{filtering} step, we propose a (I) functional  univariate filter, which is named as FUF and is an extension of the filtering stage of \cite{agostinelli2015robust} to identify functional cellwise outliers and replace them by missing components.
and, then, replaced by  missing components. 
In the \textit{imputation} step, these are then imputed through a new (II) robust functional data imputation, which is named RoMFDI and is based on the technique of \cite{branden2009robust}.
Once the imputed Phase I sample is obtained, it is used to  estimate the RoMFPCA model, which is considered as the casewise robust dimensionality reduction method, and perform the (III)  \textit{casewise robust dimensionality reduction} step.
The Hotelling's $ T^2 $ and $ SPE $ statistics of the  (IV)  monitoring strategy  as well as their control limits are, then, computed.
In Phase II, each new observation  $X_{new}$ is projected on the  RoMFPCA model  to compute the values of the Hotelling's $ T^2 $ and $ SPE $ statistics according to the model identified in Phase I and referred to as $T^2_{new}$ and $SPE_{new}$, respectively.
An alarm signal is issued if either $ T^2_{new} $ or $ SPE_{new} $  violates the corresponding control limit.
Note that, when the sample size $n$ is small compared to the number of process variables, an undesirable effect upon the performance of the RoMFCC could arise \citep{ramaker2004effect,kruger2012statistical}. 
Strictly speaking, to prevent overfitting issues that could reduce the monitoring performance of the RoMFCC, the Phase I sample is randomly split into non-overlapping sets, referred to as \textit{training} and \textit{tuning} sets.
The former is used to estimate the RoMFPCA model, whereas the latter is considered to estimate the $ T^2 $ and $ SPE $ control limits.

The RoMFPCA is presented in Section \ref{sec_RoMFPCA}. Then, Section \ref{sec_univfilter}, Section \ref{sec_dataimpu}, and, Section \ref{sec_monstr}  describe the FUF, the RoMFDI method, and the monitoring strategy, respectively. 

\subsection{Robust Multivariate Functional Principal Component Analysis}
\label{sec_RoMFPCA}

Let $L^2(\mathcal{T})$ denote the Hilbert space of square integrable functions defined on the compact set $\mathcal T \in \mathbb{R}$, with the inner product of two functions $f,g \in L^{2}\left(\mathcal{T}\right)$ given by $\langle f,g\rangle=\int_{\mathcal{T}}f\left(t\right)g\left(t\right)dt$, and the norm $\lVert \cdot \rVert=\sqrt{\langle \cdot,\cdot\rangle}$.
Let $\bm{X}=\left(X_1,\dots, X_p\right)^{T}$ be a random vector with realizations in the Hilbert space $\mathbb{H}$ of $p$-dimensional vectors of $L^2(\mathcal{T})$ functions, with the inner product of two function vectors $\mathbf{f}=\left(f_1,\dots,f_p\right)^{T}$ and $\mathbf{g}=\left(g_1,\dots,g_p\right)^{T}$ in $\mathbb{H}$ given by $\langle \mathbf{f},\mathbf{g} \rangle _{\mathbb{H}}=\sum_{j=1}^{p}\langle f_j,g_j\rangle$ and the norm $\lVert \cdot \rVert_{\mathbb{H}}=\sqrt{\langle \cdot,\cdot\rangle_{\mathbb{H}}}$.

We assume that $\bm{X}$ has mean $\bm{\mu}=\left(\mu_1,\dots,\mu_p\right)^T$, where $\mu_j(t)=\E(X_j(t))$, $j = 1,\dots, p$, $t\in\mathcal{T}$ and covariance $\bm{G}=\lbrace G_{jk}\rbrace_{1\leq j,k \leq p}$, $G_{jk}(s,t)=\Cov(X_j(s),X_k(t))$, $s,t\in \mathcal{T}$.
In what follows, differences in variability and unit of measurement among $X_1,\dots, X_p$, are taken into account by using the transformation approach of \cite{chiou2014multivariate}, that is we replace $\bm X$ with the vector of standardized  variables $\bm{Z}=\left(Z_1,\dots,Z_p\right)^T$, where $Z_j(t)=v_j(t)^{-1/2}(X_j(t)-\mu_j(t))$,  with $v_j(t)=G_{jj}(t,t)$, $j=1,\dots,p$, $t\in\mathcal{T}$.   Then, from the multivariate Karhunen-Lo\`{e}ve's theorem \citep{happ2018multivariate} it follows that
\begin{equation*}
    \bm{Z}(t)=\sum_{l=1}^{\infty} \xi_l\bm{\psi}_l(t),\quad t\in\mathcal{T},
\end{equation*}
where $\xi_l=\langle \bm{\psi}_l, \bm{Z}\rangle_{\mathbb{H}} $ are random variables, say \textit{principal components scores} or simply \textit{scores}, such that  $\E\left( \xi_l\right)=0$ and $\E\left(\xi_l \xi_m\right)=\lambda_{l}\delta_{lm}$, with $\delta_{lm}$ the Kronecker delta.
The elements of the orthonormal set $\lbrace \bm{\psi}_l\rbrace $, $\bm{\psi}_l=\left(\psi_{l1},\dots,\psi_{lp}\right)^T$, with $\langle \bm{\psi}_l,\bm{\psi}_m\rangle_{\mathbb{H}}=\delta_{lm}$, are referred to as \textit{principal components}, and are the  eigenfunctions  of the covariance $\bm{C}$  of $\bm{Z}$ corresponding to the eigenvalues $\lambda_1\geq\lambda_2\geq \dots\geq 0$.
Following the approach of \cite{ramsay2005functional}, the eigenfunctions and eigenvalues of the covariance $\bm{C}$ are estimated through a basis function expansion approach.
Specifically, we assume that each function $Z_j$ and eigenfunction $\bm \psi_l$ of $\bm{C}$ with components $\psi_{lj}$, for $j=1,\dots,p$, $l=1,\dots,\infty$, can be approximated by the following finite sums of $K$ terms
 \begin{equation}
 \label{eq_appcov}
    Z_j(t)\approx \sum_{k=1}^{K} c_{jk}\phi_{jk}(t), 
    \quad 
    \psi_{lj}(t)\approx \sum_{k=1}^{K} b_{ljk}\phi_{jk}(t), \quad t\in\mathcal{T}.
\end{equation}
That is, $Z_j$ and $\psi_{lj}$ are approximated as linear combinations of the components of a $K$-dimensional vector of basis functions $\bm{\phi}_j=\left(\phi_{j1},\dots,\phi_{jK}\right)^T$, with coefficient vectors $\bm{c}_j=\left(c_{j1},\dots,c_{jK}\right)^T$ and $\bm{b}_{lj}=\left(b_{lj1},\dots,b_{ljK}\right)^T$, respectively. 
With these assumptions, standard multivariate functional principal component analysis \citep{ramsay2005functional,chiou2014multivariate} estimates eigenfunctions and eigenvalues of the covariance $\bm{C}$  by  performing standard multivariate principal component analysis on the random vector $\bm{W}^{1/2}\bm{c}$, where $\bm{c}=\left(\bm{c}_1^T,\dots,\bm{c}_p^T\right)^T$ and $\bm{W}$ is a block-diagonal matrix with diagonal blocks  $\bm{W}_j$, $j=1,\dots,p$, whose entries are $\langle\phi_{jk_1},\phi_{jk_2}\rangle$, $k_1,k_2=1,\dots,K$. 
Then, the eigenvalues $\lambda_l$ of $\bm{C}$ are estimated by those of the covariance matrix of $\bm{W}^{1/2}\bm{c}$, whereas, the components $\psi_{l1},\dots,\psi_{lp}$ of the corresponding eigenfunction $\bm \psi_l$ are estimated through Equation \eqref{eq_appcov}
with $\bm{b}_{lj}=\bm{W}^{-1/2}\bm{u}_{lj}$, where $\bm{u}_l=\left(\bm{u}_{l1}^T,\dots,\bm{u}_{lp}^T\right)^T$ is the eigenvector of the covariance matrix of $\bm{W}^{1/2}\bm{c}$ corresponding to $\lambda_l$. 
However, it is well known that standard multivariate  principal component analysis is not robust to outliers \citep{maronna2019robust}, which obviously reflects on the functional principal component analysis by probably providing misleading results. Extending the approach of \cite{sawant2012functional} for multivariate functional data, the proposed RoMFPCA applies a robust principal component analysis alternative to the random vector $\bm{W}^{1/2}\bm{c}$. Specifically, we consider the ROBPCA approach of \cite{hubert2005robpca}, which is a computationally efficient method explicitly conceived to produce estimates with a high breakdown in high dimensional data settings and is commonly adopted in the functional context to handle a large percentage of contamination. 
Thus, given $n$ independent realizations $\bm{X}_1, \dots, \bm{X}_n$ of $\bm{X}$, dimensionality reduction is achieved by approximating  $\bm{X}_i$ through $\hat{\bm{X}}_i$, for $i=1,\dots,n$, as
\begin{equation}
\label{eq_appx}
\hat{\bm{X}}_i(t)= \hat{\bm{\mu}}(t)+\hat{\bm{D}}(t)\sum_{l=1}^{L} \hat{\xi}_{il}\hat{\bm{\psi}}_l(t) \quad t\in\mathcal{T}
\end{equation}
where $\hat{\bm{D}}$ is a diagonal matrix whose diagonal entries are robust estimates $\hat{v}_j^{1/2}$ of $v_j^{1/2}$, $\hat{\bm{\mu}}=\left(\hat{\mu}_1,\dots,\hat{\mu}_p\right)^T$ is a robust estimate of $\bm{\mu}$, $\hat{\bm{\psi}}_l$, $l=1,\dots,L$, are the first $L$  robustly estimated principal components and $\hat{\xi}_{il}= \langle \hat{\bm{\psi}}_l, \hat{\bm{Z}}_i\rangle_{\mathbb{H}}$ are the corresponding scores with robustly estimated variances $\hat{\lambda}_l$. 
The estimates  $\hat{\bm{\psi}}_l$ and $\hat{\lambda}_l$ are obtained through the $n$ realizations of $\bm{Z}_i$ estimated as $\hat{\bm Z_i} = \hat{v}_j^{-1/2} (\bm X_i - \hat{\bm{\mu}}_i)$ by using $\hat{\mu}_j$ and $\hat{v}_j$.
The robust estimates $\hat{\mu}_j$ and $\hat{v}_j$ are obtained through the scale equivariant functional $M$-estimator and the functional normalized median absolute deviation estimator proposed by  \cite{centofanti2021rofanova}.
As in the multivariate setting, $L$ is generally chosen such that the retained principal components $\hat{\bm{\psi}}_1,\dots, \hat{\bm{\psi}}_L$ explain at least a given percentage of the total variability, which is usually in the range 70-90$\%$. 
However, more sophisticated methods could be used as well (see \cite{jolliffe2011principal} for further details).

\subsection{Functional Univariate Filter}
\label{sec_univfilter}
To extend the filter of \cite{agostinelli2015robust} and \cite{leung2016robust} to univariate functional data, let us consider $n$ independent realizations $\bm{X}_1, \dots, \bm{X}_n$ of a random function $X$ whose realizations are in $L^2(\mathcal{T})$. The proposed FUF considers the functional distances
\begin{equation}
\label{eq_dfil}
    D_{i}^{\text{fil}}=\sum_{l=1}^{L^{\text{fil}}} 
    \frac{(\hat{\xi}_{il}^{\text{fil}})^2}{\hat{\lambda}_{l}^{\text{fil}}}, 
    \quad i=1,\dots,n,
\end{equation}
where the estimated scores $\hat{\xi}_{il}^{\text{fil}}= \langle \hat{{\psi}}_{l}^{\text{fil}}, \hat{{Z}}_i\rangle$, eigenvalues $\hat{\lambda}_{l}^{\text{fil}}$, principal components $\hat{{\psi}}_{j}^{\text{fil}}$, and standardized observations $\hat{{Z}}_i$  of ${X}_{i}$ are obtained by applying, with $p=1$, the proposed RoMFPCA described in Section \ref{sec_RoMFPCA} to the sample $X_1, \dots, X_n$. 
In this setting, the RoMFPCA is suitably used to represent distances among $X_i$'s and not to perform dimensionality reduction, thus $L^{\text{fil}}$ should be sufficiently large to  capture a large percentage of the total variability $\delta^{\text{fil}}$.
Let $G_n$ be the empirical cumulative distribution function (cdf) of $D_{i}^{\text{fil}}$, i.e.,
\begin{equation*}
    G_n(x)=\frac{1}{n}\sum_{i=1}^n I(D_{i}^{\text{fil}}\leq x), \quad x\geq 0,
\end{equation*}
where $I(\cdot)$ denotes the indicator function.
Then, a functional observation $X_i$ is labeled as a cellwise outlier by comparing $G_n(x)$ with $G(x)$, $x \geq 0$, where $G$ is the reference cdf for $D_{i}^{\text{fil}}$. Following \cite{leung2016robust}, we consider $D_{i}^{\text{fil}}$ distributed as a chi-squared random variable with $L^{\text{fil}}$ degrees of freedom, i.e., $D_{i}^{\text{fil}} \sim \chi^2_{L^{\text{fil}}}$. 
The proportion of flagged cellwise outliers is defined by 
\begin{equation*}
d_n=\sup_{x\geq \eta}\lbrace G(x)-G_n(x)\rbrace^{+},
\end{equation*}
where $\lbrace a\rbrace^{+}$ represents the positive part of $a$, and $\eta=G^{-1}(\alpha)$ is a large quantile of $D_{i}^{\text{fil}}$. 
As \cite{agostinelli2015robust}, we set $\alpha=0.95$.
Finally, we flag $\lfloor nd_n\rfloor$ observations with the largest functional distances $D_{i}^{\text{fil}}$ as functional cellwise outliers, where $\lfloor \cdot \rfloor$ denotes the floor function. 
From the arguments in \cite{agostinelli2015robust} and \cite{leung2016robust} the proposed FUF is consistent even when the actual distribution of $D_{i}^{\text{fil}}$ is unknown. That is, when the tail of $G$ is heavier than or equal to that of the actual unknown distribution, it will flag a cellwise outlier asymptotically correctly.

\subsection{Robust Multivariate Functional Data Imputation}
\label{sec_dataimpu}

Let us consider the setting where the $n$ independent realizations $\bm{X}_i = \left(X_{i1},\dots,X_{ip}\right)^T$, $i=1,\dots,n$, of the multivariate functional quality characteristic $\bm{X}$ may present missing components, i.e., for some $i$, at least one among $X_{i1},\dots,X_{ip}$ is missing. 
Let us assume that there is a set $S_c$  of $c < n$ realizations with no missing component.
As the robust imputation approach of \cite{branden2009robust}, the RoMFDI is a sequential imputation method. 
It starts from any observation $\bm{X}_{\underline{i}}\notin S_c$ having the smallest number, say $s$, of missing components.
For notational convenience and without loss of generality, let us arrange the $p$ components of $\bm{X}_{\underline{i}} = ((\bm{X}_{\underline{i}}^{\text{m}})^T, (\bm{X}_{\underline{i}}^{\text{o}})^T)^T$ such that the first $s$ missing components are in the vector $\bm{X}_{\underline{i}}^{\text{m}}$, while the remaining $p-s$ observed ones are in $\bm{X}_{\underline{i}}^{\text{o}}$.
The standardized version of $\bm{X}_{\underline{i}}$ is partitioned as $\hat{\bm{Z}}_{\underline{i}} = ((\hat{\bm{Z}}_{\underline{i}}^{\text{m}})^T, (\hat{\bm{Z}}_{\underline{i}}^{\text{o}})^T)^T$, where $\hat{\bm{Z}}_{\underline{i}}^{\text{m}}$ and $\hat{\bm{Z}}_{\underline{i}}^{\text{o}}$ are the standardized versions of $\bm{X}_{\underline{i}}^{\text{m}}$ and $\bm{X}_{\underline{i}}^{\text{o}}$, respectively. 
By Equation \eqref{eq_appcov}, $\hat{\bm{Z}}_{\underline{i}}$ is uniquely identified by the $Kp$-dimensional coefficient vector $\bm{c}_{\underline{i}} = ((\bm{c}_{\underline{i}}^{\text{m}})^T, (\bm{c}_{\underline{i}}^{\text{o}})^T)^T$ related to the basis expansions of its $p$ components, where $K$ is the number of basis functions as in Equation \eqref{eq_appcov}.
Therefore, the imputation of $\bm{X}_{\underline{i}}^{\text{m}}$ reduced to the imputation of $\bm{c}_{\underline{i}}^{\text{m}}$.
We do this by minimizing the distance of $\bm{X}_{\underline{i}}$ from the space generated by the observations in $S_c$
\begin{equation}
\label{eq_imp1}
    \sum_{l=1}^{L^{\text{imp}}} \frac{(\hat{\xi}_{\underline{i}l}^{\text{imp}})^2}{\hat{\lambda}_{l}^{\text{imp}}},
\end{equation}
where $\hat{\xi}_{\underline{i}l}^{\text{imp}} = \langle \hat{\bm{\psi}}_l^{\text{imp}}, \hat{\bm{Z}}_{\underline i}\rangle_{\mathbb{H}}$, with principal components $\hat{\bm{\psi}}_{l}^{\text{imp}} = (\hat \psi_{l1}^{\text{imp}}, \dots, \psi_{lp}^{\text{imp}})^T$ and corresponding eigenvalues $\hat \lambda_l^{\text{imp}}$ obtained by applying RoMFPCA on the complete realizations in $S_c$, as in Section \ref{sec_RoMFPCA}. 
The number $L^{\text{imp}}$ of components is chosen sufficiently large to capture the desired percentage of the total variability $\delta^{\text{imp}}$. 
Since principal components $\hat{\psi}_{lj}^{\text{imp}}$ are uniquely identified by their basis coefficients $\hat{\bm{b}}_{lj}$, $l=1,\dots, L^{\text{imp}}$, $j=1,\dots, p$, which can be arranged into the matrix $\hat{\bm{B}}$ having columns $\hat{\bm{b}}_l=\left(\hat{\bm{b}}_{l1}^T,\dots,\hat{\bm{b}}_{lp}^T\right)^T$, the objective function \eqref{eq_imp1} can be calculated as $\bm{c}_{\underline{i}}^T \bm{C} \bm{c}_{\underline{i}}$, where the $Kp \times Kp$ matrix $\bm{C}$ is
\begin{equation*}
    \bm{C}=\bm{W}\hat{\bm{B}}\hat{\bm{\Lambda}}^{-1}\hat{\bm{B}}^T\bm{W}
    =
    \begin{pmatrix}
    \bm{C}^{\text{m,m}} & \bm{C}^{\text{m,o}} \\
    (\bm{C}^{\text{m,o}})^T & \bm{C}^{\text{o,o}} \\
    \end{pmatrix},
\end{equation*}
where $\bm{W}$ is the block-diagonal matrix defined in Section \ref{sec_RoMFPCA} and $\hat{\bm{\Lambda}}$ the diagonal matrix  whose diagonal entries are $\hat{\lambda}_{l}^{\text{imp}}$, $l=1,\dots,L^{\text{imp}}$.
The upper left $Ks \times Ks$ block $\bm{C}^{\text{m,m}}$ of $\bm C$ contains the elements of $\bm{C}$ corresponding to the missing components, while $\bm{C}^{\text{m,o}}$ represents the part of $\bm C$ with the missing components in the rows and the observed components in the columns. 
The solution that minimizes the objective function \eqref{eq_imp1} with respect to $\bm{c}_{\underline{i}}^{\text{m}}$ is
\begin{equation}
\label{eq_modim}
    \hat{\bm{c}}_{\underline{i}}^{\text{m}}= \argmin_{\bm{c}_{\underline{i}}^{\text{m}}} \left( \left( \bm{c}_{\underline{i}}^{\text{m}} \right)^T, \left( \bm{c}_{\underline{i}}^{\text{o}} \right)^T \right)^T \bm C \left( \left( \bm{c}_{\underline{i}}^{\text{m}} \right)^T, \left( \bm{c}_{\underline{i}}^{\text{o}} \right)^T\right)^T = -\left( \bm{C}^{\text{m,m}}\right)^{+}\bm{C}^{\text{m,o}}\bm{c}_{\underline{i}}^{\text{o}},
\end{equation}
where $(\bm{C}^{\text{m,m}})^{+}$ is the Moore-Penrose inverse of $\bm{C}^{\text{m,m}}$.

In order to overcome the correlation bias issue typical of  deterministic imputation approaches \citep{little2019statistical,van2018flexible}, we propose a stochastic imputation method by imputing $\bm{c}_{\underline{i}}^{\text{m}}$ with 
\begin{equation}
\label{eq_imperr}
    \bm{c}_{\underline{i}}^{\text{imp}}=\hat{\bm{c}}_{\underline{i}}^{\text{m}} + \bm{\varepsilon}_{\underline{i}},
\end{equation}
where $\bm{\varepsilon}_{\underline{i}}$ is a multivariate normal random variable with mean zero and covariance matrix robustly estimated (e.g., using the Rocke type estimator \citep{rocke1996robustness}) based on the regression residuals of the coefficient vectors of the missing component on those of the observed components using the observations in $S_c$ through the model in Equation \eqref{eq_modim}.
Accordingly, the components of $\hat{\bm{Z}}_{\underline{i}}^{\text{m}}$ are imputed with $\hat{\bm{Z}}_{\underline{i}}^{\text{imp}}$, whose $j$-th component is
\begin{equation*}
\hat{Z}_{\underline{i}j}^{\text{imp}} (t)
=
\left(\bm{c}_{\underline{i}j}^{\text{imp}}\right)^T\bm{\phi}_j(t),\quad  t\in\mathcal{T}, \quad j=1, \dots, s,
\end{equation*}
where $\bm{c}_{\underline{i}j}^{\text{imp}}$ is the vector of the elements of $\bm{c}^{\text{imp}}_{\underline{i}}$ corresponding to the $j$-th component of $\hat{\bm{Z}}_{\underline{i}}$ and $\bm{\phi}_{j}$ is defined as in Equation \eqref{eq_appcov}. 
The imputed values of $\bm{X}_{\underline{i}}^\text{m}$ are finally obtained by unstandardizing $\hat{\bm{Z}}_{\underline{i}}^{\text{imp}}$, and $\bm{X}_{\underline{i}}$ is then added to $S_c$.
The whole imputing step described for $\bm{X}_{\underline{i}}$ is repeated iteratively until all missing observations are imputed.
Observations where all components are missing, i.e., $s=p$, are trivially removed from the sample because their imputation does not provide any additional information for the analysis.

However, similarly to \cite{branden2009robust}, if the cardinality of $S_c$ at the first iteration is sufficiently large, we suggest not updating the RoMFPCA model at each iteration. 
To take into account the increased noise due to single imputation, the proposed RoMFDI can be easily included in a multiple  imputation framework \citep{van2018flexible,little2019statistical}.
It is worth explicitly noting that, due to the presence of the stochastic component $\bm{\varepsilon}_{\underline{i}}$ in Equation \eqref{eq_imperr}, the imputed data set is not deterministically assigned. 
Therefore, by performing several times the RoMFDI in the imputation step of the RoMFCC implementation, the corresponding multiple estimated RoMFPCA models could be combined by averaging the robustly estimated  covariance functions, thus performing a multiple imputation strategy as suggested by \cite{van2007two}.

\subsection{The Monitoring Strategy}
\label{sec_monstr}
The last (IV) element of the proposed RoMFCC implementation relies on the consolidated monitoring strategy for  a multivariate functional quality characteristic $\bm{X}$ based on  Hotelling's $ T^2 $ and $ SPE $ control charts. The former  assesses the stability of $ \bm{X} $  in the finite dimensional space spanned by the first principal components identified through the RoMFPCA (Section \ref{sec_RoMFPCA}), whereas, the latter monitors changes along directions in the complement space.  
Specifically, the Hotelling's $ T^2 $ statistic for $\bm{X}$ is defined as 
\begin{equation*}
	T^2=\sum_{l=1}^{L^{\text{mon}}}\frac{(\xi_{l}^{\text{mon}})^2}{\lambda_{l}^{\text{mon}}},
\end{equation*}
where $\lambda_{l}^{\text{mon}} $ are the variances of the scores $ \xi_{l}^{\text{mon}}=\langle \bm{\psi}_{l}^{\text{mon}}, \bm{Z}\rangle_{\mathbb{H}}$, $\bm{Z}$ is the vector of standardized variable corresponding to $\bm{X}$ and $\bm{\psi}_{l}^{\text{mon}}$ is the vector of principal components as defined in Section \ref{sec_RoMFPCA}.
 The number $L^{\text{mon}}$  is  chosen such that the retained principal components  explain at least a given percentage $\delta^{\text{mon}}$ of the total variability.
The statistic $ T^2$ is the standardized squared distance from the center of the orthogonal projection of  $\bm{Z}$ onto the principal component space spanned by  $ \bm{\psi}_{1}^{\text{mon}},\dots,\bm{\psi}_{L^{\text{mon}}}^{\text{mon}}$.
Whereas, the distance between $ \bm{Z} $  and its orthogonal projection onto the principal component space is measured through the $ SPE $ statistic, defined  as 
\begin{equation*}
	SPE=||\bm{Z}-\hat{\bm{Z}}||_{\mathbb{H}}^2,
\end{equation*}
where  $ \hat{\bm{Z}}=\sum_{l=1}^{L^{\text{mon}}}\xi_{l}^{\text{mon}}\bm{\psi}_{l}^{\text{mon}}$. 

Under the assumption of multivariate normality of $\xi_{l}^{\text{mon}}$, which is approximately true by the central limit theorem \citep{nomikos1995multivariate}, the control limits of the Hotelling's $ T^2 $ control chart can be obtained by considering the $ (1-\alpha^*) $ quantiles of chi-squared distribution with $L^{\text{mon}}$ degrees of freedom \citep{johnson2014applied}.
Whereas, the control limit for the $ SPE $ control chart  can be computed by using the following equation \citep{jackson1979control}
\begin{equation*}
CL_{SPE,\alpha^*}=\theta_1\left[\frac{c_{\alpha^*}\sqrt{2\theta_2h_0^2}}{\theta_1}+1+\frac{\theta_2h_0(h_0-1)}{\theta_1^2}\right]^{1/h_0}
\end{equation*}
where $c_{\alpha^*}$ is the $ (1-\alpha^*) $-quantile of the standard normal distribution, $h_0=1-2\theta_1\theta_3/3\theta_2^2$, $\theta_j=\sum_{l=L^{\text{mon}}+1}^{\infty}(\lambda_{l}^{\text{mon}})^j$, $j=1,2,3$.
Note that $ \alpha^* $ should be appropriately chosen to control the family wise error rate (FWER) denoted by $\alpha$. 
We use the \v{S}id\'ak correction  $\alpha^{*}=1-\left(1-\alpha\right)^{1/2}$ \citep{vsidak1967rectangular}, which is exact for independent tests.

\section{Simulation Study}
\label{sec_sim}
The performance of the RoMFCC in identifying mean shifts of the multivariate functional quality characteristic is assessed through a Monte Carlo simulation of two scenarios with different Phase I sample contamination and data generation process (detailed in Supplementary Material A) inspired by typical behaviors of DRCs appeared  in the real-case study of Section \ref{sec_real}. 
Specifically, in Scenario 1 and Scenario 2, the Phase I sample is contaminated by functional cellwise and casewise outliers, respectively, with a contamination probability equal to 0.05  in both cases.
Supplementary Material B reports  additional results obtained  with  a contamination probability equal to 0.1.
For each scenario, two contamination models, referred to as Out-E and Out-P, with three increasing contamination levels, referred to as C1, C2, and C3, are considered to mimic typical outlying DRCs. Specifically, Out-E  mimics a splash weld (expulsion) caused by excessive welding current, while Out-P resembles the phase shift of the peak time caused by an increased force applied to the electrode used in the welding process \citep{xia2019online}.
A further Scenario 0, simulates a Phase I sample  not contaminated by any type of outliers.
The Phase II sample is generated with two types of OC conditions at 4 different severity levels $SL = \lbrace 1, 2, 3, 4 \rbrace$, referred to as OC-E and OC-P coherently with contamination models Out-E and Out-P, respectively.

The proposed RoMFCC framework is compared with several natural competing approaches grouped into control charts for multivariate non-functional and functional data. The first group consists of control charts for multivariate data built on the mean vector of each component of multivariate functional data observations. 
In this group, we consider ($i$) the classical \textit{multivariate}  Hotelling's $T^2$ control chart, referred to as MCC; ($ii$) its \textit{iterative} variant, referred to as iterMCC, where outliers detected by the control chart in Phase I are iteratively removed and control limits are revised until all data are assumed to be IC; ($iii$) the \textit{multivariate robust} control chart proposed by \cite{chenouri2009multivariate} and referred to as RoMCC. 
In the second group, we consider two  approaches recently appeared in the profile monitoring literature, namely ($iv$) the \textit{multivariate functional control chart} and referred to as MFCC, proposed by \cite{capezza2020control}; ($v$) its \textit{iterative} variant, referred to as iterMFCC, where, as before, outliers detected in Phase I are iteratively removed until all data are assumed to be IC.

The RoMFCC is implemented as described in Section \ref{sec_method} with $\delta_{fil}=\delta_{imp}=0.999$ and $\delta_{mon}=0.7$, and, to take into account the noise increase  due to single imputation, five differently imputed datasets are generated through RoMFDI.
While data are observed through noisy discrete values, each component of the generated quality characteristic observations is obtained by  Equation \eqref{eq_appcov} with $K=10$ cubic B-splines estimated through the spline smoothing approach with a roughness penalty on the second derivative \citep{ramsay2005functional}.
For each Phase I sample generation scenario and Phase II OC condition setting, 50 simulation runs are performed. 
Each run considers a Phase I random sample of 4000  observations. In particular,  for MFCC, iterMFCC and RoMFCC, 1000 are used as a training set, and the remaining 3000 are used as a tuning set. 
The Phase II sample is  also composed of a random sample of 4000 observations.
The RoMFCC and the performance of the competing method are assessed through the true detection rate (TDR) and the false alarm rate (FAR), which are estimated as the average proportion, over the simulation runs, of points that fall outside the control limits whilst the process is, respectively, OC or IC.
The latter are hereinafter referred to as mean TDR and mean FAR, respectively.
The mean FAR should be as similar as possible to the overall type I error probability $\alpha$ considered to obtain the control limits  and set equal to 0.05 in this simulation study, whereas the mean TDR should be as large as possible.

\begin{figure}[h]
	\caption{Mean FAR ($ SL=0 $) or TDR ($ SL\neq 0 $) achieved by MCC, iterMCC, RoMCC, MFCC, iterMFCC and RoMFCC for each OC condition (OC-E and OC-P) as a function of the severity level $SL$ in Scenario 0.
	All the functional approaches (namely MFCC, iterMFCC, RoMFCC) plot equally and above the non-functional approaches (namely MCC, iterMCC, RoMCC) that also plot indistinctly.}
	
	\label{fi_results_1}
	\centering
		\begin{tabular}{cc}
		\hspace{0.6cm}\textbf{\footnotesize{OC-E}}&\hspace{0.6cm}\textbf{\footnotesize{OC-P}}\\
			\includegraphics[width=.25\textwidth]{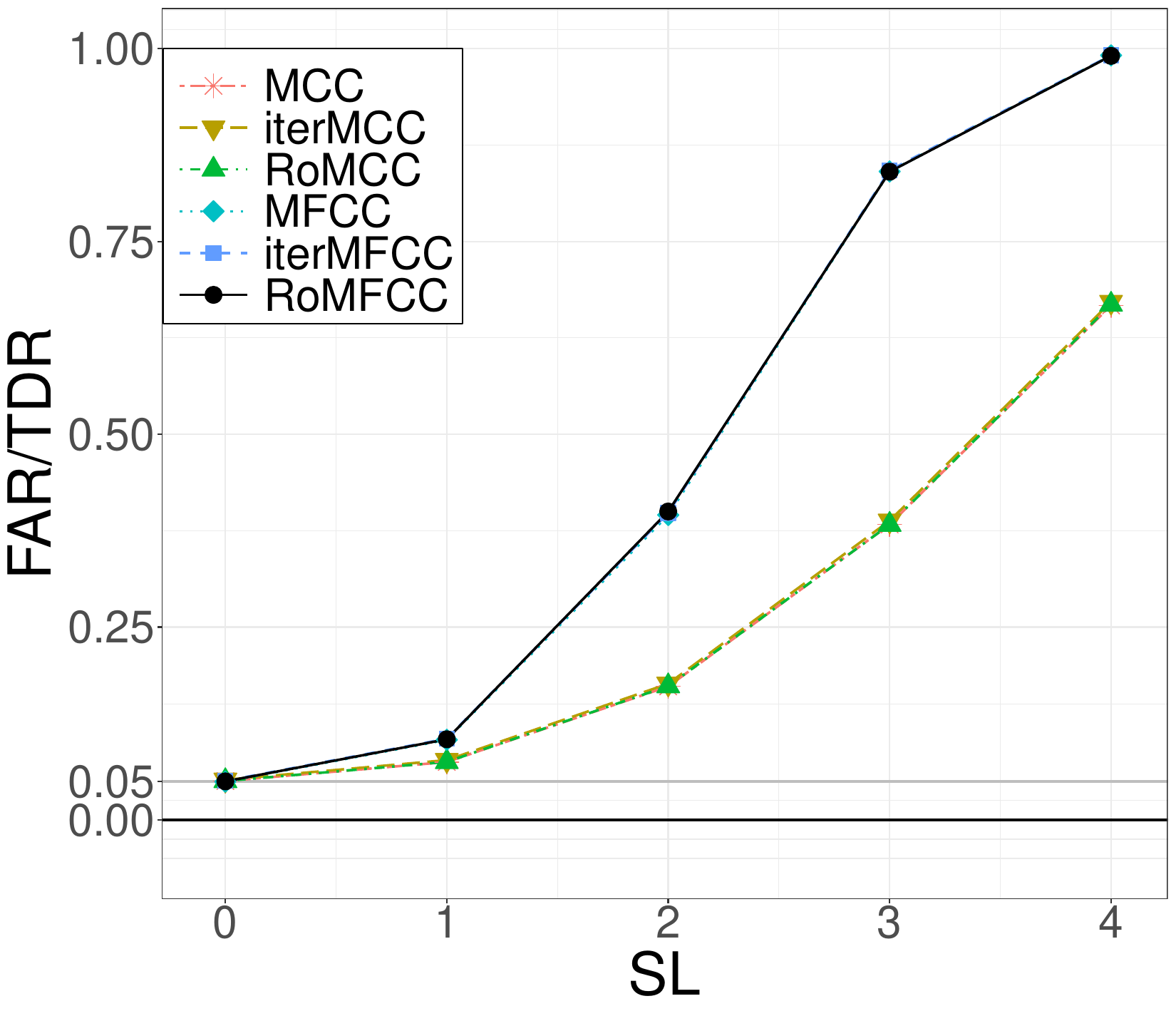}&\includegraphics[width=.25\textwidth]{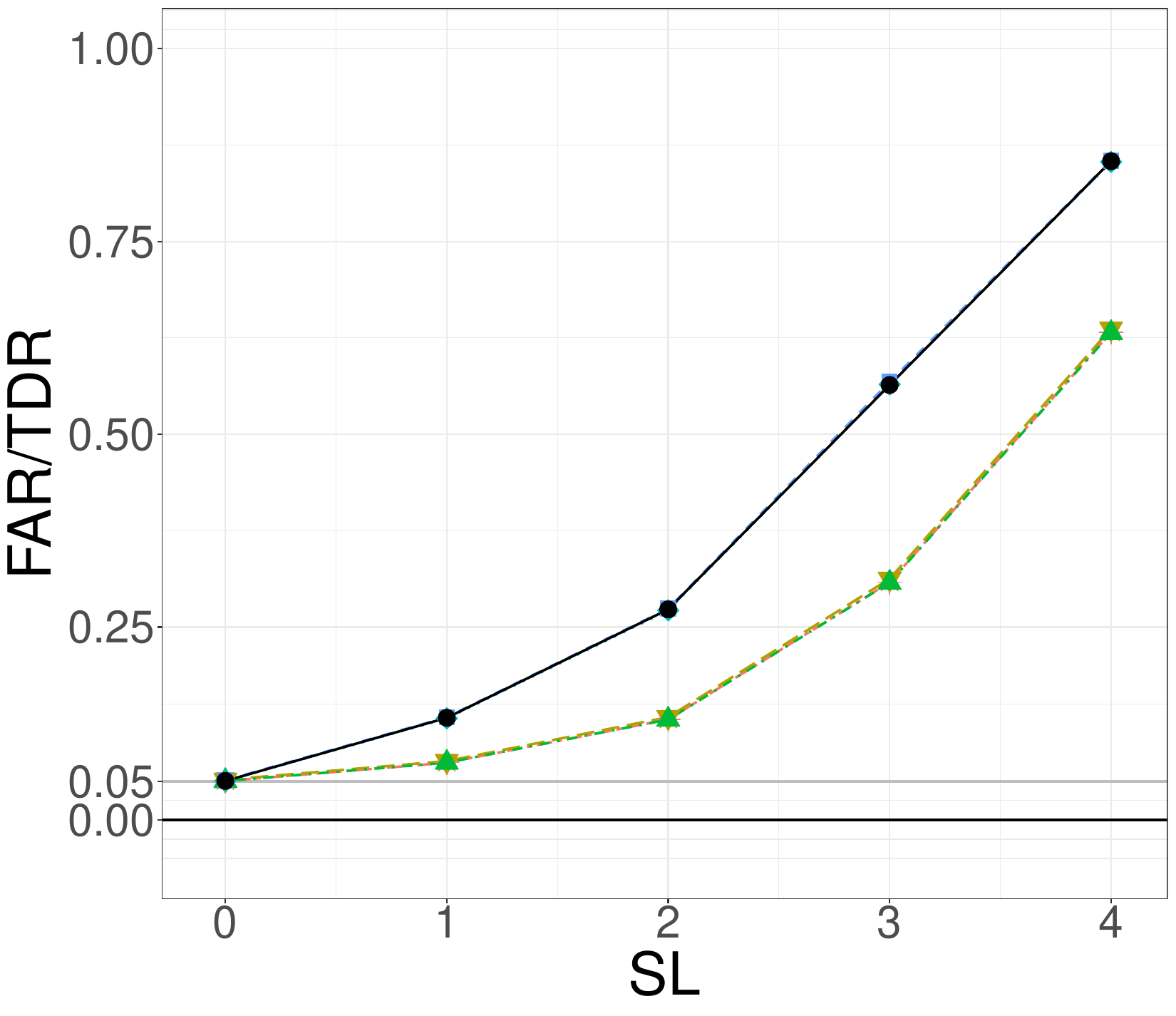}
	\end{tabular}
	
		\vspace{-.5cm}
\end{figure}

\begin{figure}[h]
	\caption{Mean FAR ($ SL=0 $) or TDR ($ SL\neq 0 $) achieved in Phase II by MCC, iterMCC, RoMCC, MFCC, iterMFCC and RoMFCC for each contamination level (C1, C2, and C3), OC condition (OC-E and OC-P) as a function of the severity level $SL$ with contamination model Out-E and Out-P and contamination probability 0.05 in Scenario 1.}
	
	\label{fi_results_2}
	
	\centering
		\hspace{-2.05cm}
	\begin{tabular}{cM{0.24\textwidth}M{0.24\textwidth}M{0.24\textwidth}M{0.24\textwidth}}
			&\multicolumn{2}{c}{\hspace{0.12cm} \textbf{\large{Out-E}}}&	\multicolumn{2}{c}{\hspace{0.12cm} \textbf{\large{Out-P}}}\\
		&\hspace{0.6cm}\textbf{\footnotesize{OC-E}}&\hspace{0.6cm}\textbf{\footnotesize{OC-P}}&\hspace{0.5cm}\textbf{\footnotesize{OC-E}}&\hspace{0.5cm}\textbf{\footnotesize{OC-P}}\\
		\textbf{\footnotesize{C1}}&\includegraphics[width=.25\textwidth]{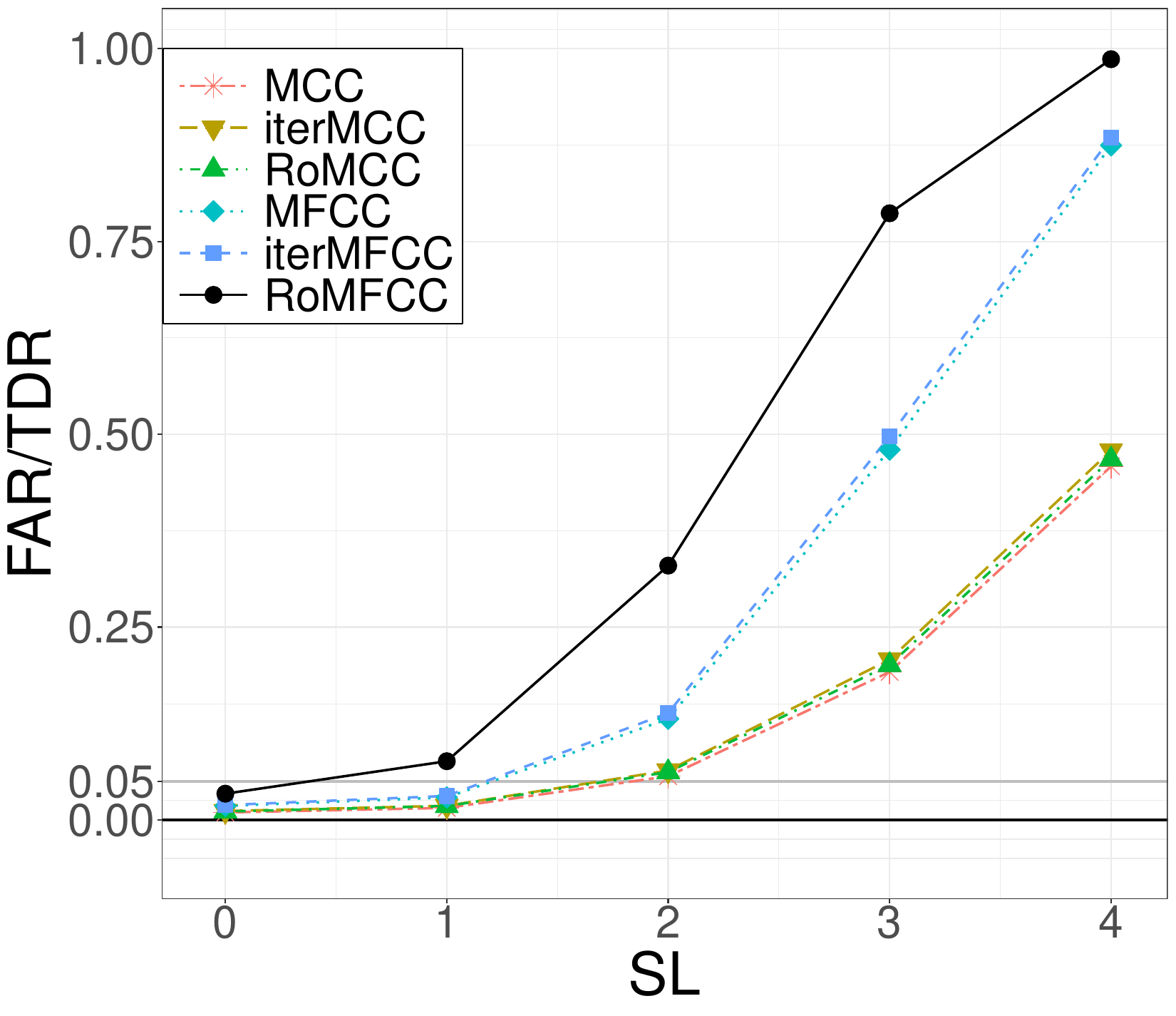}&\includegraphics[width=.25\textwidth]{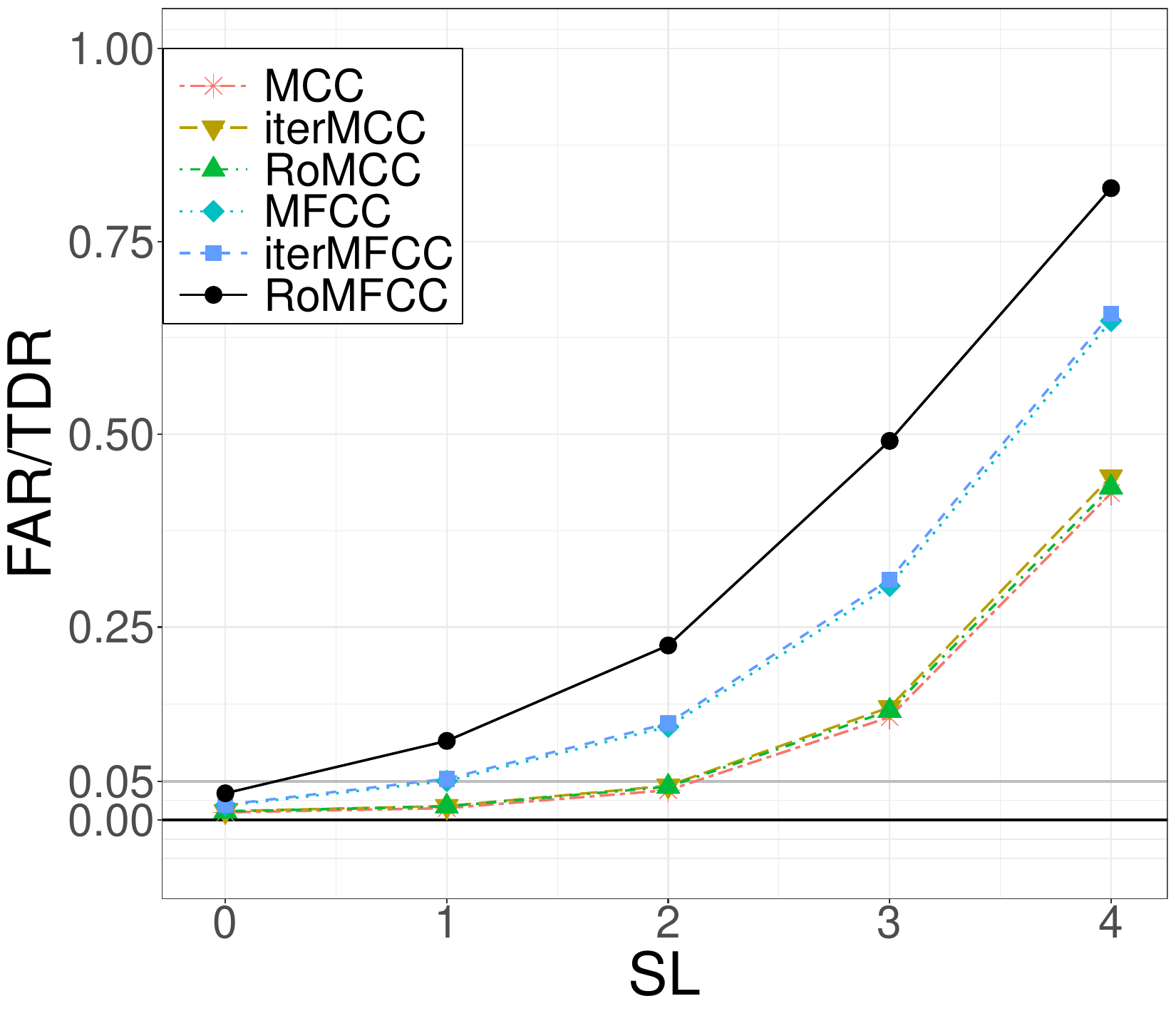}&\includegraphics[width=.25\textwidth]{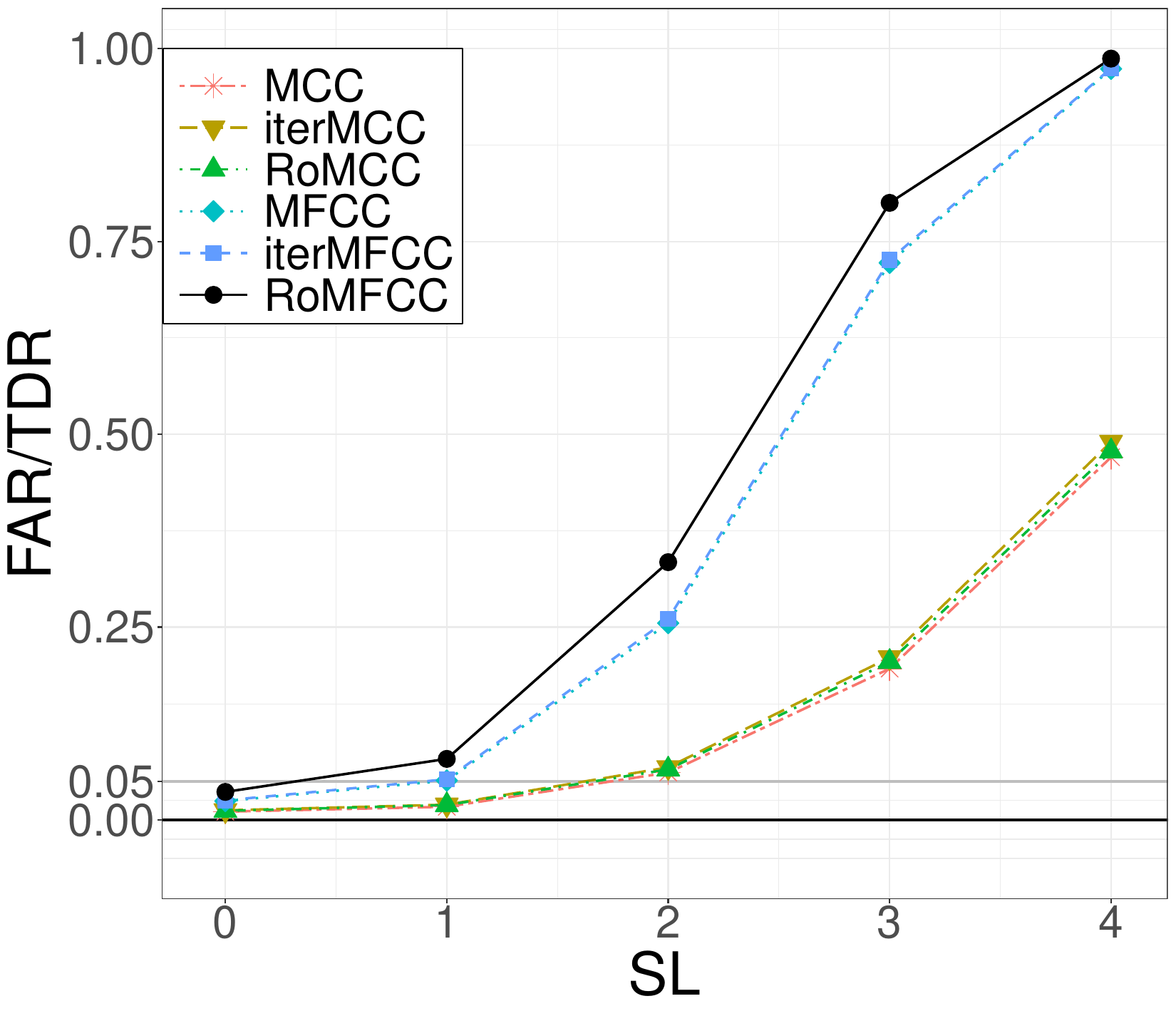}&\includegraphics[width=.25\textwidth]{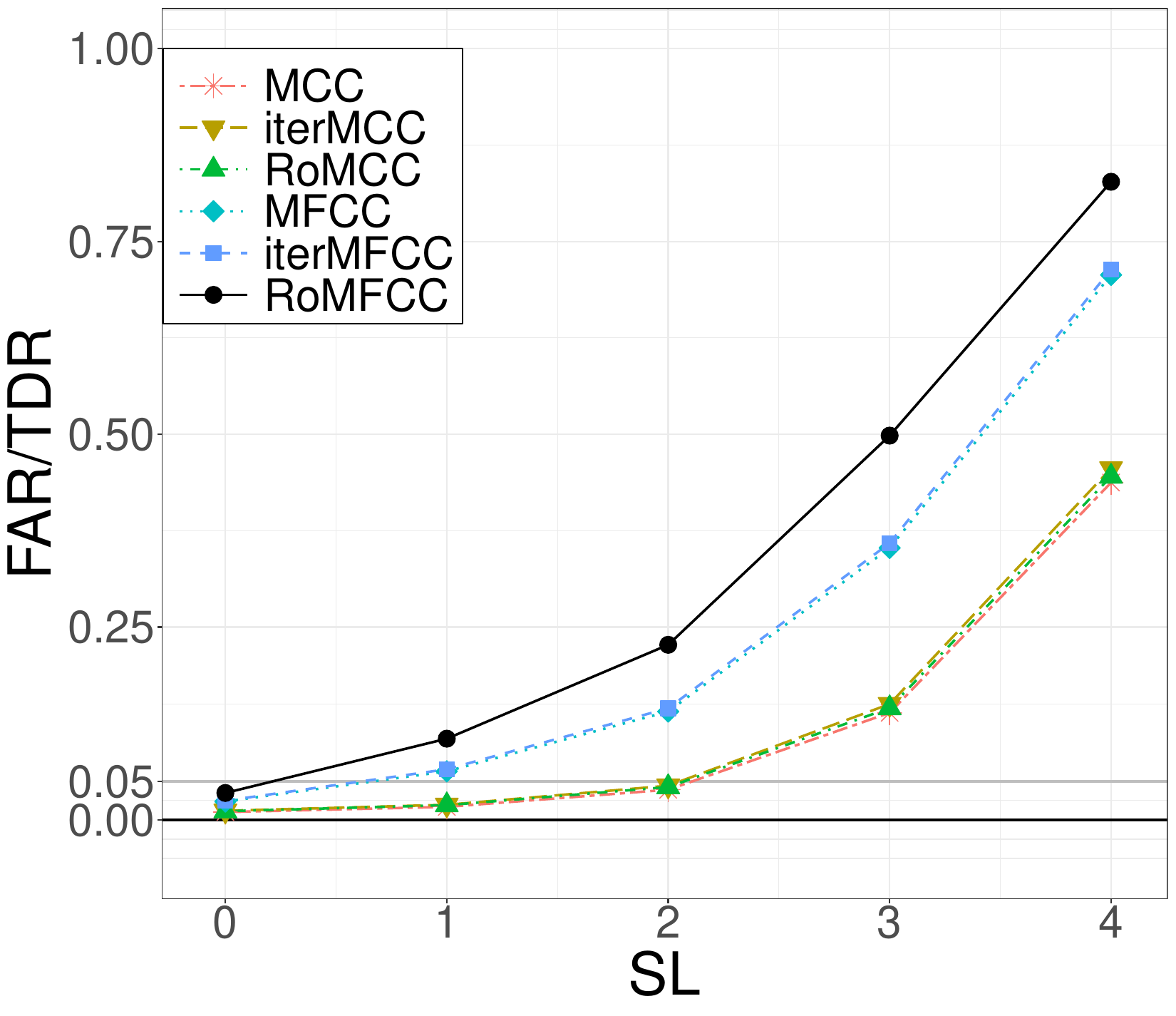}\\
		\textbf{\footnotesize{C2}}&\includegraphics[width=.25\textwidth]{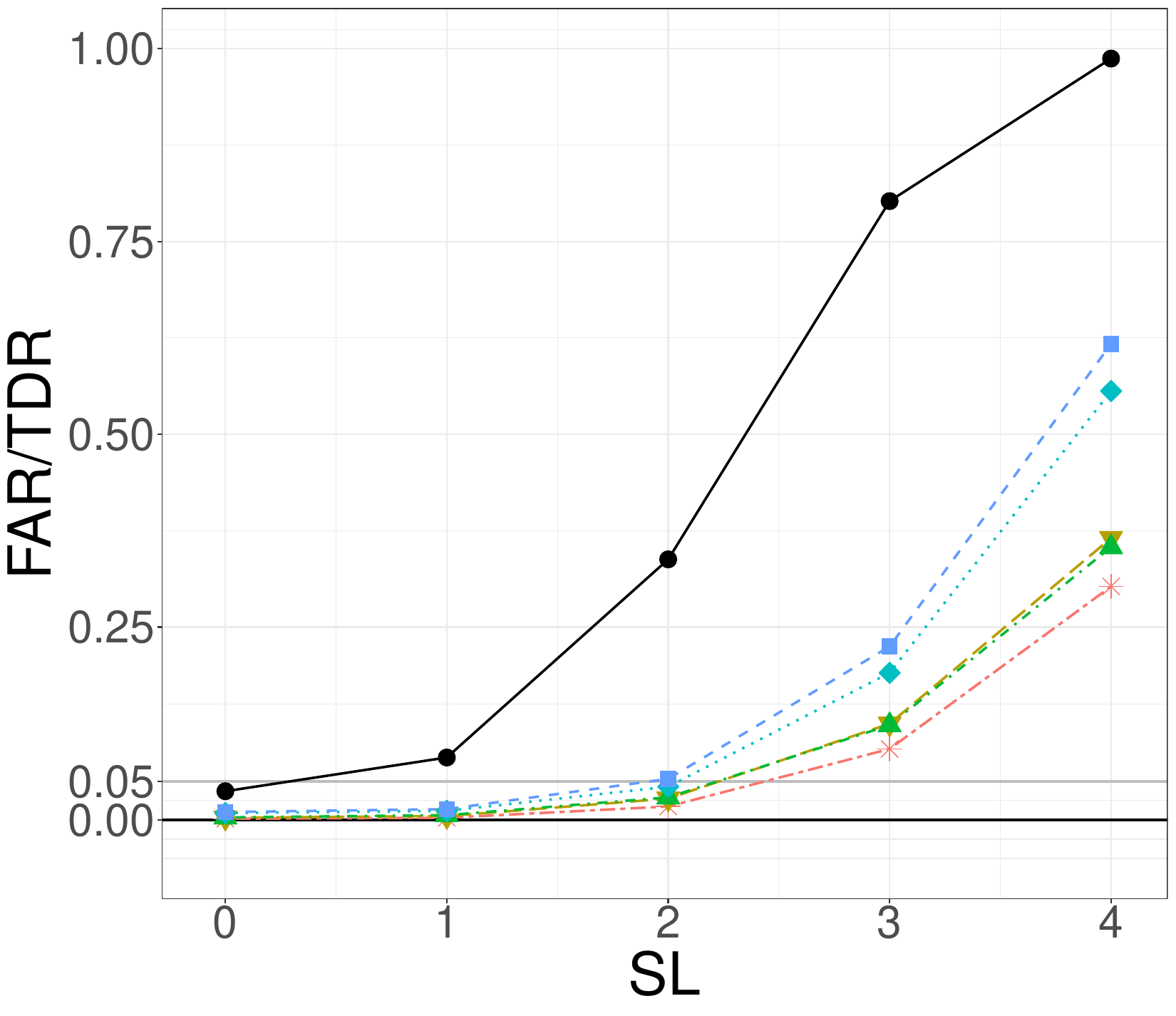}&\includegraphics[width=.25\textwidth]{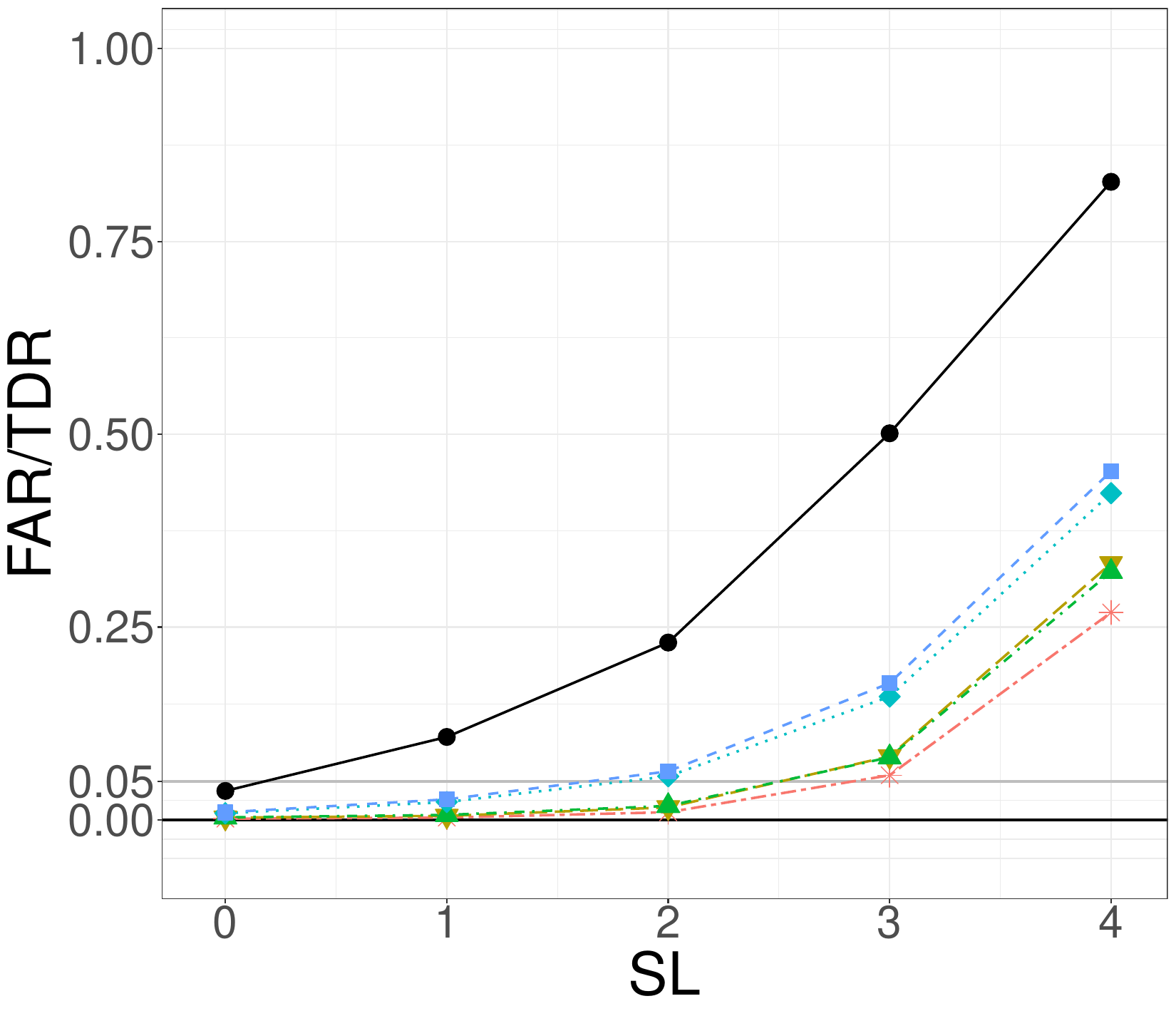}&\includegraphics[width=.25\textwidth]{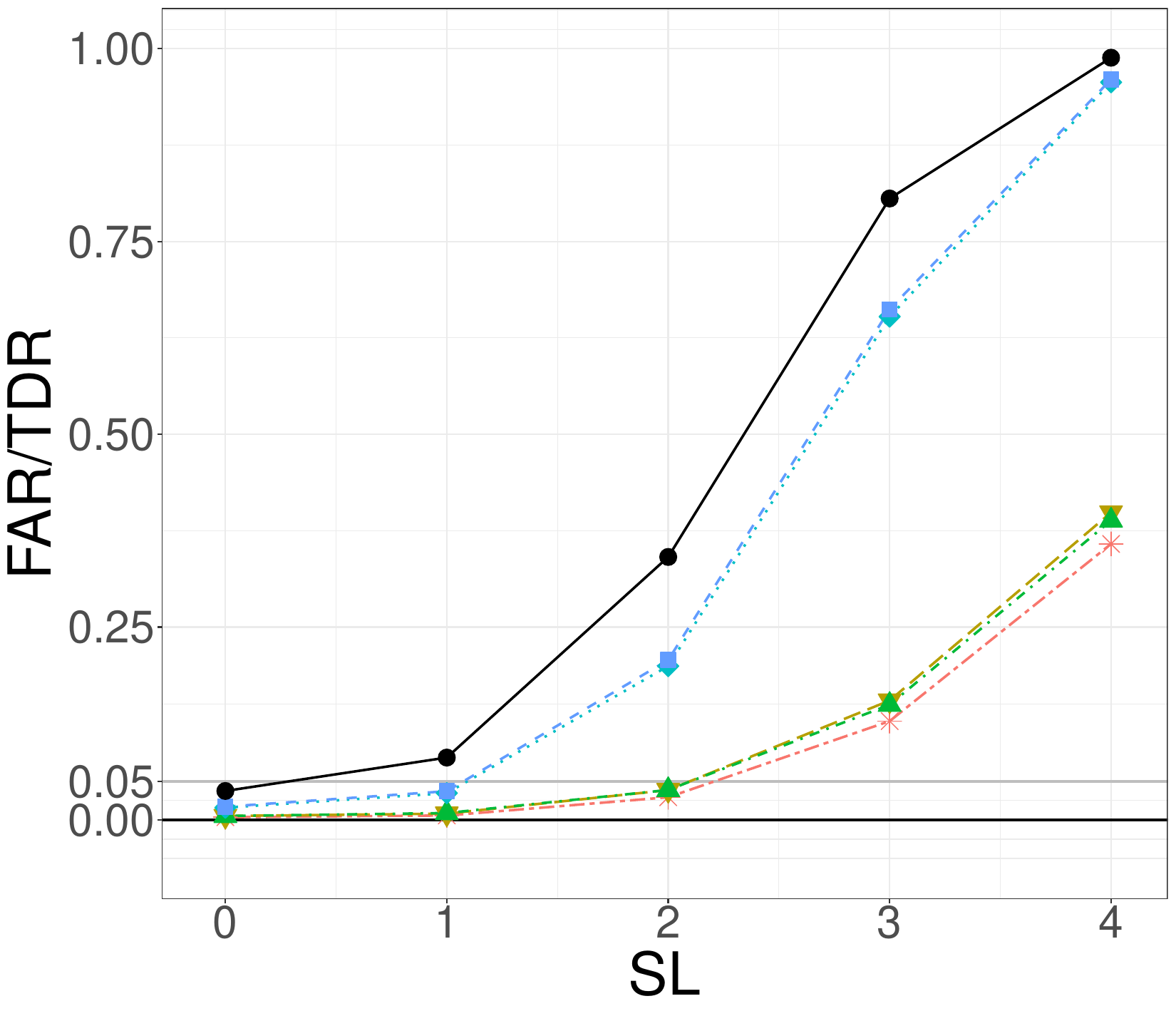}&\includegraphics[width=.25\textwidth]{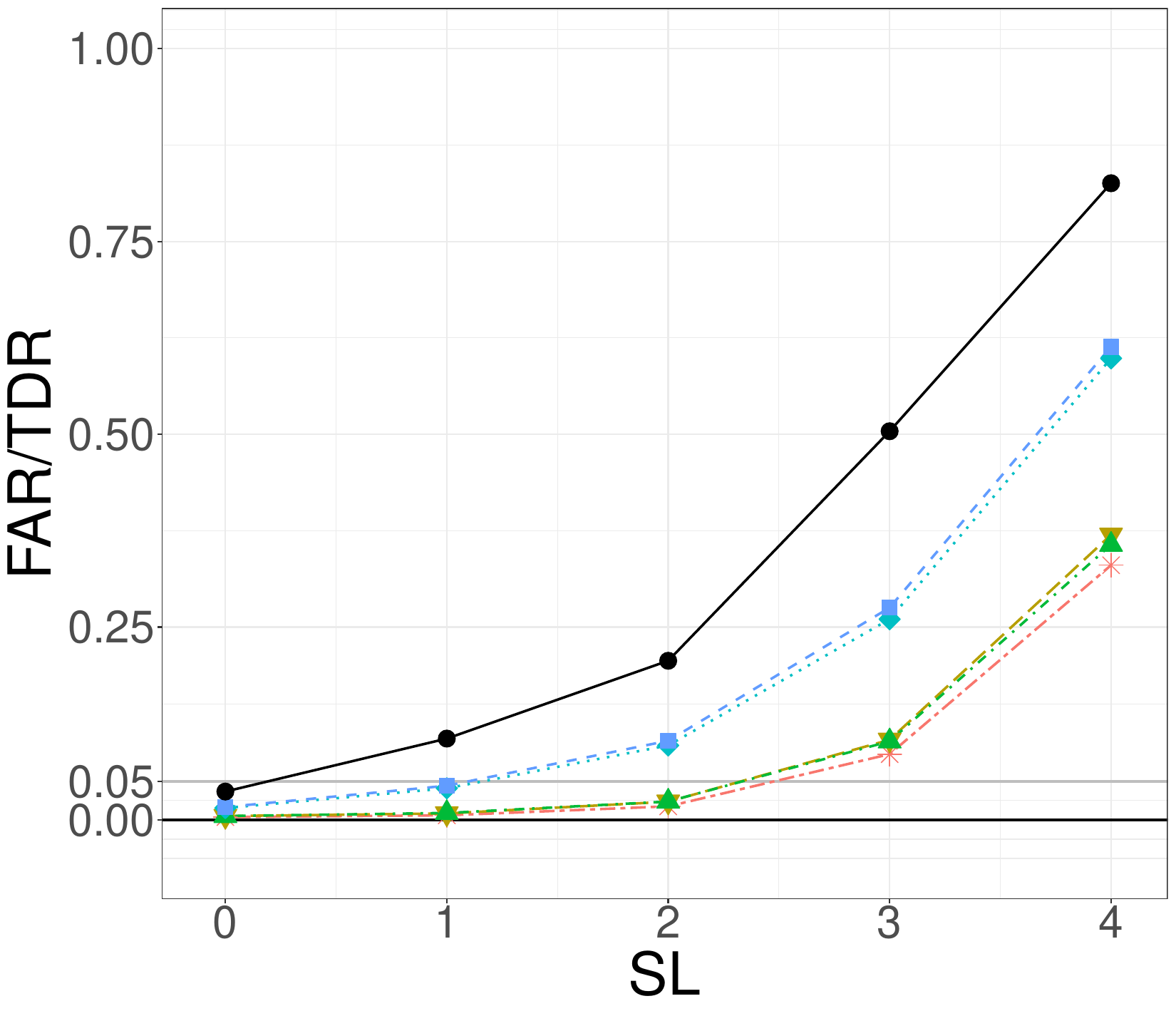}\\
		\textbf{\footnotesize{C3}}&\includegraphics[width=.25\textwidth]{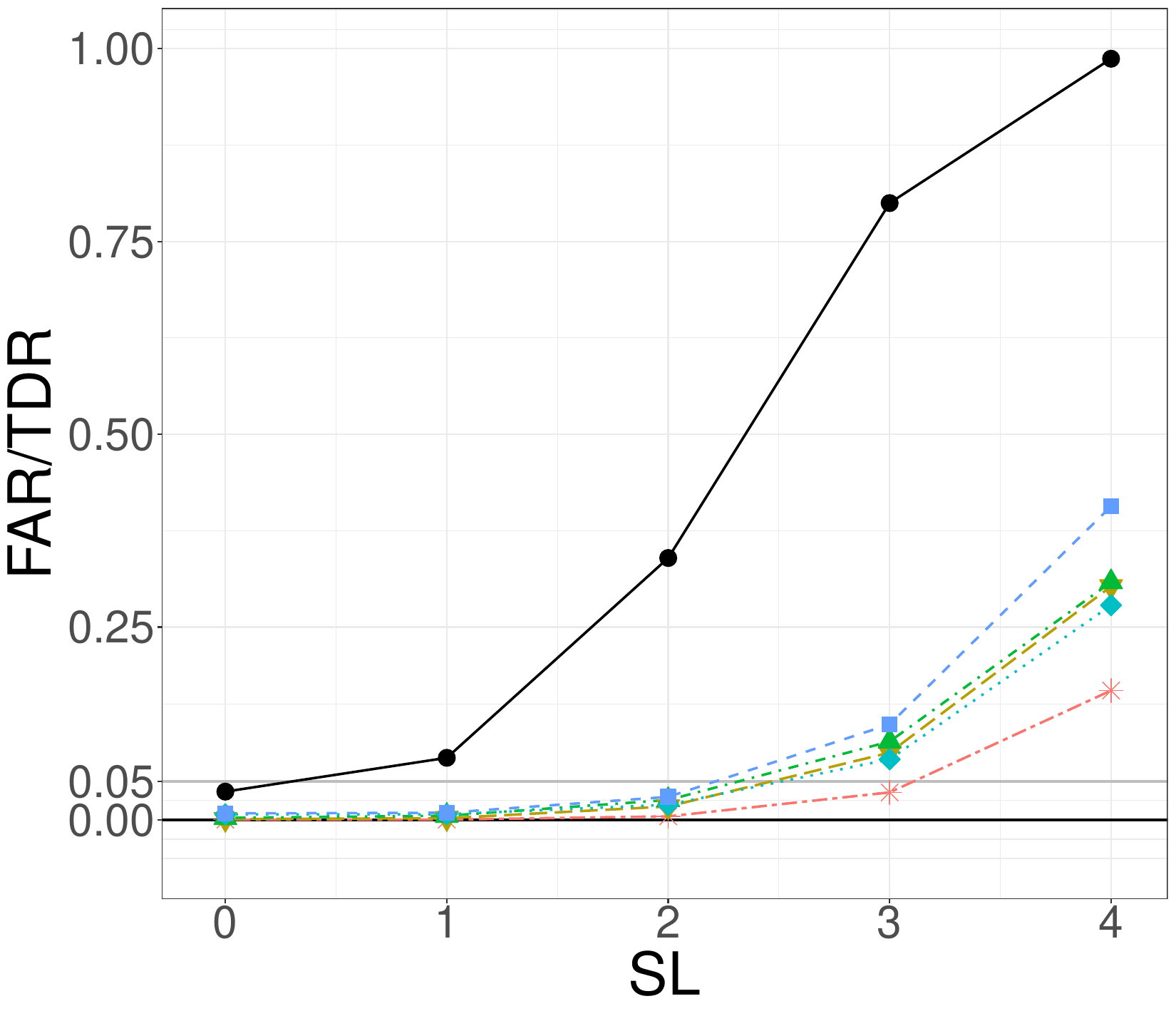}&\includegraphics[width=.25\textwidth]{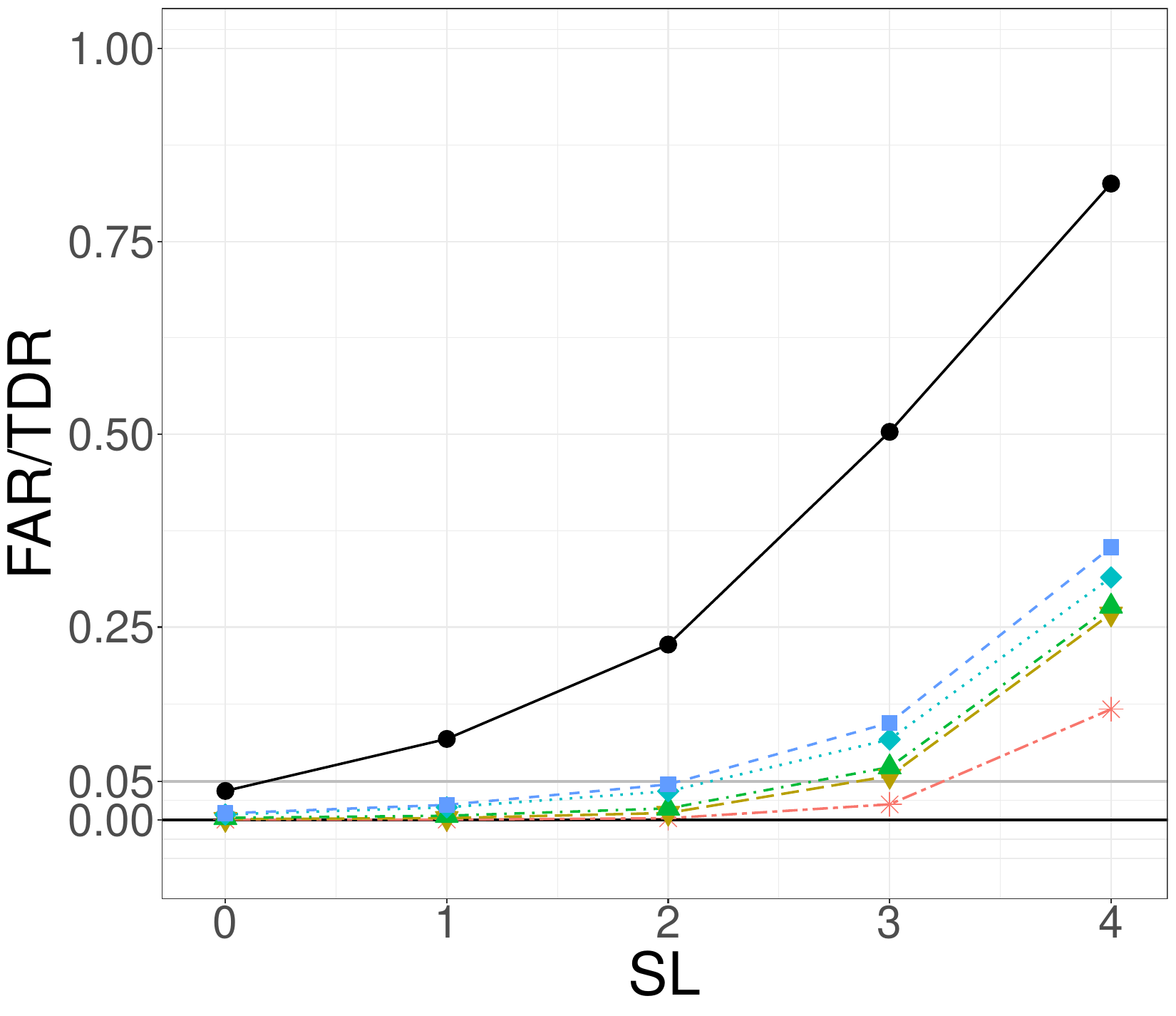}&\includegraphics[width=.25\textwidth]{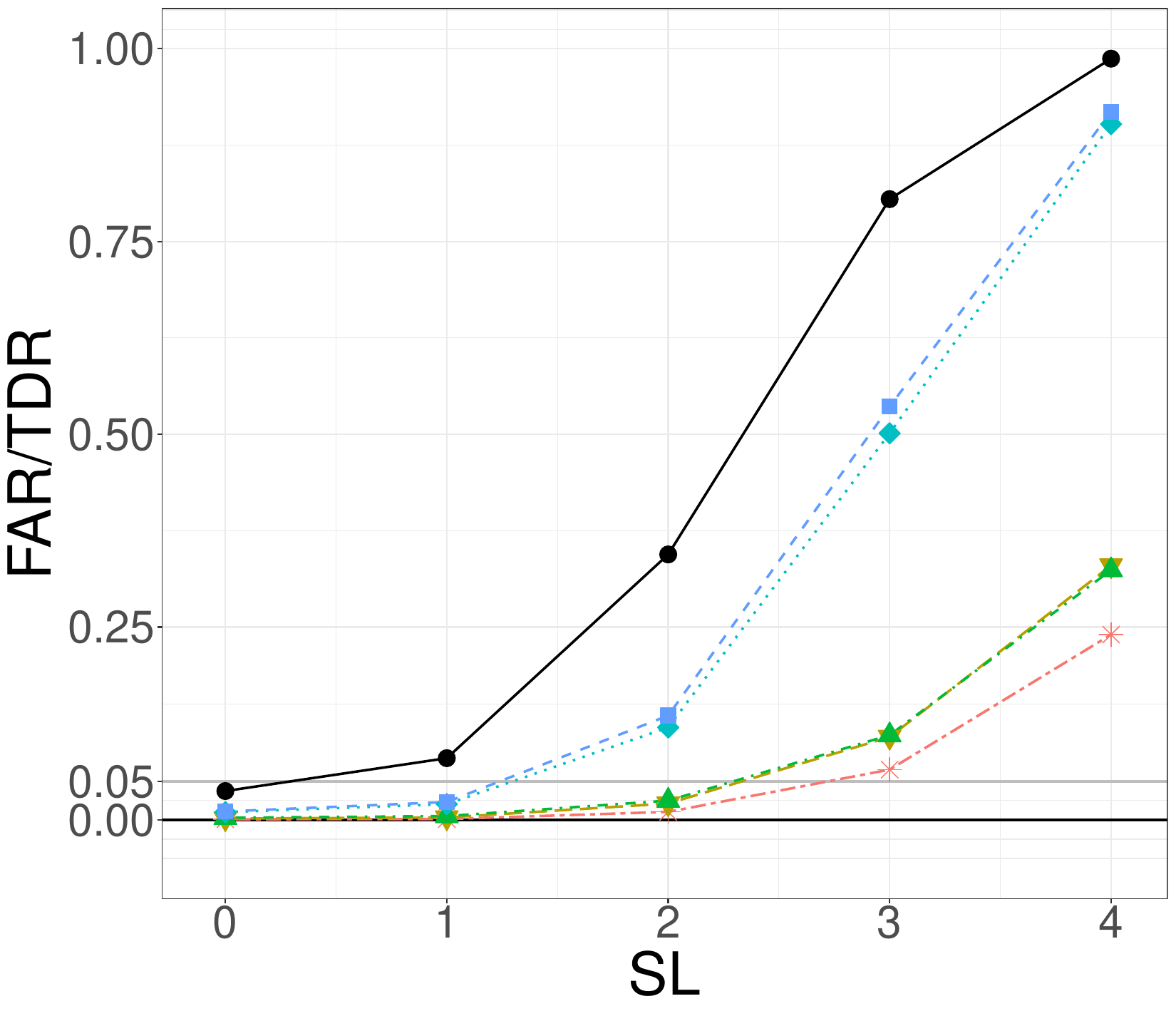}&\includegraphics[width=.25\textwidth]{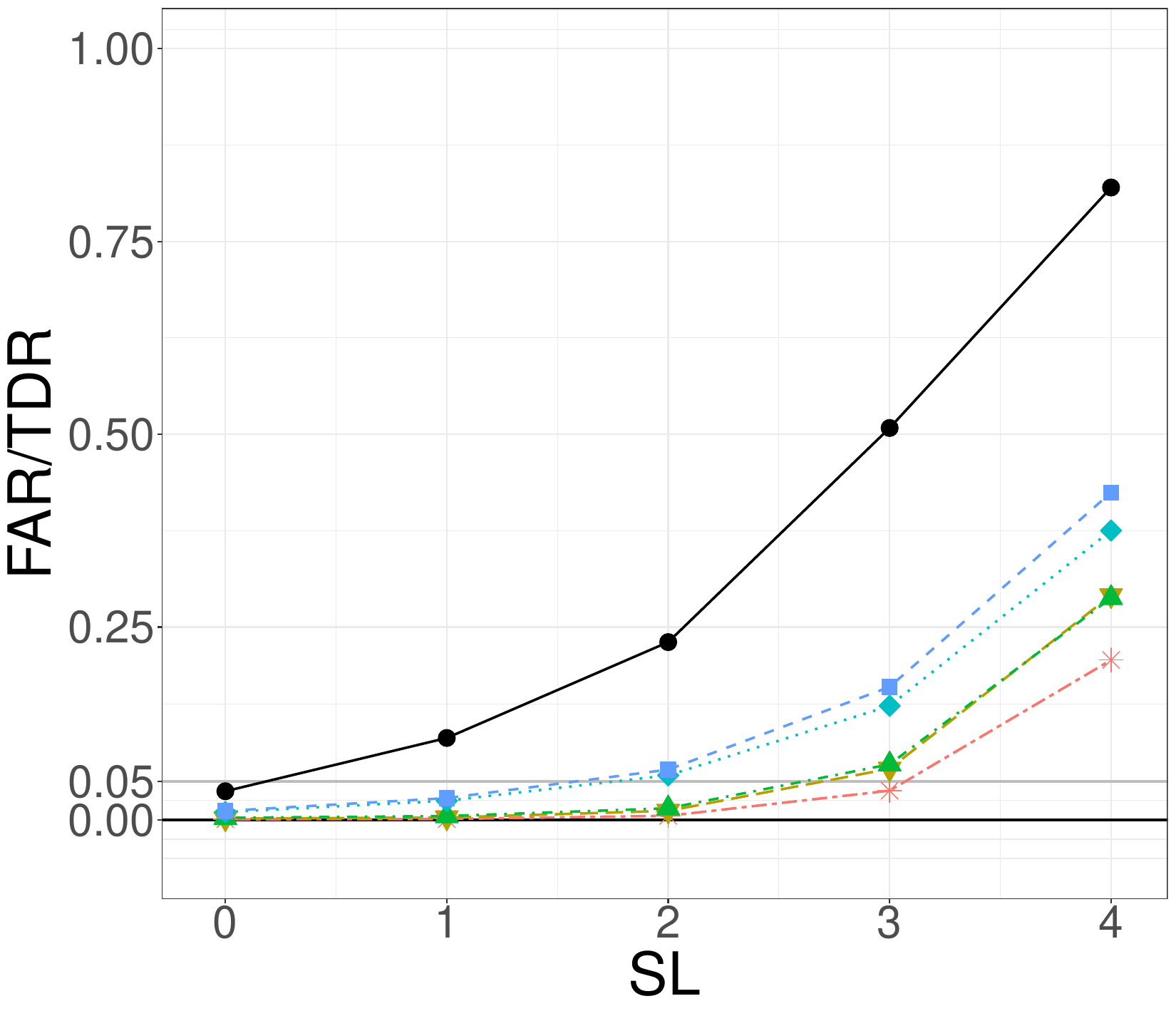}\\
	\end{tabular}
	\vspace{-.5cm}
\end{figure}

\begin{figure}[h]
	\caption{Mean FAR ($ SL=0 $) or TDR ($ SL\neq 0 $) achieved in Phase II by MCC, iterMCC, RoMCC, MFCC, iterMFCC and RoMFCC for each contamination level (C1, C2 and C3), OC condition (OC-E and OC-P) as a function of the severity level $SL$ with contamination model Out-E and Out-P and contamination probability 0.05 in Scenario 2.}
	
	\label{fi_results_3}
	
	\centering
		\hspace{-2.05cm}
	\begin{tabular}{cM{0.24\textwidth}M{0.24\textwidth}M{0.24\textwidth}M{0.24\textwidth}}
			&\multicolumn{2}{c}{\hspace{0.12cm} \textbf{\large{Out-E}}}&	\multicolumn{2}{c}{\hspace{0.12cm} \textbf{\large{Out-P}}}\\
		&\hspace{0.6cm}\textbf{\footnotesize{OC-E}}&\hspace{0.6cm}\textbf{\footnotesize{OC-P}}&\hspace{0.5cm}\textbf{\footnotesize{OC-E}}&\hspace{0.5cm}\textbf{\footnotesize{OC-P}}\\
		\textbf{\footnotesize{C1}}&\includegraphics[width=.25\textwidth]{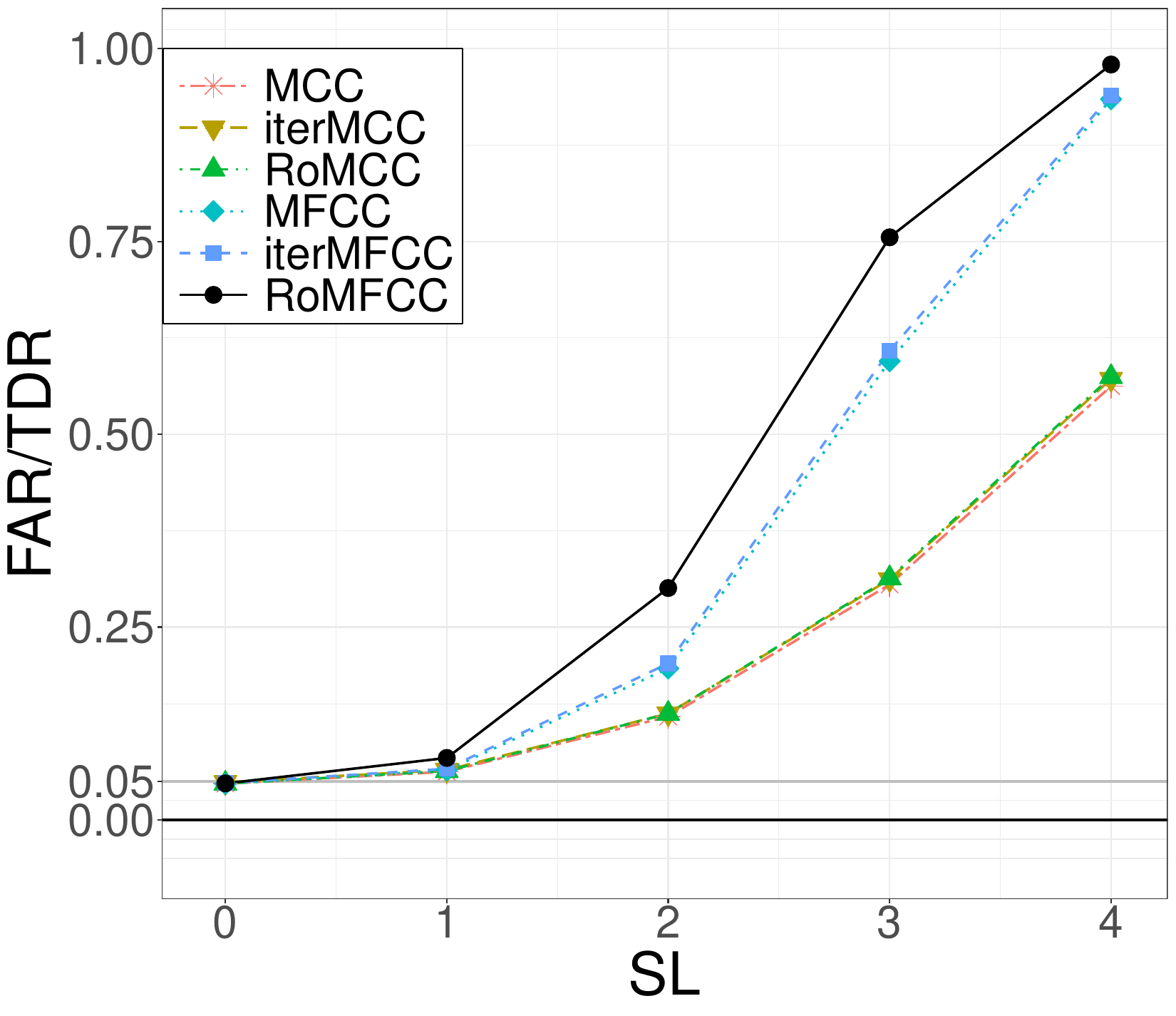}&\includegraphics[width=.25\textwidth]{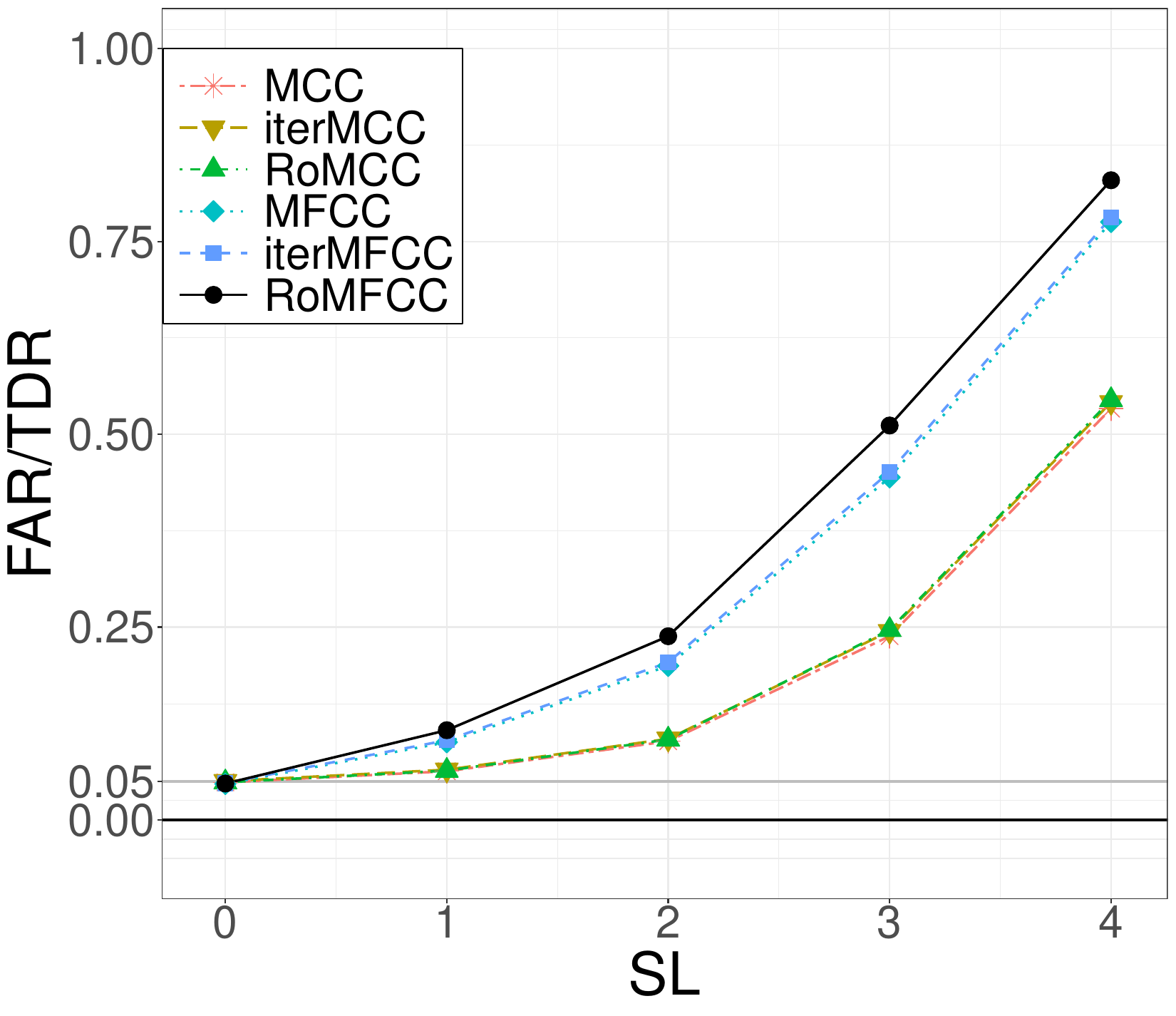}&\includegraphics[width=.25\textwidth]{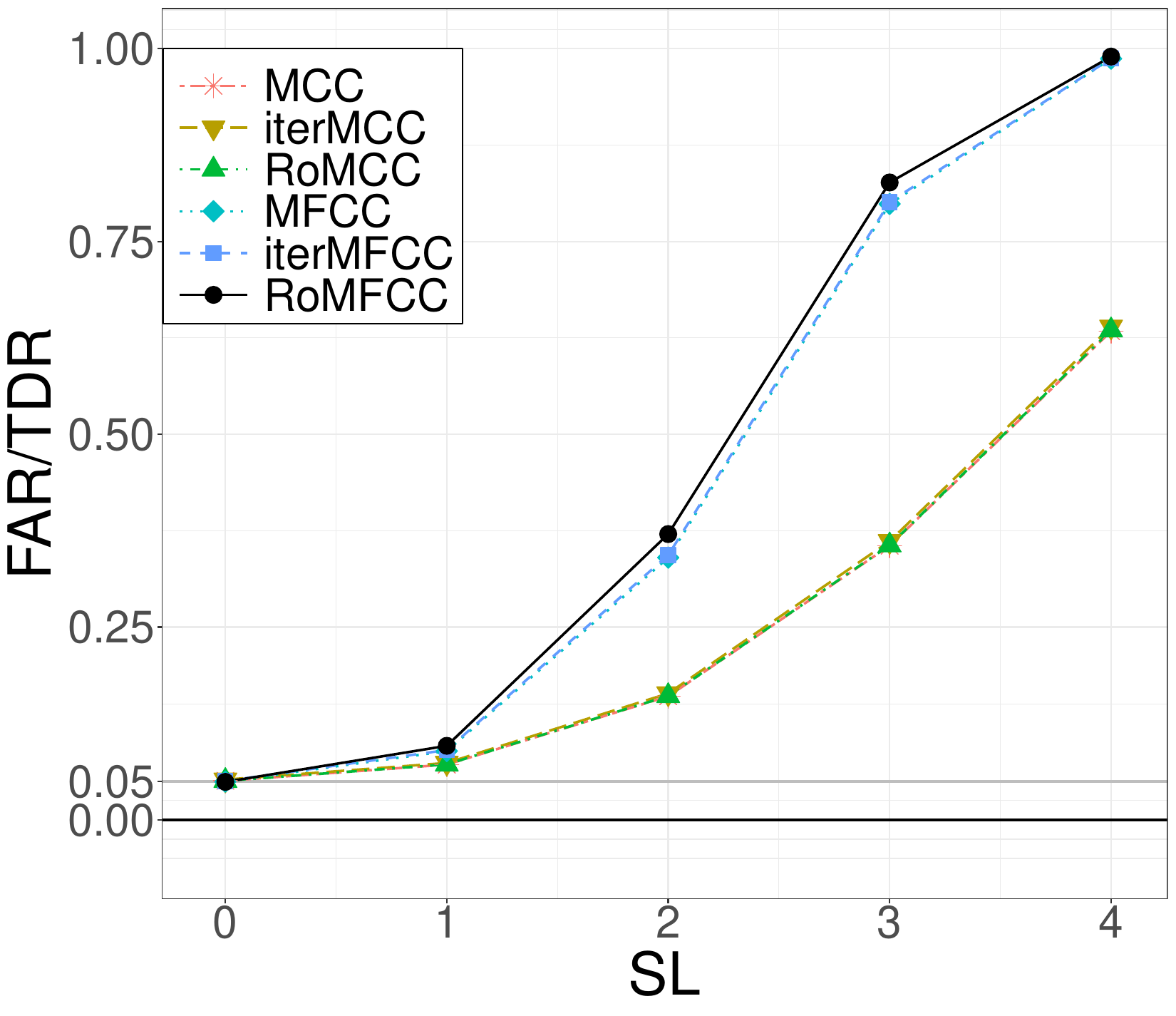}&\includegraphics[width=.25\textwidth]{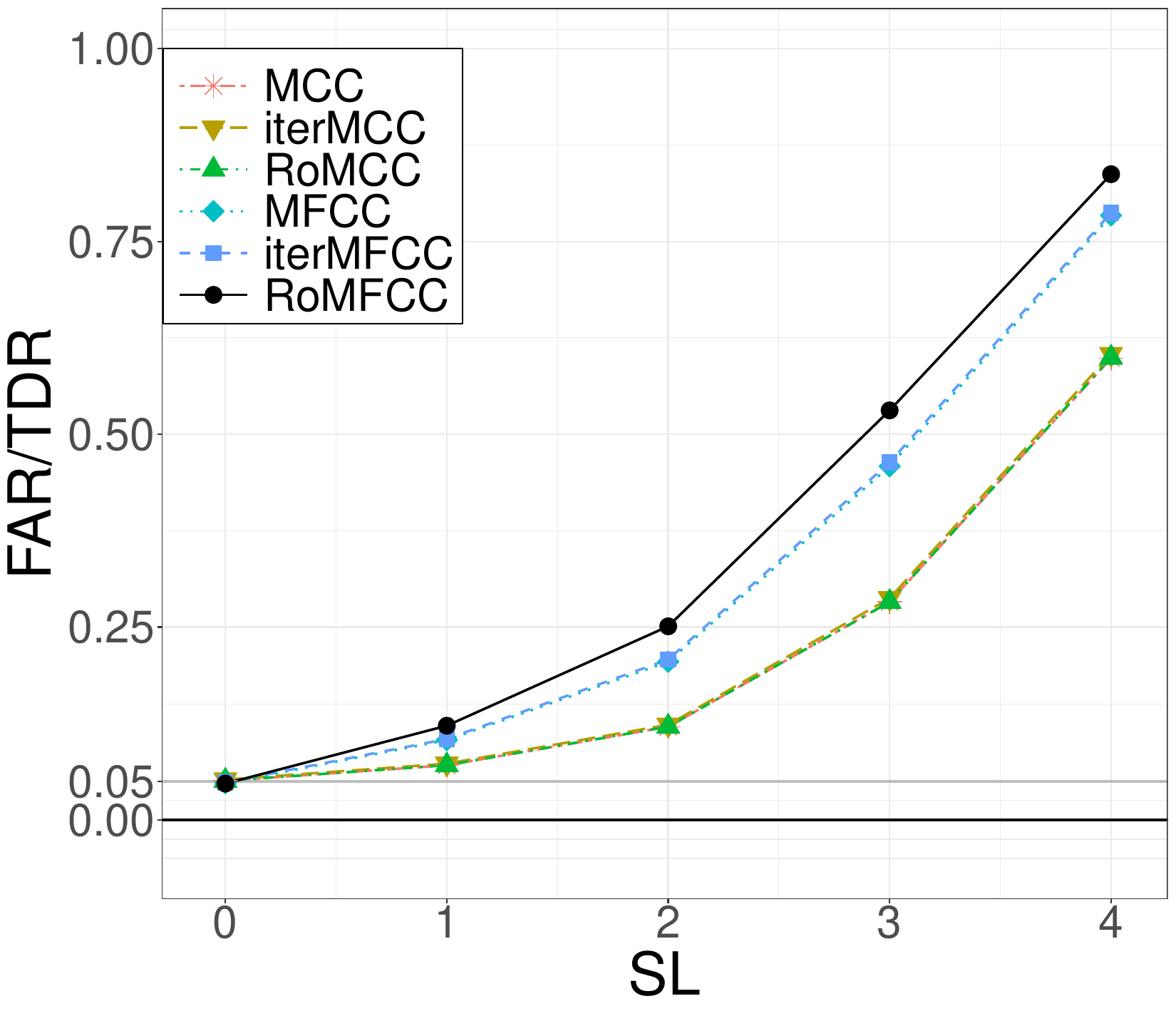}\\
		\textbf{\footnotesize{C2}}&\includegraphics[width=.25\textwidth]{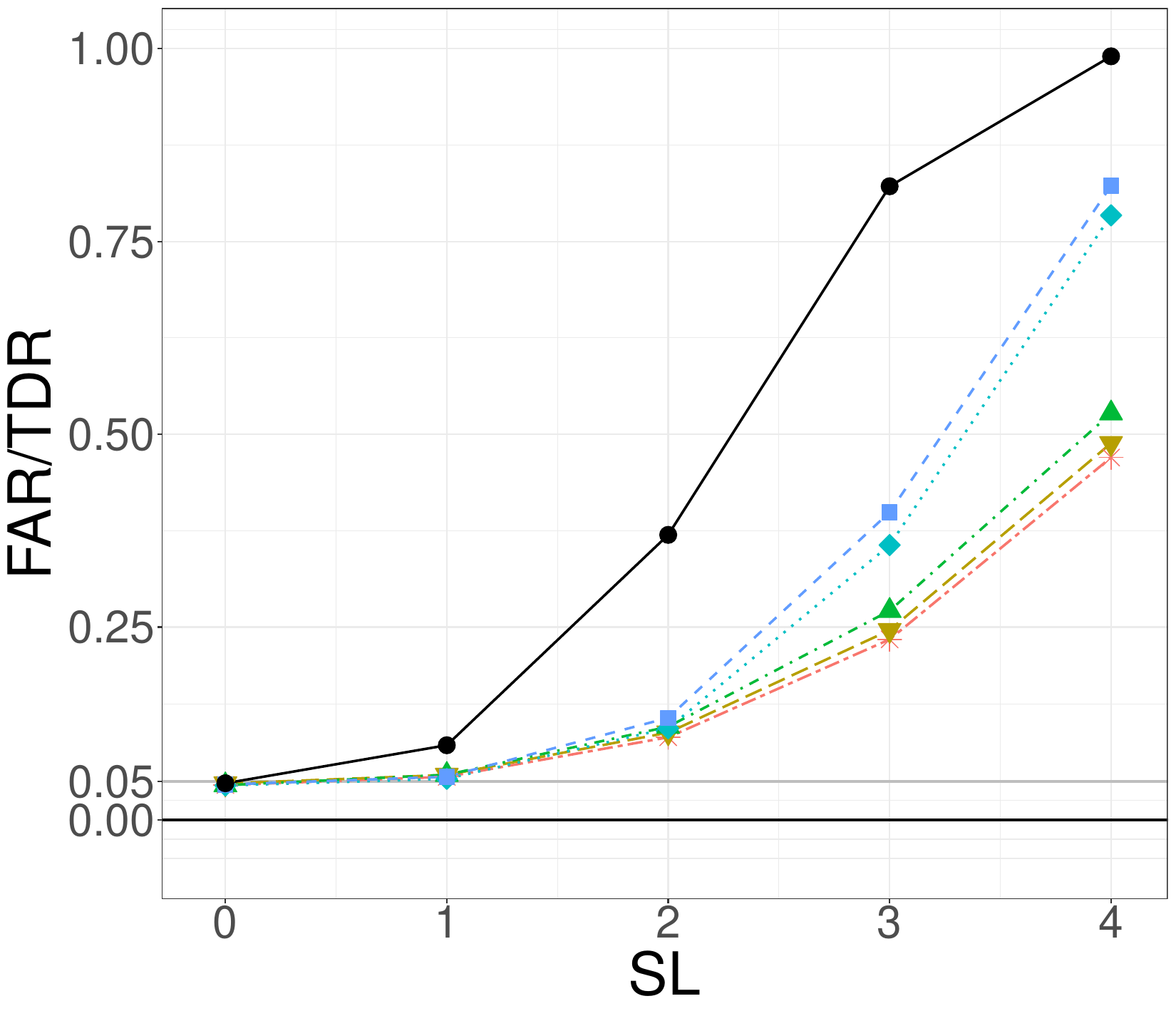}&\includegraphics[width=.25\textwidth]{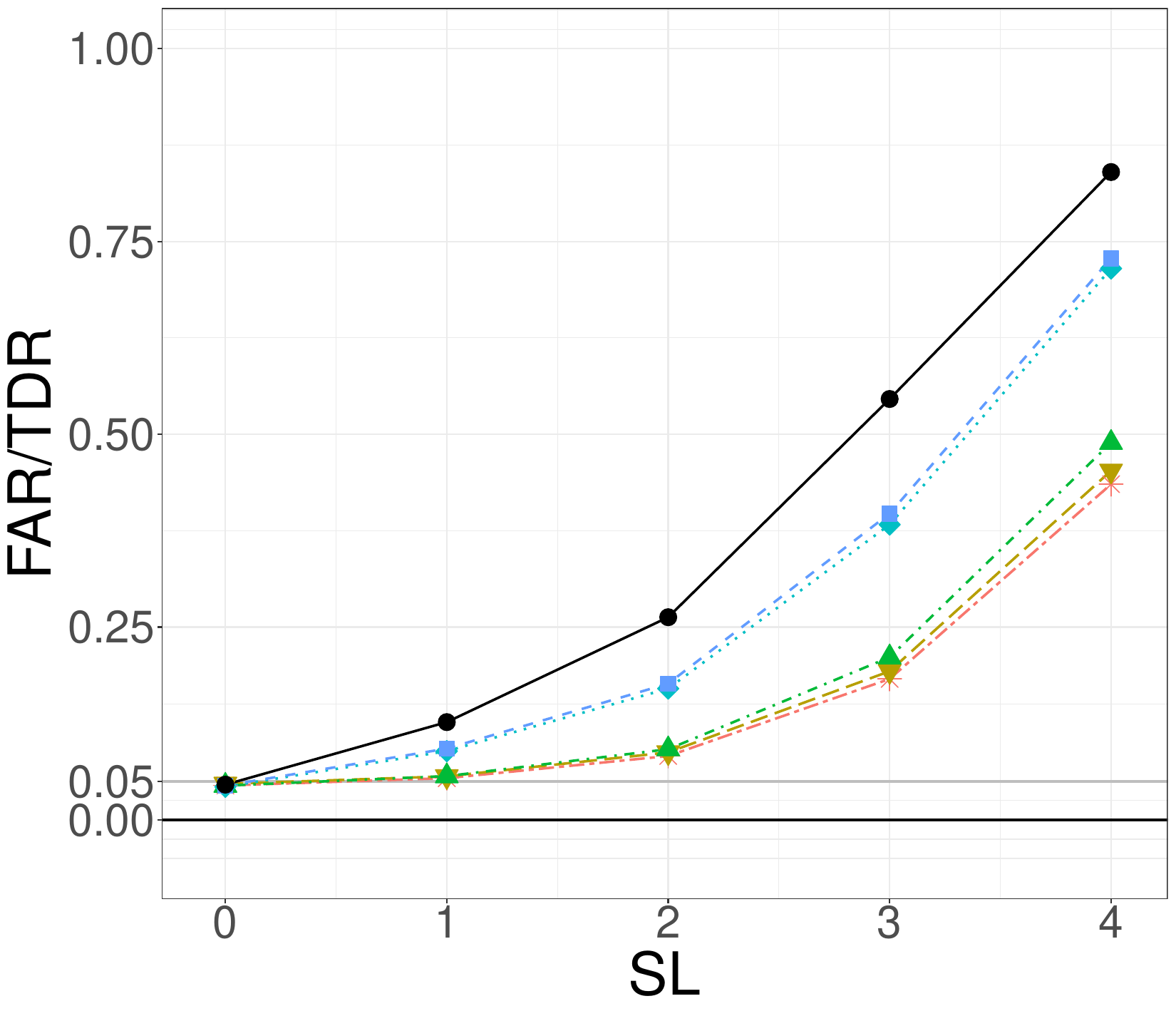}&\includegraphics[width=.25\textwidth]{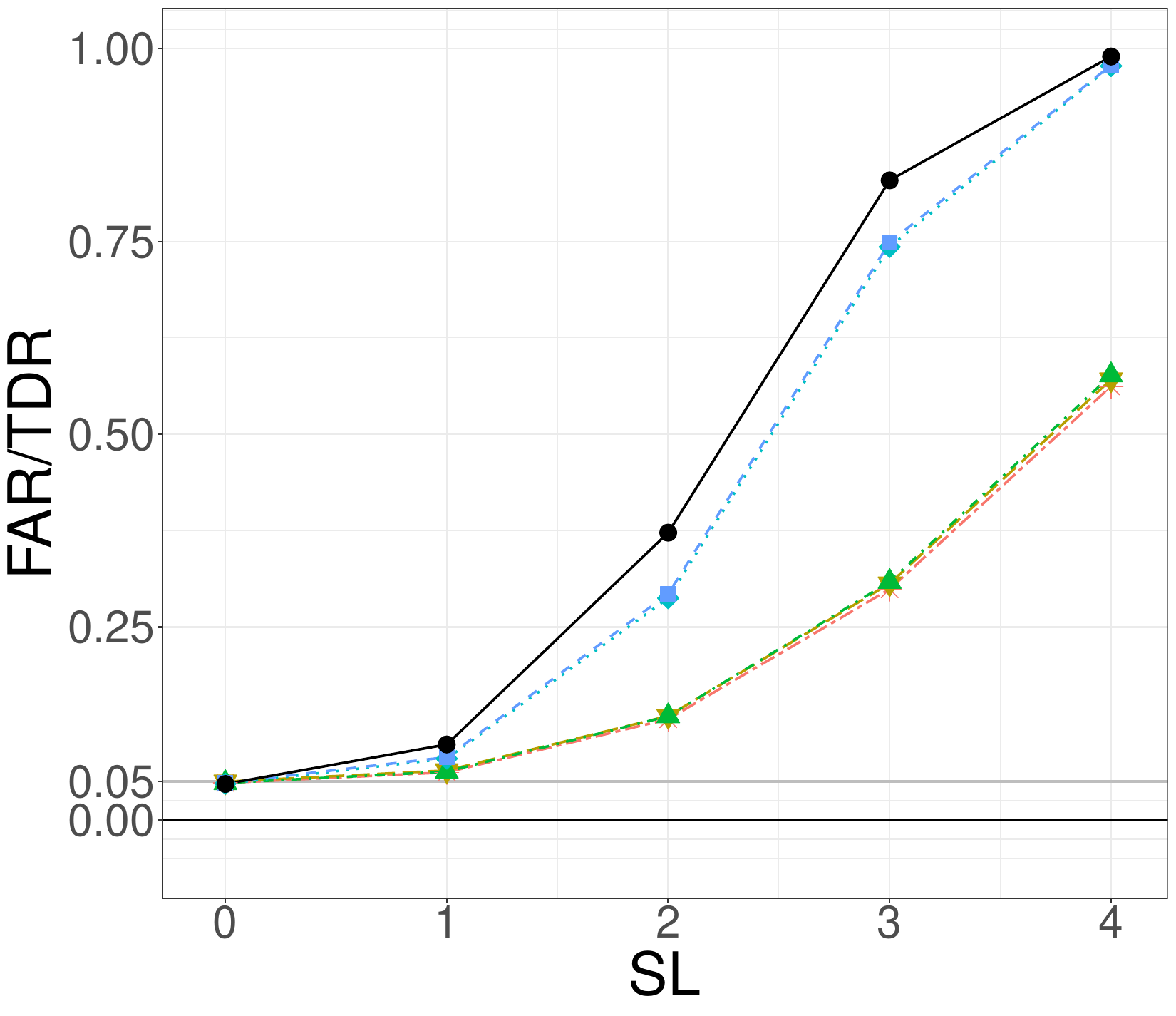}&\includegraphics[width=.25\textwidth]{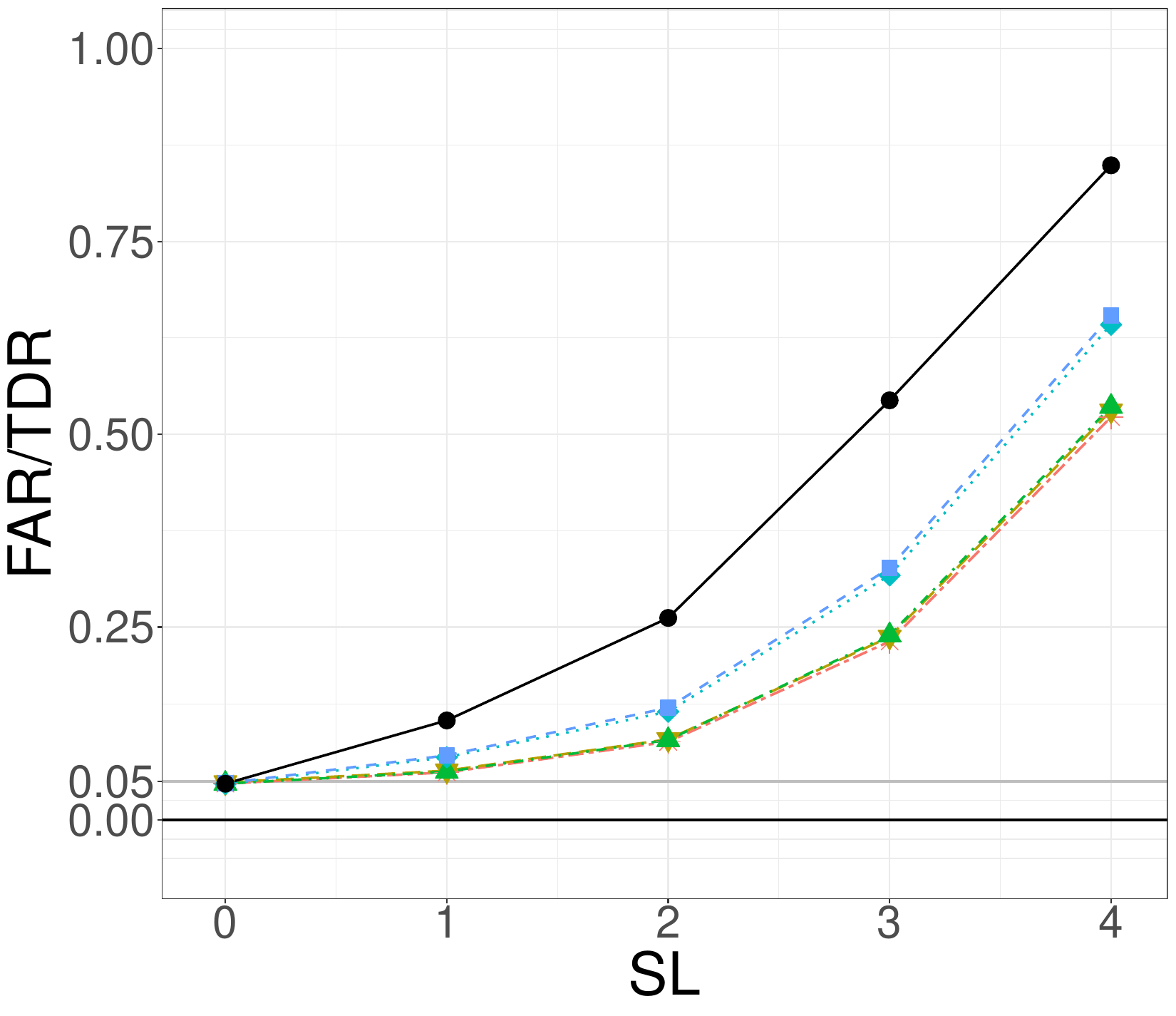}\\
		\textbf{\footnotesize{C3}}&\includegraphics[width=.25\textwidth]{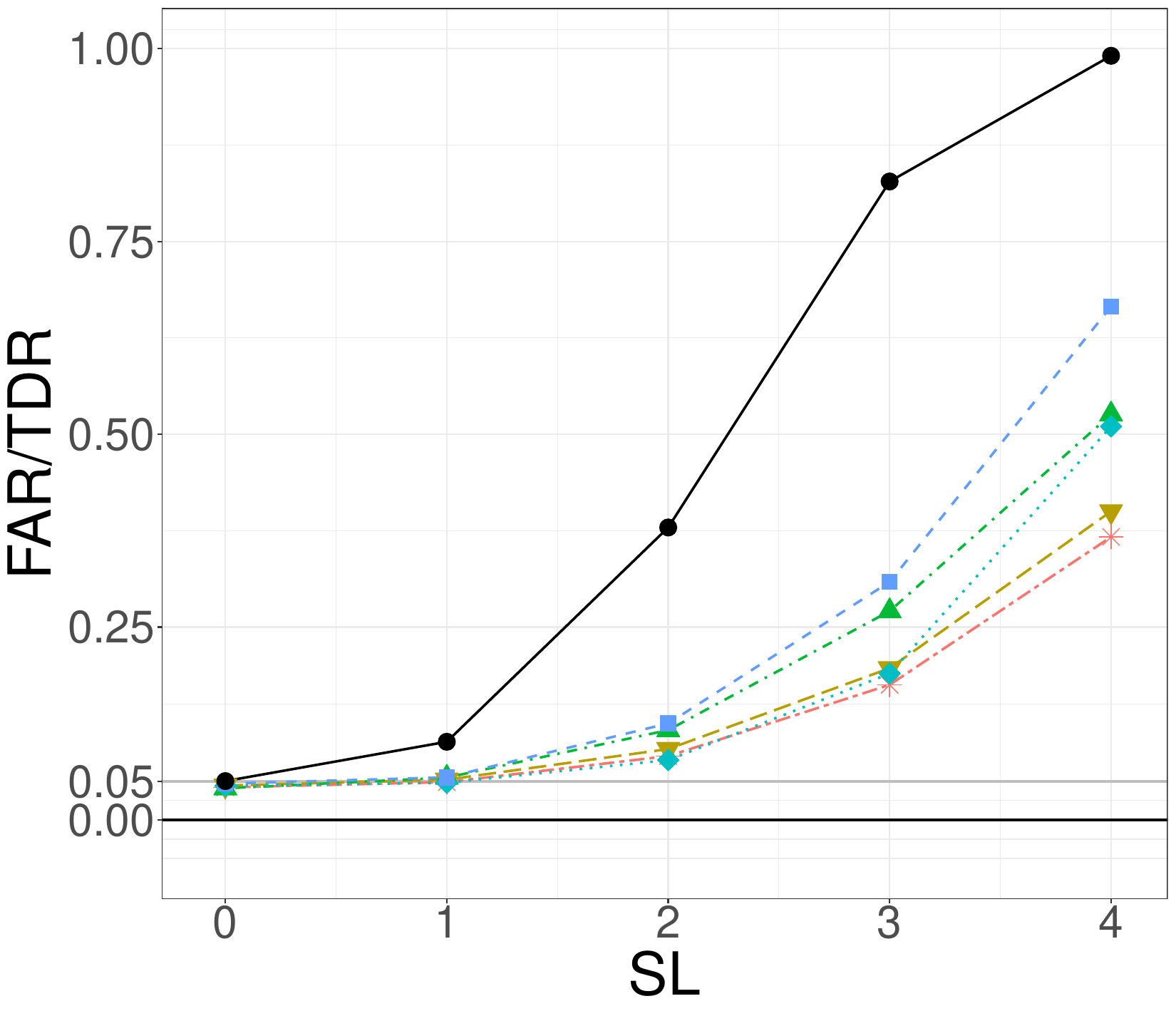}&\includegraphics[width=.25\textwidth]{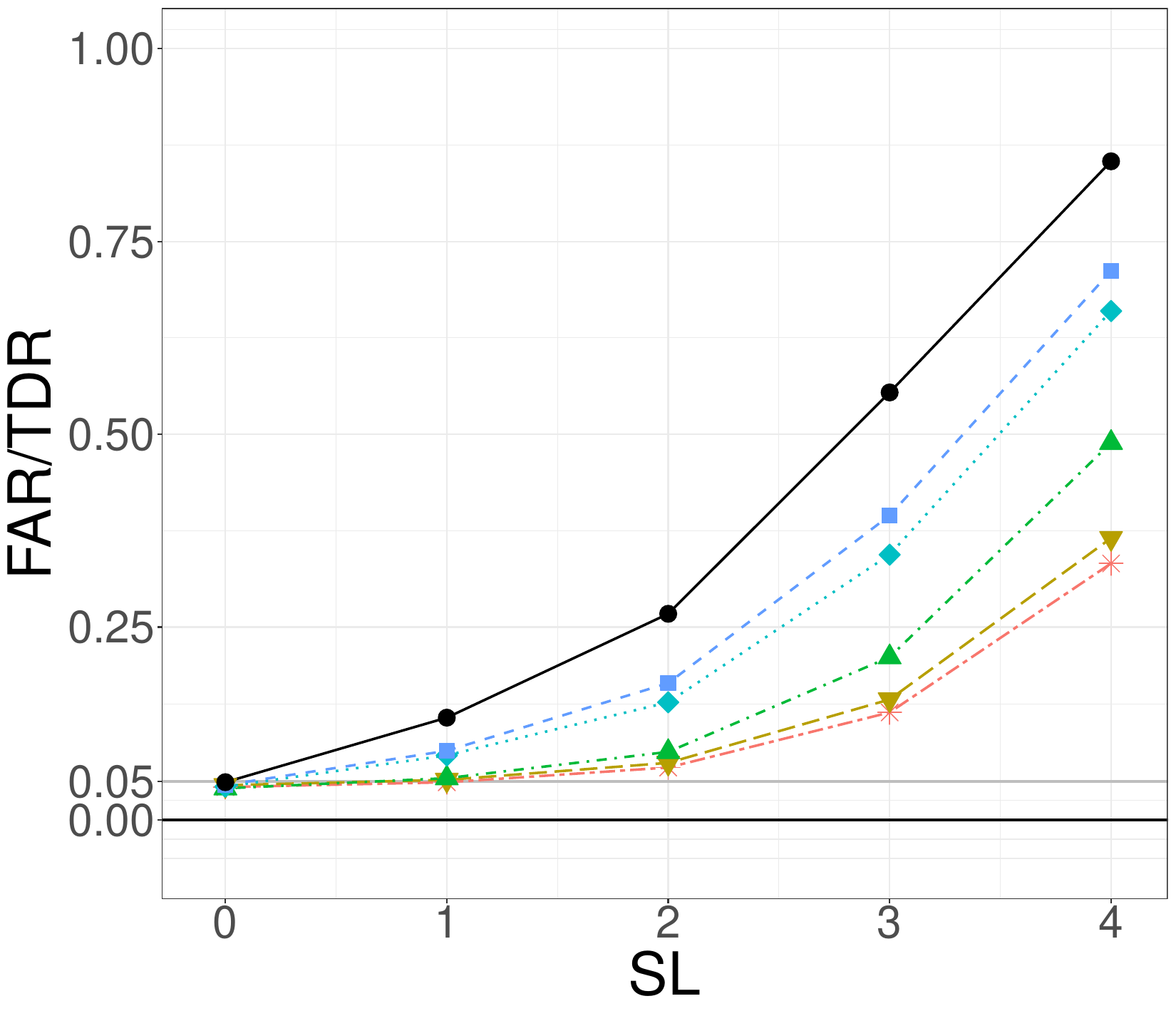}&\includegraphics[width=.25\textwidth]{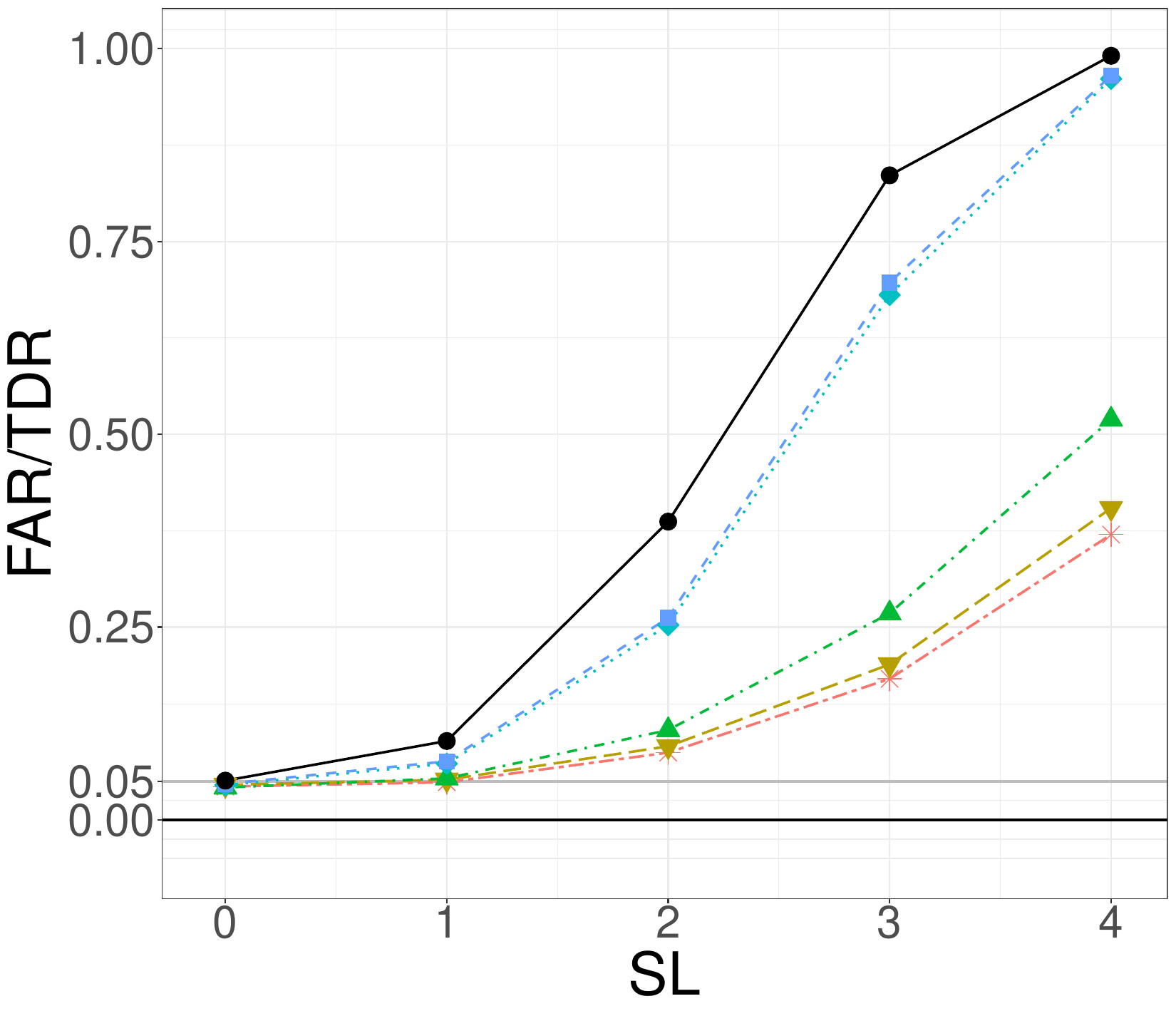}&\includegraphics[width=.25\textwidth]{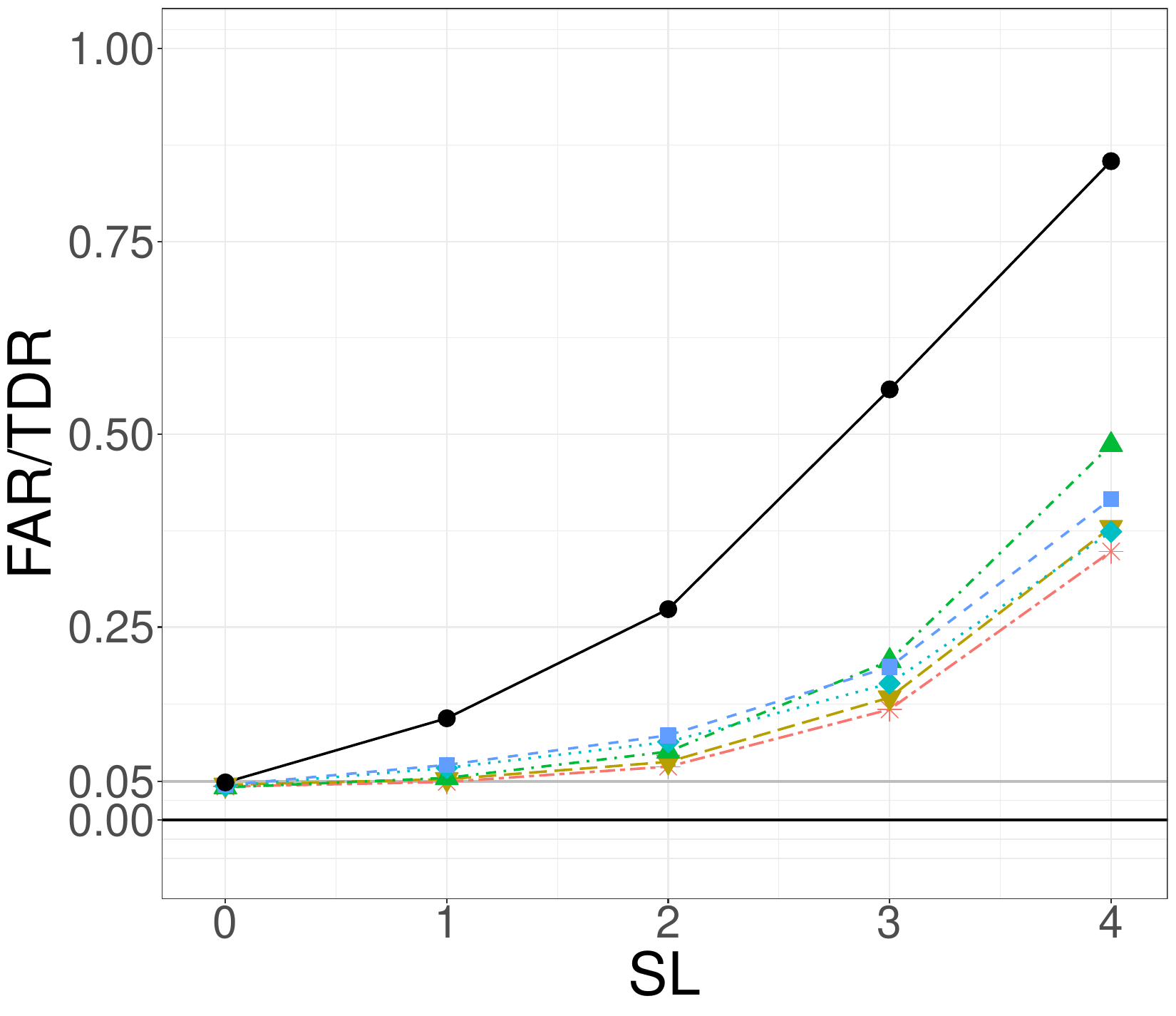}\\
	\end{tabular}
	\vspace{-.5cm}
\end{figure}

Figure \ref{fi_results_1} through \ref{fi_results_3} display for  Scenario 0, Scenario 1, and Scenario 2, respectively,  the mean FAR and TDR as a function of the severity level $SL$, at each OC condition type, contamination model and contamination level, as reported in Table \ref{tab_sim}.
\begin{table}[]
    \centering
     \resizebox{\columnwidth}{!}{
\begin{tabular}{clcccclcc}
\toprule
         &  & \multicolumn{4}{c}{Phase I}                                                                          &  & \multicolumn{2}{c}{Phase II}      \\ \cline{1-1} \cline{3-6} \cline{8-9} 
Scenario &  &\bigcell{c}{ Phase I\\ contamination type}   & \bigcell{c}{Contamination \\ probability} & \bigcell{c}{Contamination \\ model} & \bigcell{c}{Contamination \\ level} &  & OC condition type & SL            \\\cline{1-1} \cline{3-6} \cline{8-9} 
0        &  & No contamination             & -                         & -                   & -                   &  & OC-E, OC-P        & 0, 1, 2, 3, 4 \\[0.2cm]
1        &  & \bigcell{c}{Functional cellwise\\ outliers} & 0.05, 0.01                & Out-E, Out-P        & C1, C2, C3          &  & OC-E, OC-P        & 0, 1, 2, 3, 4 \\ [0.2cm]
2        &  & \bigcell{c}{Functional casewise\\ outliers} & 0.05, 0.01                & Out-E, Out-P        & C1, C2, C3          &  & OC-E, OC-P        & 0, 1, 2, 3, 4 \\ \bottomrule
\end{tabular}
}
    \caption{Phase I generation scenarios and Phase II OC condition settings for the comparison of RoMFCC with competing methods.}
    \label{tab_sim}
\end{table}
For Scenario 0, where the Phase I sample is not contaminated by outliers,  Figure \ref{fi_results_1} shows that all the approaches that are able to  account for the functional nature of the data (namely MFCC, iterMFCC, RoMFCC) achieve the same performance for both OC condition types. 
Although this scenario should be not favourable for approaches specifically designed to deal with outliers,  iterMFCC and RoMFCC perform equal to  MFCC.
All the non-functional approaches (namely MCC, iterMCC, RoMCC) show instead worse performance than the functional counterparts and   no significant  differences among themselves.

In Scenario 1, the proposed RoMFCC is shown by Figure \ref{fi_results_2}  to outperform all the competing methods for each contamination model, level and OC condition type. 
In particular, the larger the contamination level, the more the RoMFCC outperforms  the competing methods.
In fact, the performance of RoMFCC is insensitive to contamination models and levels, whereas the performance of the competing methods decreases with the contamination level for each contamination model, especially for Out-E.
Regarding the non-robust functional methods (namely MFCC and iterMFCC), it turns out that the iterMFCC, which is expected to be the best competitor, does not really outperform MFCC. 
Moreover, their advantages over non-functional approaches decreases with the contamination level.
The unsatisfactory performance of the non-functional methods reveals their inadequacy in jointly dealing with the functional nature of the data cellwise outliers. 

As in Scenario 1,  Figure \ref{fi_results_3} shows that the RoMFCC turns out to be the best method also in Scenario 2, although the performance difference  with the competing methods is sometimes less pronounced.
This is expected because in this scenario the Phase I sample is only contaminated by functional casewise outliers, which is the only contamination type  the competing methods are  designed to be robust against. 
Note that the performance of RoMFCC is in fact practically unaffected by  outlier contamination  also in this scenario.
This completes the proof of the superiority  of the proposed method in dealing not only with  cellwise but also casewise outliers.

\section{Real-Case Study}
\label{sec_real}


To demonstrate the potential of the proposed RoMFCC in practical situations, a real-case study in the automotive industry is presented henceforth.
As introduced in Section \ref{sec_intro}, it addresses the issue of monitoring the quality of the RSW process, which is an autogenous welding process in which two overlapping conventional steel galvanized sheets are joint together, without the use of any filler material \citep{zhang2011resistance}. 

guarantees the structural integrity and solidity of welded items \citep{martin}.

Joints are formed by applying pressure to the weld area from two opposite sides by means of two copper electrodes.
Voltage applied to the electrodes generates a current flowing between them through the material.
The electrical current flows because the resistance offered by metals causes significant heat generation (Joule effect) that increases the metal temperature at the faying surfaces of the work pieces up to the melting point.
Finally, due to the mechanical pressure of the electrodes, the molten metal of the jointed metal sheets  cools and solidifies, forming the so-called weld nugget \citep{raoelison}.
Further details on how the typical behaviour of a DRC is related to the physical and metallurgical development of a spot weld are provided in \cite{capezza2021functional_clustering}.


Data analyzed in this study are courtesy of Centro Ricerche Fiat and are recorded during automotive body-in-white lab tests.
This stage is generally characterized by a large number of spot welds with different characteristics, e.g, the  thickness and material of the sheets to be joined together and the welding time.
Specifically, this real-case study focuses on the  monitoring of ten spot welds made by only one welding machine.
In particular, for each item we consider the multivariate functional quality characteristic represented by the  vector of  ten DRCs relative to the same ten spot welding points, normalized on the time domain $[0, 1]$. The data set contains  a total number of 1839 items and resistance measurements were collected at a regular grid of points equally spaced by 1 ms.

\begin{figure}
\begin{center}
\includegraphics[width=.5\textwidth]{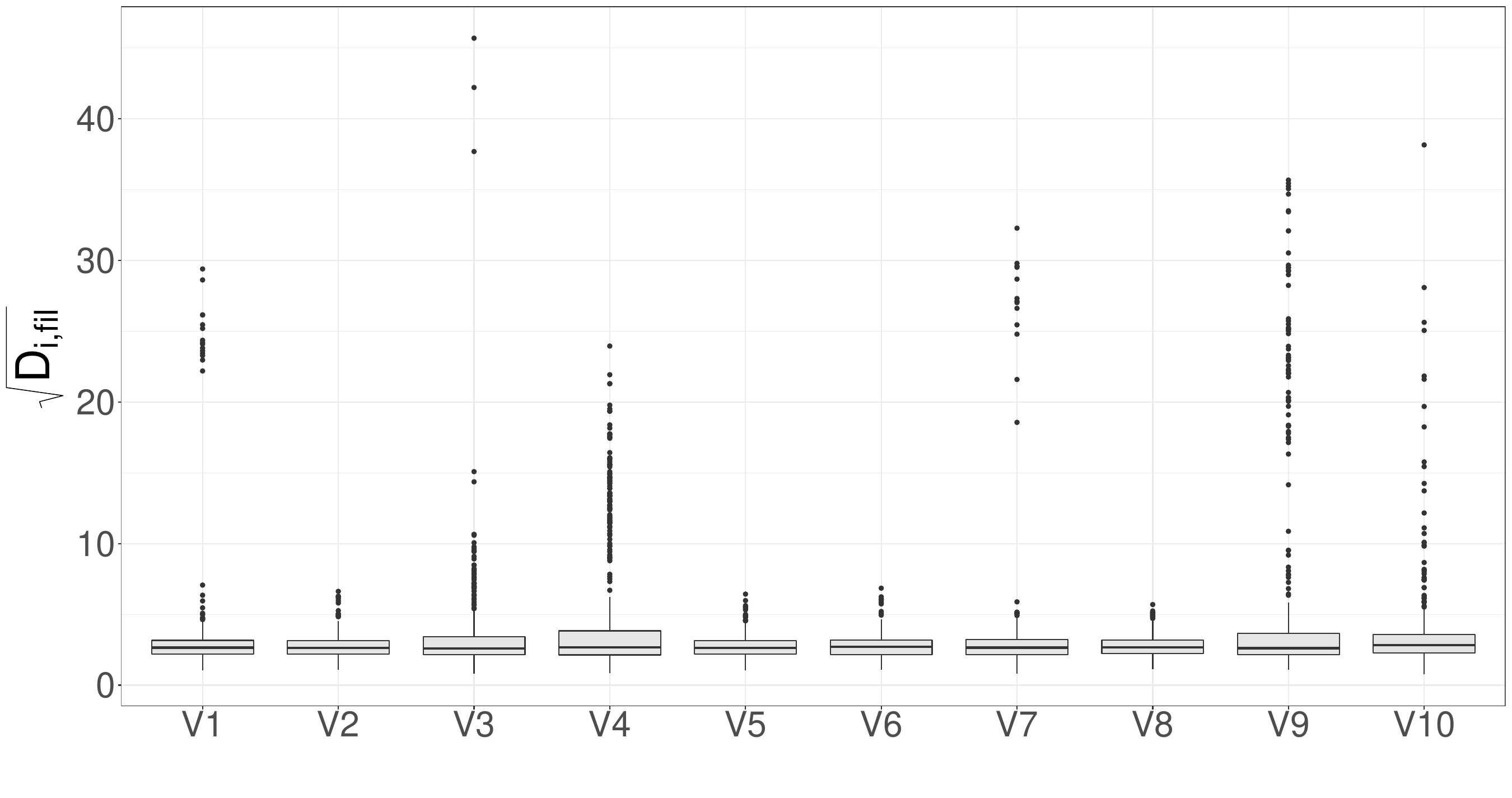}
\caption{Boxplot of  the  functional distance square roots $\sqrt{D_{i,fil}}$ (Equation \eqref{eq_dfil})  obtained from the FUF applied  on the 460 Phase I observation of the training set.}
\label{fig_boxplot}
\end{center}
\end{figure}

The RSW process quality is directly affected by electrode wear that leads to changes  in electrical, thermal and mechanical contact conditions at electrode and metal sheet interfaces \citep{manladan2017review}.
Foir this reason, electrodes are subjected to periodical tip dressing.
Thus, a critical issue is the swift identification of DRCs mean shifts caused by electrode wear as a criterion to guide the electrode tip dressing program.
To this aim, 919 observations of the multivariate functional quality characteristic corresponding to spot welds made immediately before electrode tip dressing are used to form the Phase I sample. 
The remaining 920 observations are used in Phase II to compare the in-line monitoring performance of the proposed RoMFCC with that of competing methods.
The RoMFCC is implemented as in Section \ref{sec_sim}, where 460 Phase I observations, which are randomly selected without remittance, are used as training set and the remaining 459 ones as tuning set.
Figure \ref{fig_boxplot} shows the boxplot of the  functional distance square root $\sqrt{D_{i,fil}}$ (Equation \eqref{eq_dfil})  obtained from the FUF applied  on the 460 Phase I observation of the training set.
This figure confirms the outlier contamination in the training set, as already illustrated by Figure \ref{fig_drc} for 100 randomly sampled DRCs.


\begin{figure}
	\centering
	\includegraphics[width=0.9\textwidth]{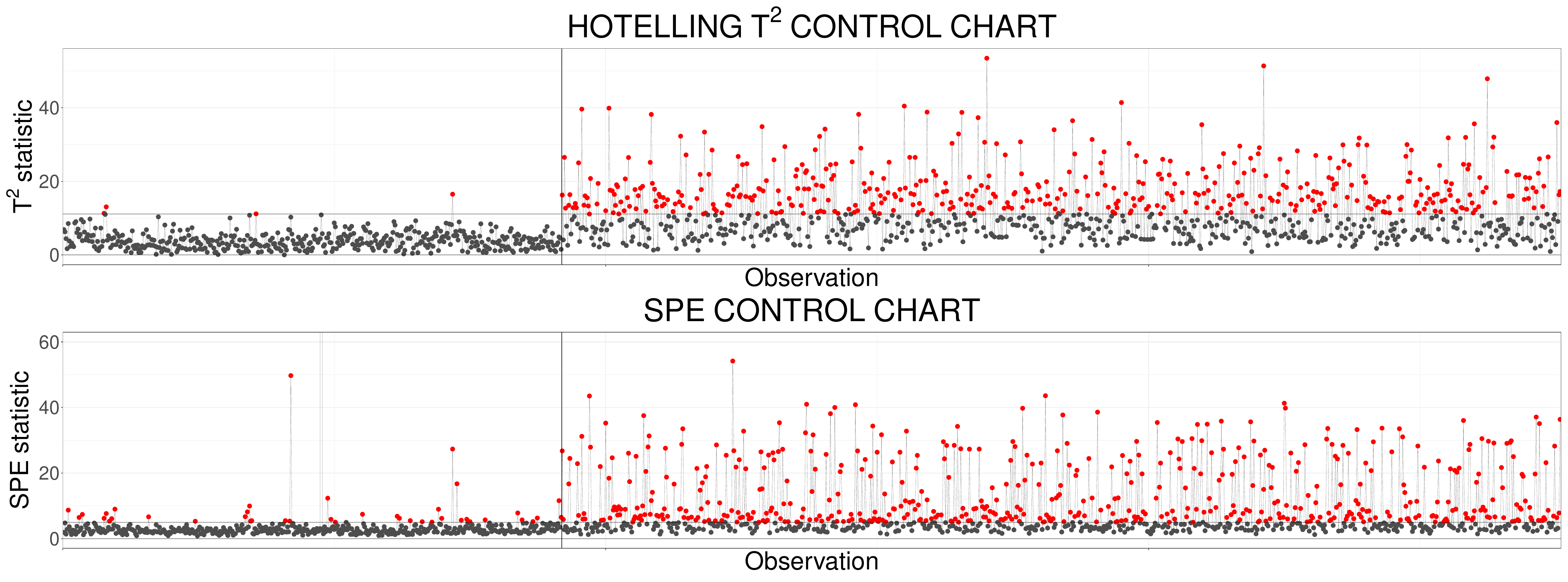}
	\caption{Hotelling's $ T^2 $ and $ SPE $   control charts of the RoMFCC framework for the real-case study data set. The vertical line  separates the monitoring statistics calculated for the tuning set, on the left,  and the Phase II sample, on the right, while the control limits are shown as horizontal lines.}
	\label{fig_ccreal}
  \vspace{-0.5cm}
\end{figure}

Figure \ref{fig_ccreal} shows the Hotelling's $ T^2 $ and $SPE$   control charts of the RoMFCC framework for the real-case study data set.
The vertical line  separates the monitoring statistics calculated for the tuning set, on the left,  and the Phase II sample on the right, while the control limits are shown as horizontal lines. 
Note that the number of tuning set observations that plot above the control limits is expected because these points may include functional casewise outlier not filtered out by the FUF.
In Phase II,  the RoMFCC signals 72.3\% of the observations as OC, which reflects  the proposed method performance in tracking mean shifts caused by  electrode wear.


Finally,  the proposed method is compared with the competing methods presented in the simulation study of Section \ref{sec_sim} through the estimated TDR, denoted as $\widehat{TDR}$, on  the Phase II sample, as shown in Table \ref{tab_arlreal}.
As expected from the results in Section \ref{sec_sim}, the considered non-functional approaches (MCC, iterMCC and RoMCC) show worse performance than the functional counterparts (MFCC, iterMFCC and RoMFCC) because  they are not able  to satisfactorily capture the functional nature of the data.
Morevoer, robust approaches (RoMCC and RoMFCC) outperform the non-robust counterparts, that is the $\widehat{TDR}$ achieved by RoMCC is larger than MCC and iterMCC, whereas the $\widehat{TDR}$ achieved by RoMFCC is larger than MFCC and iterMFCC.
The uncertainty of $\widehat{TDR}$ is quantified through a bootstrap analysis \citep{efron1994introduction}. 
Table \ref{tab_arlreal} reports the mean of the empirical bootstrap distribution of $\widehat{TDR}$, denoted by $\overline{TDR}$,  and the corresponding bootstrap 95\% confidence interval (CI)  for each monitoring method. 
\begin{table}[]
    \centering
     \resizebox{.5\columnwidth}{!}{
    \begin{tabular}{ccccc}
    \toprule
        & & $\widehat{TDR}$ & $\overline{TDR}$ & CI\\
        \midrule
        \multirow{3}{*}{Non functional}&
       MCC & 0.336 & 0.335 & [0.305,0.368]\\
        & iterMCC & 0.462 & 0.461 & [0.428,0.496]\\
        & RoMCC & 0.513 & 0.512 & [0.481,0.547]\\
        \midrule
        \multirow{3}{*}{Functional}
        & MFCC & 0.541 & 0.541 & [0.511,0.574]\\
       & iterMFCC & 0.632 & 0.632 & [0.595,0.664]\\
        & RoMFCC & 0.723 & 0.723 & [0.695,0.753]\\
        \bottomrule
    \end{tabular}
    }
    \caption{Estimated TDR values, denoted as $\widehat{TDR}$, on  the Phase II sample, mean of the empirical bootstrap distribution of $\widehat{TDR}$, denoted by $\overline{TDR}$,  and the corresponding bootstrap 95\% confidence interval (CI)  for each monitoring method. }
    \label{tab_arlreal}
\end{table}
Bootstrap 95\% confidence intervals achieved by the RoMFCC are strictly above those of all considered monitoring approaches.
Therefore,  the proposed RoMFCC stands out as the best method to promptly identify OC conditions in the considered RWS process characterized by a Phase I sample contaminated by functional outliers.

\section{Conclusions}
\label{sec_conclusions}

In this paper, we propose a new robust framework for the statistical process monitoring of multivariate functional data, referred to as \textit{robust multivariate functional control chart} (RoMFCC).

To the best of the authors' knowledge, the RoMFCC  is the first statistical process monitoring (SPM) framework for multivariate functional quality characteristic that is robust to functional casewise and cellwise outliers.
Indeed, methods already present in the literature either apply robust approaches to multivariate scalar features extracted from the profiles or use diagnostic approaches on the multivariate functional data to iteratively remove outliers. However, the former  are not able to capture the functional nature of the data, while the latter  are not able to deal with functional cellwise outliers.


The performance of the RoMFCC framework is assessed through an extensive Monte Carlo simulation study and is compared with several competing monitoring methods for multivariate scalar data and multivariate functional data.
The ability  to estimate the distribution of the data without removing observations  allows the  RoMFCC to outperform the competitors in all the considered settings and to represent the only alternative in high-dimensional scenarios where  most of the competing methods may even fail.
The proposed method is suitable to monitor industrial processes where many functional variables are available and are possibly contaminated by outliers,  as anomalies in the data acquisition and data collected during a fault in the process.

The practical applicability of the proposed method is illustrated through the real-case study that motivated this research and addressed the issue of monitoring the quality of a resistance spot-welding (RSW) process in the automotive industry through  multivariate observations of the dynamic resistance curves as the quality characteristic of interest. Also in this real-case study, the RoMFCC  outperforms all the considered competitors in the identification of out-of-control state of the RSW process due to the excessive wear of the electrode used.

\section*{Supplementary Materials}
The Supplementary Materials contain additional details about  the data generation process in the simulation study (A), additional simulation results (B), as well as the R code  to reproduce graphics and results over competing methods in the simulation study.

\bibliographystyle{apalike}
\setlength{\bibsep}{5pt plus 0.2ex}
{\small
\spacingset{1}
\bibliography{ref}
}
\end{document}


\def\spacingset#1{\renewcommand{\baselinestretch}%
{#1}\small\normalsize} \spacingset{1}


\if0\blind
{
  
\title{Supplementary Materials to ``Robust Multivariate Functional Control Chart''}
\author[]{Christian Capezza}
\author[]{Fabio Centofanti \thanks{Corresponding author. e-mail: \texttt{fabio.centofanti@unina.it}}}
\author[]{Antonio Lepore}
\author[]{Biagio Palumbo}

\affil[]{Department of Industrial Engineering, University of Naples Federico II, Piazzale Tecchio 80, 80125, Naples, Italy}
\setcounter{Maxaffil}{0}
\renewcommand\Affilfont{\itshape\small}
\date{}
\maketitle
} \fi

\if1\blind
{
  \bigskip
  \bigskip
  \bigskip
  \begin{center}
    {\LARGE\bf Supplementary Materials to ``Robust Multivariate Functional Control Chart''}
\end{center}
  \medskip
} \fi

\vfill
\hfill {\tiny technometrics tex template (do not remove)}

\newpage
\spacingset{1.45} 
\appendix
\numberwithin{equation}{section}

\section{Details on Data Generation in the Simulation Study}
\label{sec_appB}
The data generation process is inspired by the real-case study  in Section 4 and mimics typical behaviours of DRCs in a RSW process.
The data correlation structure is generated similar to \cite{centofanti2020functional,chiou2014multivariate}.
The compact domain $\mathcal{T}$ is  set, without loss of generality, equal to $\left[0,1\right]$ and the number components $p$ is set equal to 10.
The eigenfunction set $\lbrace \bm{\psi}_i\rbrace $ is generated by considering the correlation function $\bm{G}$ through the following steps.
\begin{enumerate}
\item Set the diagonal elements $G_{ll}$, $l=1,\dots,p$ of $\bm{G}$ as the \textit{Bessel} correlation function of the first kind \citep{abramowitz1964handbook}. The general form of the correlation function and parameter used  are listed in Table \ref{ta_corf}. Then, calculate the eigenvalues $\lbrace\eta_{lk}^{X}\rbrace$ and  the corresponding eigenfunctions $\lbrace\vartheta_{lk}\rbrace$, $k=1,2,\dots$,  of $G_{ll}$, $l=1,\dots,p$.
\item Obtain the cross-correlation function $G_{lj}$, $l,j=1,\dots,p$ and $l\neq j$, by
\begin{equation}
G_{lj}\left(t_1,t_2\right)=\sum_{k=1}^{\infty}\frac{\eta_{k}}{1+|l-j|}\vartheta_{lk}\left(t_{1}\right)\vartheta_{jk}\left(t_{2}\right)\quad t_1,t_2\in\mathcal{T}.
\end{equation}
\item Calculate the eigenvalues $\lbrace\lambda_i\rbrace$ and the corresponding eigenfunctions  $\lbrace \bm{\psi}_i\rbrace $ through the spectral decomposition of $\bm{G}=\lbrace G_{lj}\rbrace_{l,j=1,\dots,p}$, for $i=1,\dots,L^{*}$.
\end{enumerate}
\begin{table}
\caption{Bessel correlation function and parameter for data generation in the simulation study.}
\label{ta_corf}

\centering
\resizebox{0.5\textwidth}{!}{
\begin{tabular}{ccc}
\toprule
&$\rho$&$\nu$\\
\midrule
$J_{v}\left(z\right)=\binom{|z|/\rho}{2}^{\nu}\sum_{j=0}^{\infty}\frac{\left(-\left(|z|/\rho\right)^{2}/4\right)^{j}}{j!\Gamma\left(\nu+j+1\right)}$&0.125&0\\[.35cm]

\bottomrule
\end{tabular}}
\end{table}
Further,  $L^{*}$ is set equal to $10$.
Let   $\bm{Z}=\left(Z_1,\dots,Z_p\right)$ as
\begin{equation}
\bm{Z}=\sum_{i=1}^{L^{*}}\xi_i\bm{\psi}_i.
\end{equation}
with $\bm{\xi}_{L^{*}}=\left(\xi^{X}_1,\dots,\xi^{X}_{L^{*}}\right)^{T}$   generated by means of a  multivariate normal distribution with covariance $\Cov\left(\bm{\xi}_{L^{*}}^{X}\right)=\bm{\Lambda^{X}}=\diag\left(\lambda_1,\dots,\lambda_{L^{*}}\right)$.

Furthermore, let  the mean process $m$  
\begin{multline}
    m(t)= 0.2074 + 0.3117\exp(-371.4t) +0.5284(1 - \exp(0.8217t))\\ -423.3\left[1 + \tanh(-26.15(t+0.1715)) \right]\quad t\in\mathcal{T}.
\end{multline}
Note that, the mean function $m$  is generated to resemble a typical DRC through the phenomenological model for the RSW process  presented in \cite{schwab2012improving}.
Then, the  contamination models $C_E$ and $C_P$, which mimics a splash weld (expulsion) and phase shift of the peak time, are respectively defined as 
\begin{equation}
    C_E(t)=\min\Big\lbrace 0, -2M_E(t-0.5)\Big\rbrace \quad t\in\mathcal{T},
\end{equation}
and
\begin{multline}
    C_P(t)= -m(t)-(M_P/20)t + 0.2074\\ + 0.3117\exp(-371.4h(t)) +0.5284(1 - \exp(0.8217h(t)))\\ -423.3\left[1 + \tanh(-26.15(h(t)+0.1715))\right]  \quad t\in\mathcal{T},
\end{multline}
where $h:\mathcal{T}\rightarrow\mathcal{T}$ transforms the temporal dimension $t$ as follows 
\begin{equation}
    h(t)= \begin{cases} 
    t & \text{if } t\leq 0.05 \\
	\frac{0.55-M_P}{0.55}t-(1+\frac{0.55-M_P}{0.55})0.05 & \text{if } 0.05< t\leq 0.6 \\
		\frac{0.4+M_P}{0.4}t+1-\frac{0.4+M_P}{0.4} & \text{if }  t> 0.6, \\
	\end{cases}
\end{equation} 
and $M_E$ and $M_P$ are  contamination sizes.
Then, $\bm{X}=\left(X_1,\dots,X_p\right)^T$ is obtained as follows
\begin{equation}
\label{eq_modgen}
\bm{X}\left(t\right)=  \bm{m}(t) +\bm{Z}\left(t\right)\sigma+\bm{\varepsilon}\left(t\right) +B_E\bm{C}_E(t)+B_P\bm{C}_P(t) \quad t\in \mathcal{T},
\end{equation}
where $\bm{m}$ is a $p$ dimensional vector with components  equal to $m$, $\sigma>0$, $\bm{\varepsilon}=\left(\varepsilon_1,\dots,\varepsilon_p\right)^T$, where $\varepsilon_i$ are  white noise functions such that for each $ t \in \left[0,1\right] $, $ \varepsilon_i\left(t\right) $ are  normal random varaibles with zero mean and standard deviation $ \sigma_e $, and $B_{CaE}$ and $B_{CaP}$ are two independent random
variables following  Bernoulli distributions with parameters $p_{CaE}$ and $p_{CaP}$, respectively.
Moreover, $\bm{C}_E=\left(B_{1,CeE}C_E,\dots,B_{p,CeE}C_E\right)^T$ and $\bm{C}_P=\left(B_{1,CeP}C_P,\dots,B_{p,CeP}C_P\right)^T$, where $\lbrace B_{i,CeE}\rbrace$ and $\lbrace B_{i,CeP}\rbrace$ are two sets of  independent random variables following  Bernoulli distributions with parameters $p_{CeE}$ and $p_{CeP}$, respectively. 

Then, the Phase I and Phase II samples are generated through Equation \eqref{eq_modgen} by considering the parameters listed in Table \ref{ta_1} and Table \ref{ta_2}, respectively, with $\sigma_e=0.0025$ and $\sigma=0.01$.
The parameter $\tilde{p}$ is the probability of contamination that for the data analysed in Section 3 and Supplementary Material B is set equal to 0.05 and 0.1, respectively. 
\begin{table}
	\caption{Parameters used to generate the Phase I sample  in Scenario 1 and Scenario 2 of the simulation study.}
	\label{ta_1}
	
	\centering
	\resizebox{\textwidth}{!}{
		\begin{tabular}{cM{0.1\textwidth}M{0.1\textwidth}M{0.1\textwidth}M{0.1\textwidth}M{0.1\textwidth}M{0.1\textwidth}M{0.1\textwidth}M{0.1\textwidth}}
			\toprule
			&\multicolumn{4}{c}{Scenario 1}&	\multicolumn{4}{c}{Scenario 2}\\
			\cmidrule(lr){2-5}\cmidrule(lr){6-9}
			&\multicolumn{2}{c}{Out-E}&\multicolumn{2}{c}{Out-P}&\multicolumn{2}{c}{Out-E}&\multicolumn{2}{c}{Out-P}\\
			\cmidrule(lr){2-5}\cmidrule(lr){6-9}
			&\multicolumn{2}{c}{\specialcell{$p_{CaE}=p_{CaP}=1$,\\ $p_{CeE}=\tilde{p}$, $p_{CeP}=0$}}  &\multicolumn{2}{c}{\specialcell{$p_{CaE}=p_{CaP}=1$,\\ $p_{CeE}=0$, $p_{CeP}=\tilde{p}$}}
			&\multicolumn{2}{c}{\specialcell{$p_{CaE}=\tilde{p}$, $p_{CaP}=0$,\\ $p_{CeE}=p_{CeP}=1$}}
			&\multicolumn{2}{c}{\specialcell{$p_{CaE}=0$, $p_{CaP}=\tilde{p}$, \\$p_{CeE}=p_{CeP}=1$}}\\
			\cmidrule(lr){2-3}\cmidrule(lr){4-5}\cmidrule(lr){6-7}\cmidrule(lr){8-9}
			&$M_E$&$M_P$&$M_E$&$M_P$&$M_E$&$M_P$&$M_E$&$M_P$\\
			\cmidrule(lr){2-3}\cmidrule(lr){4-5}\cmidrule(lr){6-7}\cmidrule(lr){8-9}
			C1&0.04 &0.00&0.00&0.40&0.02 &0.00&0.00 &0.20\\
		    C2&0.06 & 0.00&0.00&0.45&0.03 &0.00&0.00 &0.30\\
		    C3&0.08 & 0.00&0.00&0.50&0.04 &0.00&0.00 &0.40\\

			\bottomrule
	\end{tabular}}
\end{table}
\begin{table}
	\caption{Parameters used to generate the Phase II sample  for OC-E and OC-P and severity level $SL = \lbrace 0, 1, 2, 3, 4 \rbrace$ in the simulation study.}
	\label{ta_2}
	
	\centering
	\resizebox{0.5\textwidth}{!}{
		\begin{tabular}{cM{0.1\textwidth}M{0.1\textwidth}M{0.1\textwidth}M{0.1\textwidth}}
			\toprule
		
			&\multicolumn{2}{c}{OC-E}&\multicolumn{2}{c}{OC-P}\\
			\cmidrule(lr){2-5}
		    &\multicolumn{2}{c}{\specialcell{$p_{CaE}=1$, $p_{CaP}=0$,\\ $p_{CeE}=1$, $p_{CeP}=0$}}
			&\multicolumn{2}{c}{\specialcell{$p_{CaE}=0$, $p_{CaP}=1$,\\ $p_{CeE}=0$, $p_{CeP}=1$}}
			\\
			\cmidrule(lr){2-3}\cmidrule(lr){4-5}
			$SL$&$M_E$&$M_P$&$M_E$&$M_P$\\
			\cmidrule(lr){2-3}\cmidrule(lr){4-5}
			0&0.00 &0.00&0.00&0.00\\
		    1&0.01 & 0.00&0.00&0.20\\
		    2&0.02 & 0.00&0.00&0.27\\
		    3&0.03 & 0.00&0.00&0.34\\
		    4&0.04 & 0.00&0.00&0.40\\

			\bottomrule
	\end{tabular}}
\end{table}
Moreover, in Scenario 0 of the simulation study data are generated thorugh $p_{CaE}=p_{CaP}=p_{CeE}=p_{CeP}=0$.
Finally, the generate data are assumed to be discretely observed at 100 equally spaced time points over the domain $\left[0,1\right]$.

\section{Additional Simulation Results}
\begin{figure}[h]
	\caption{Mean FAR ($ SL=0 $) or TDR ($ SL\neq 0 $) achieved by MCC, iterMCC, RoMCC, MFCC, MFCCiter and RoMFCC for each contamination level (C1, C2 and C3), OC condition (OC-E and OC-P) as a function of the severity level $SL$ with contamination model Out-E and Out-P in Scenario 1 for $\tilde{p}=0.1$.}
	
	\label{fi_results_2_1}
	
	\centering
		\hspace{-2.5cm}
	\begin{tabular}{cM{0.24\textwidth}M{0.24\textwidth}M{0.24\textwidth}M{0.24\textwidth}}
			&\multicolumn{2}{c}{\hspace{0.12cm} \textbf{\large{Out-E}}}&	\multicolumn{2}{c}{\hspace{0.12cm} \textbf{\large{Out-P}}}\\
		&\hspace{0.6cm}\textbf{\footnotesize{OC-E}}&\hspace{0.6cm}\textbf{\footnotesize{OC-P}}&\hspace{0.5cm}\textbf{\footnotesize{OC-E}}&\hspace{0.5cm}\textbf{\footnotesize{OC-P}}\\
		\textbf{\footnotesize{C1}}&\includegraphics[width=.25\textwidth]{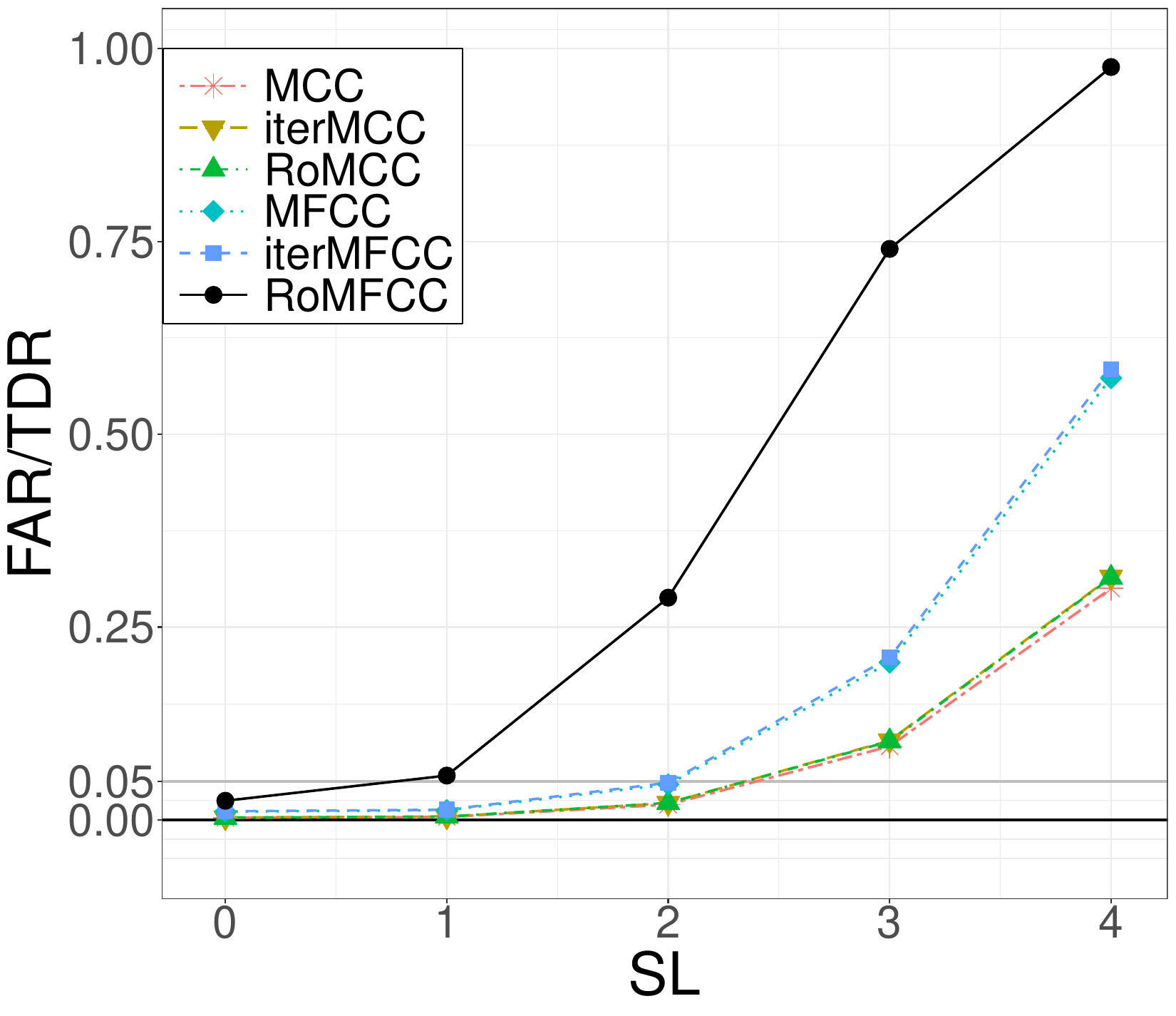}&\includegraphics[width=.25\textwidth]{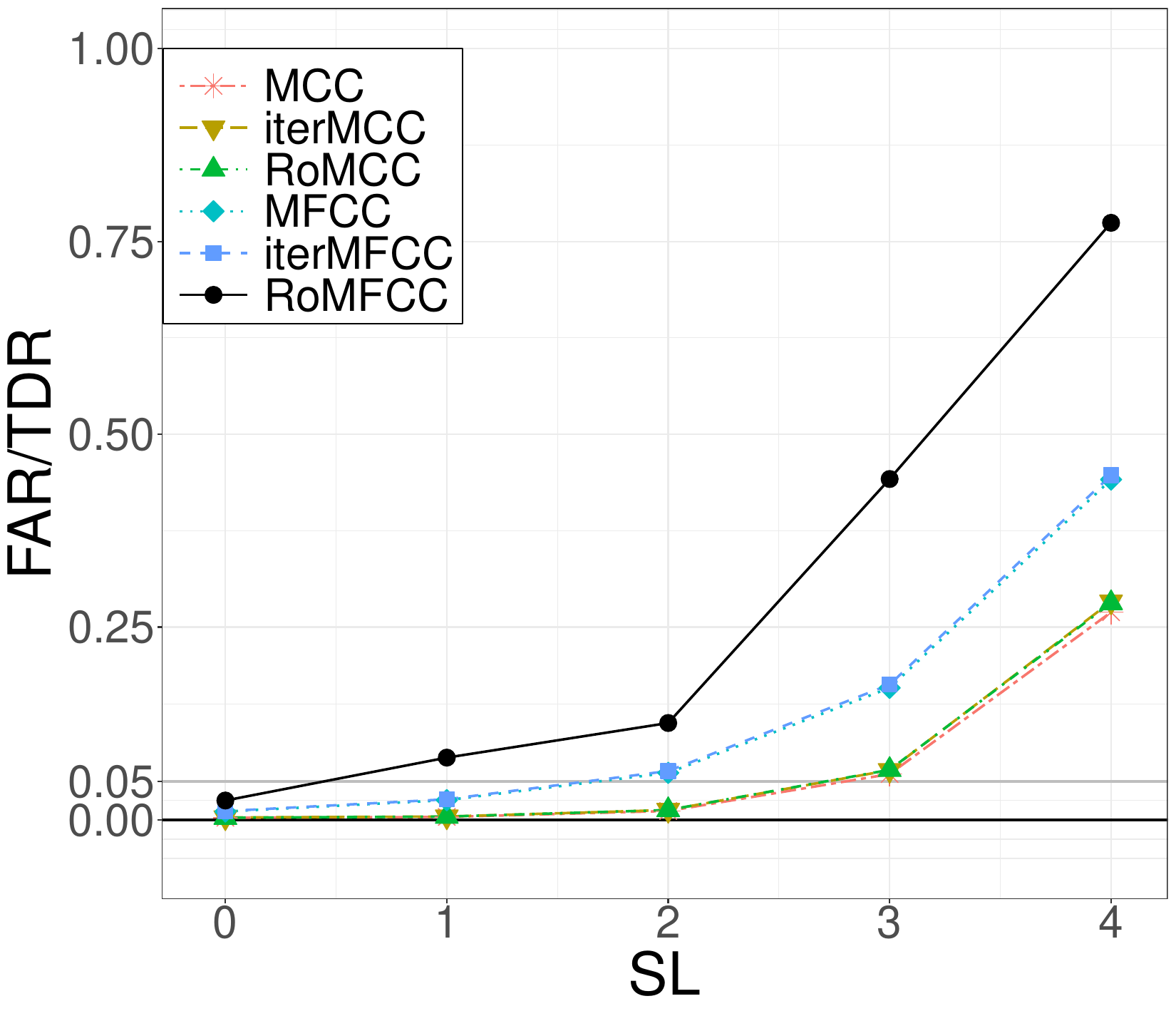}&\includegraphics[width=.25\textwidth]{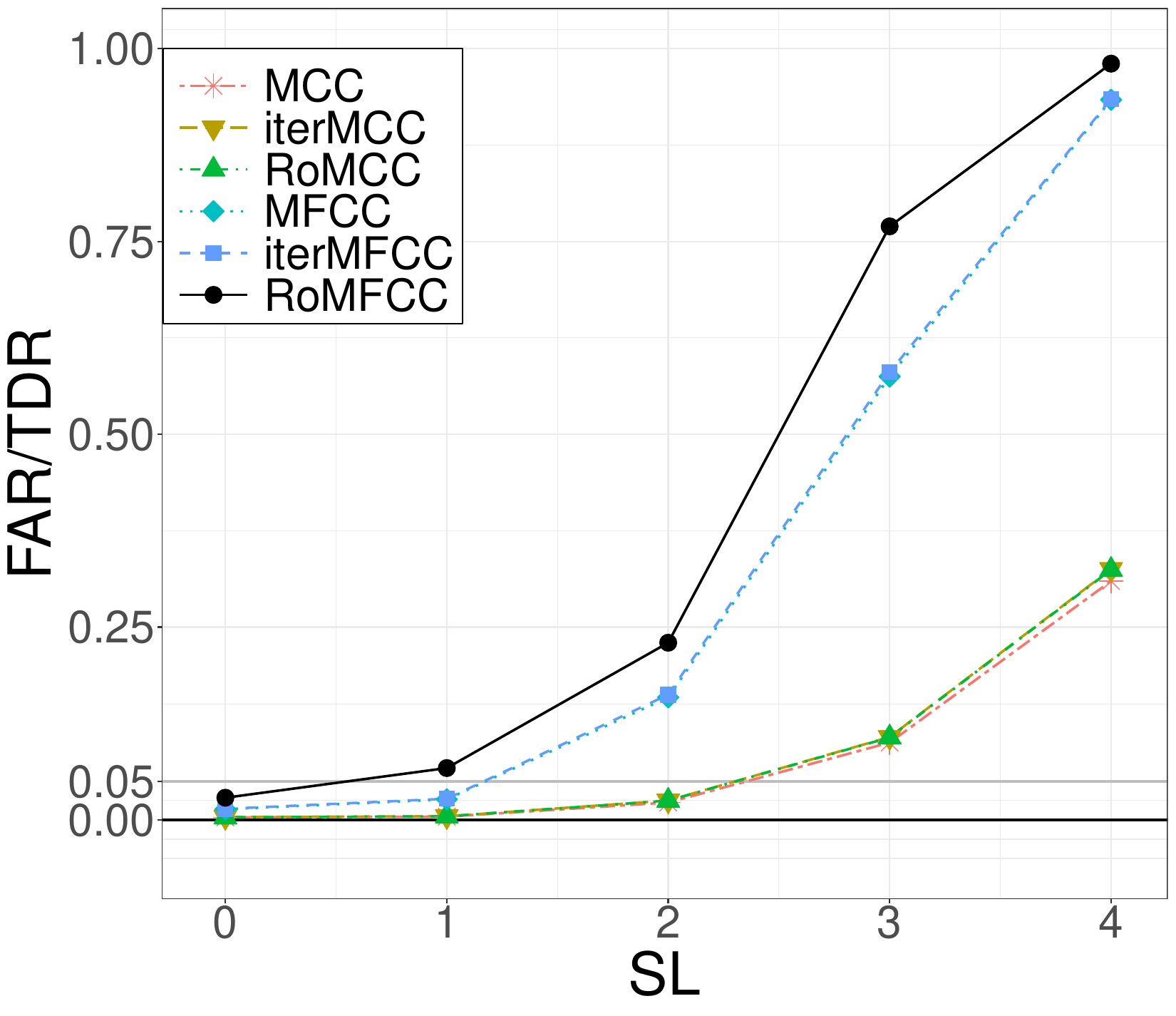}&\includegraphics[width=.25\textwidth]{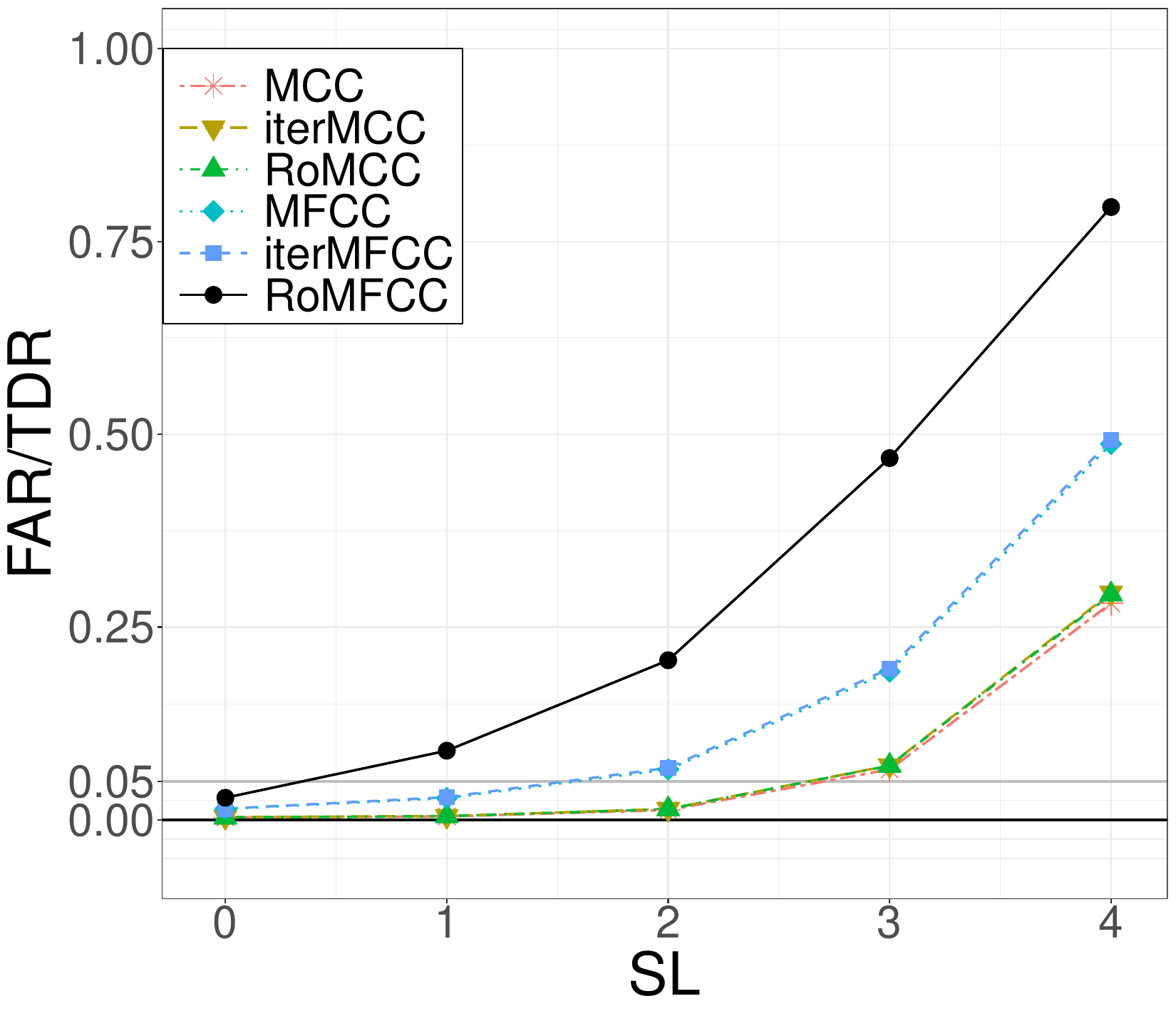}\\
		\textbf{\footnotesize{C2}}&\includegraphics[width=.25\textwidth]{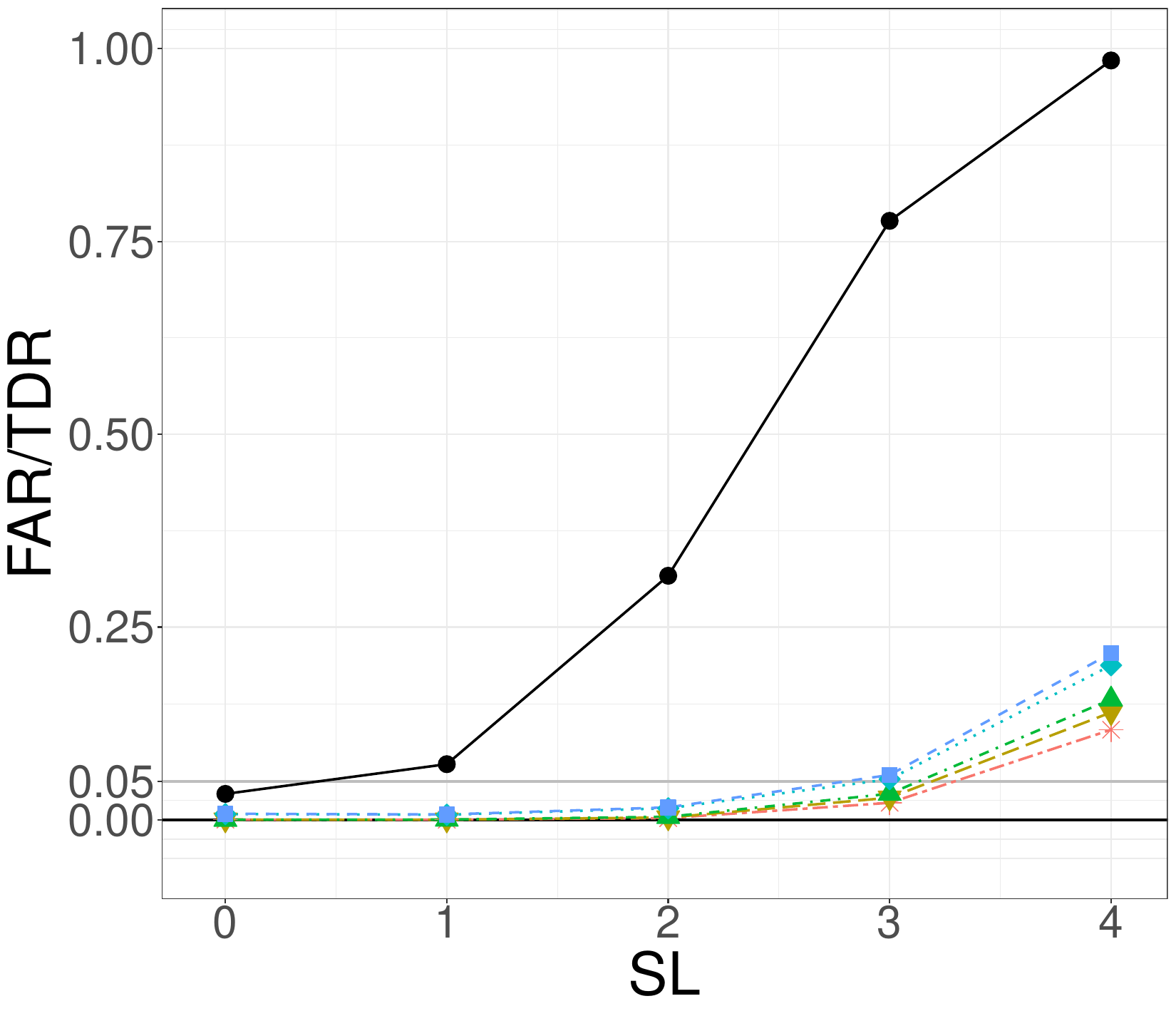}&\includegraphics[width=.25\textwidth]{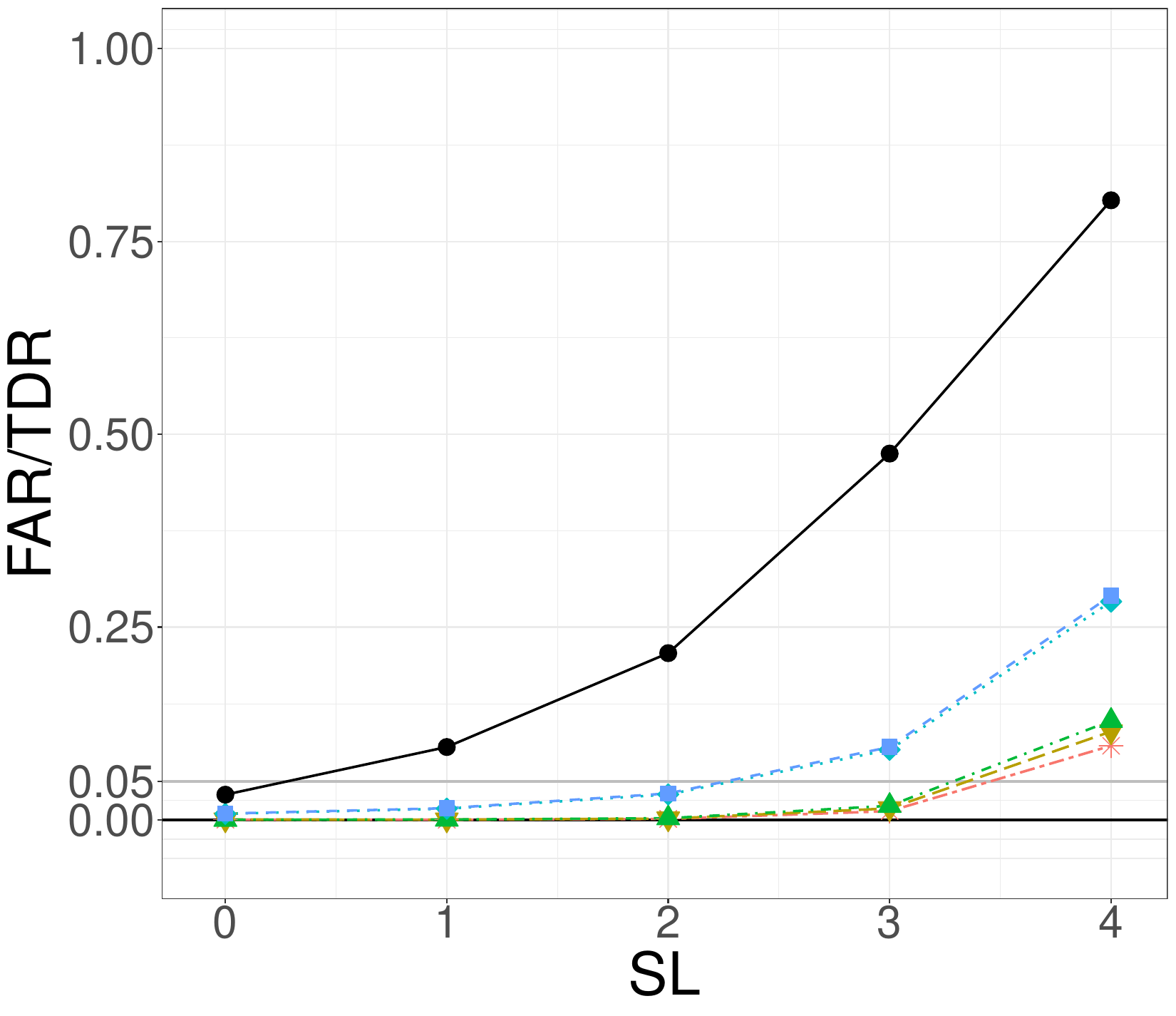}&\includegraphics[width=.25\textwidth]{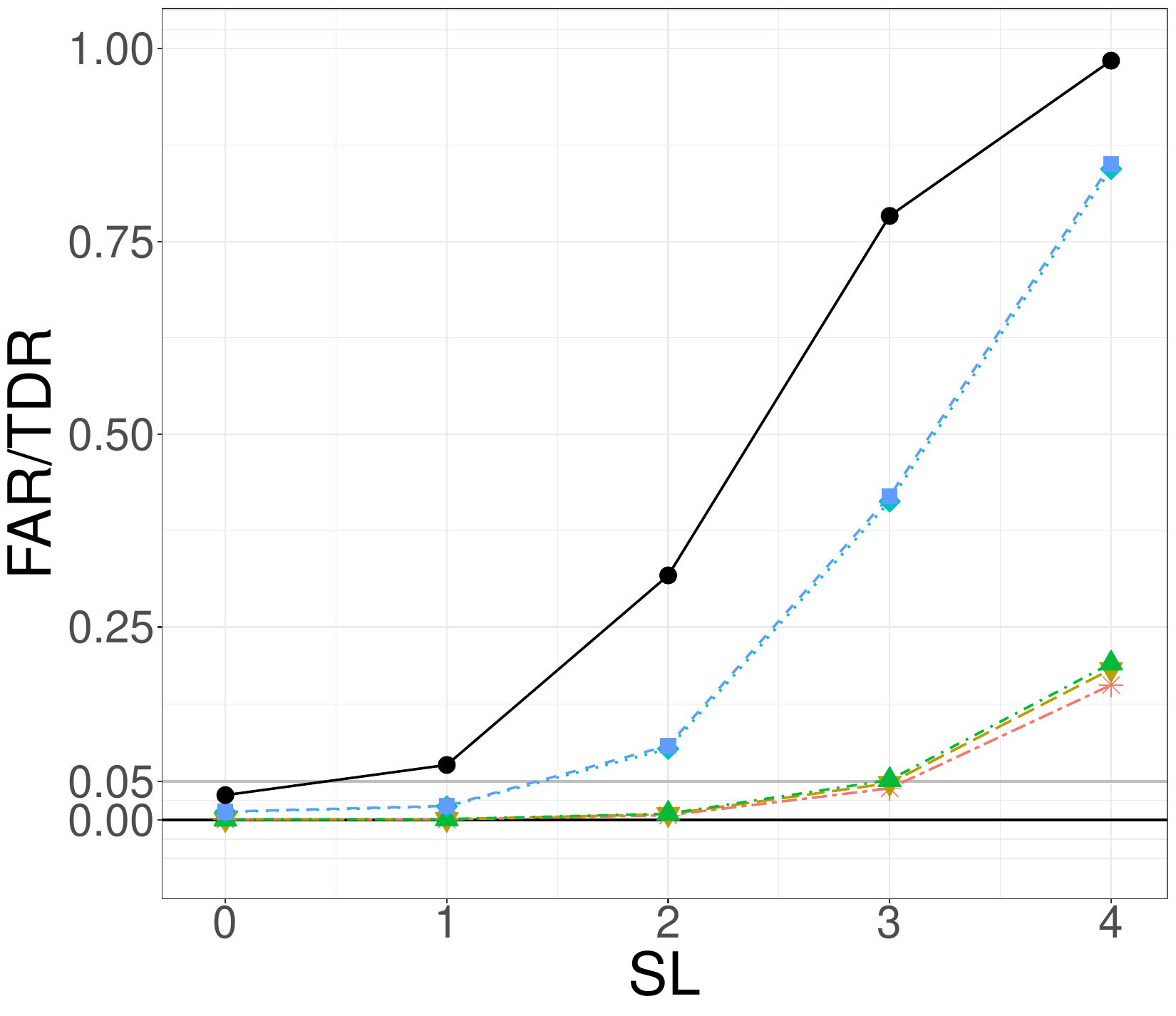}&\includegraphics[width=.25\textwidth]{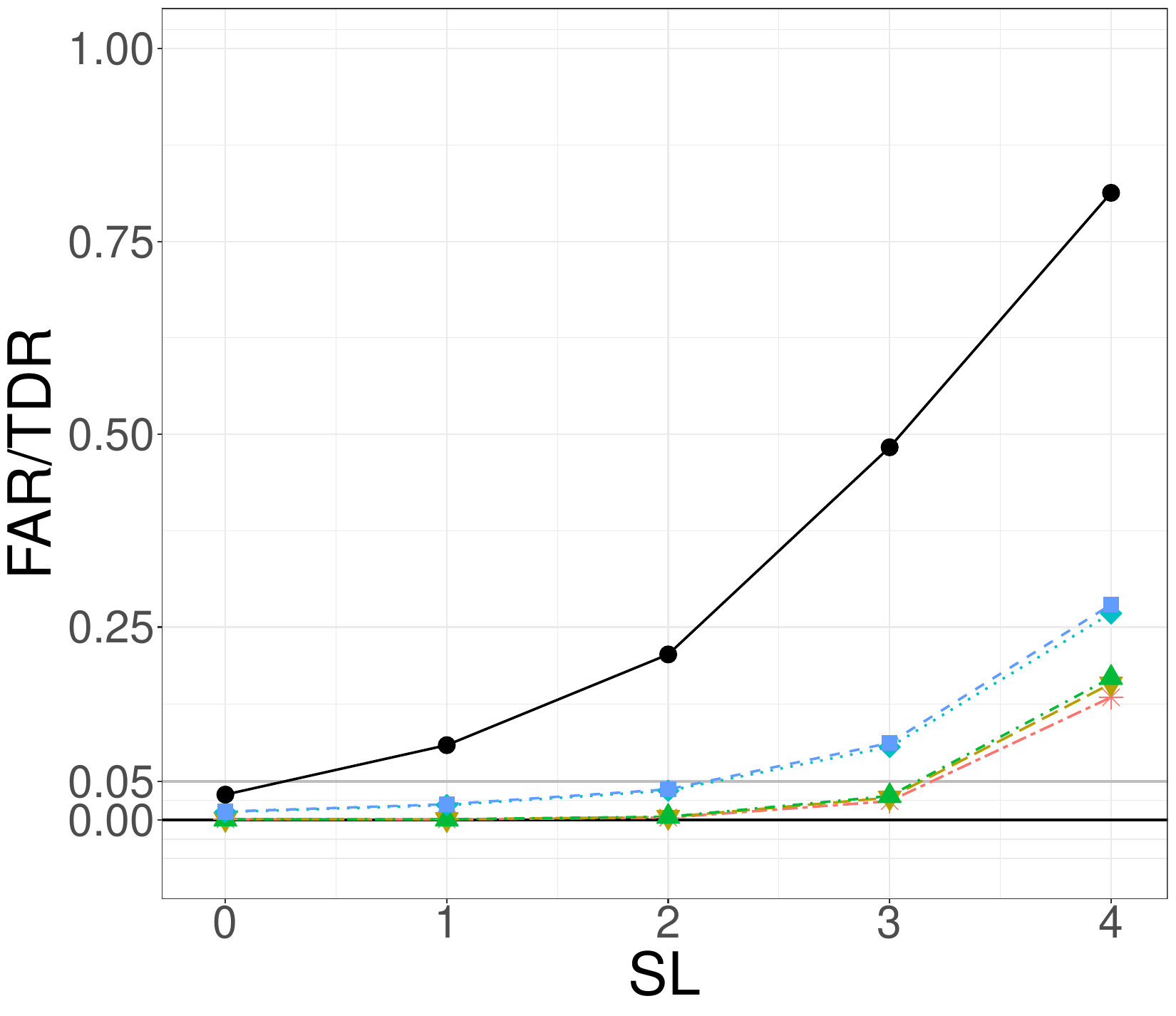}\\
		\textbf{\footnotesize{C3}}&\includegraphics[width=.25\textwidth]{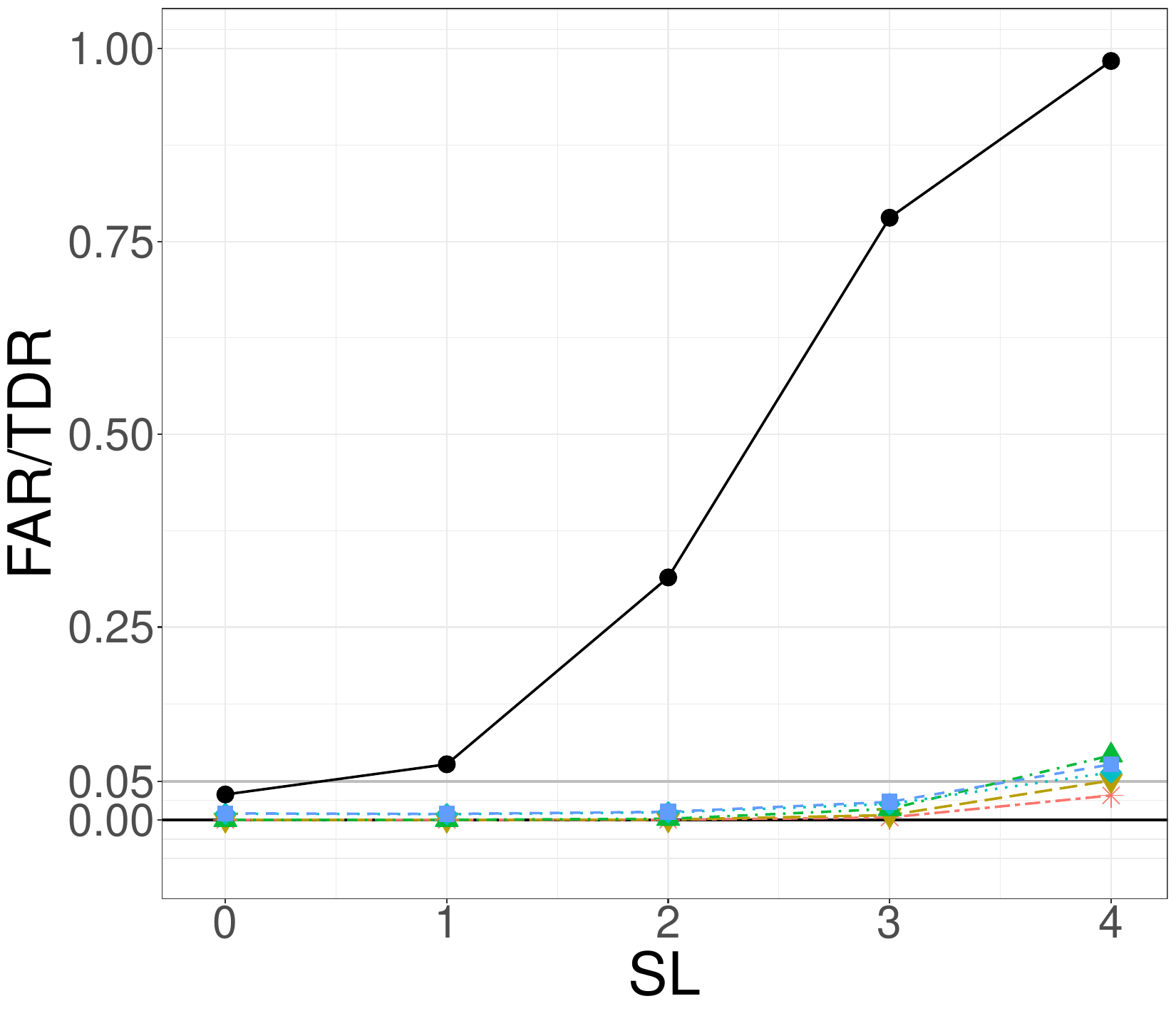}&\includegraphics[width=.25\textwidth]{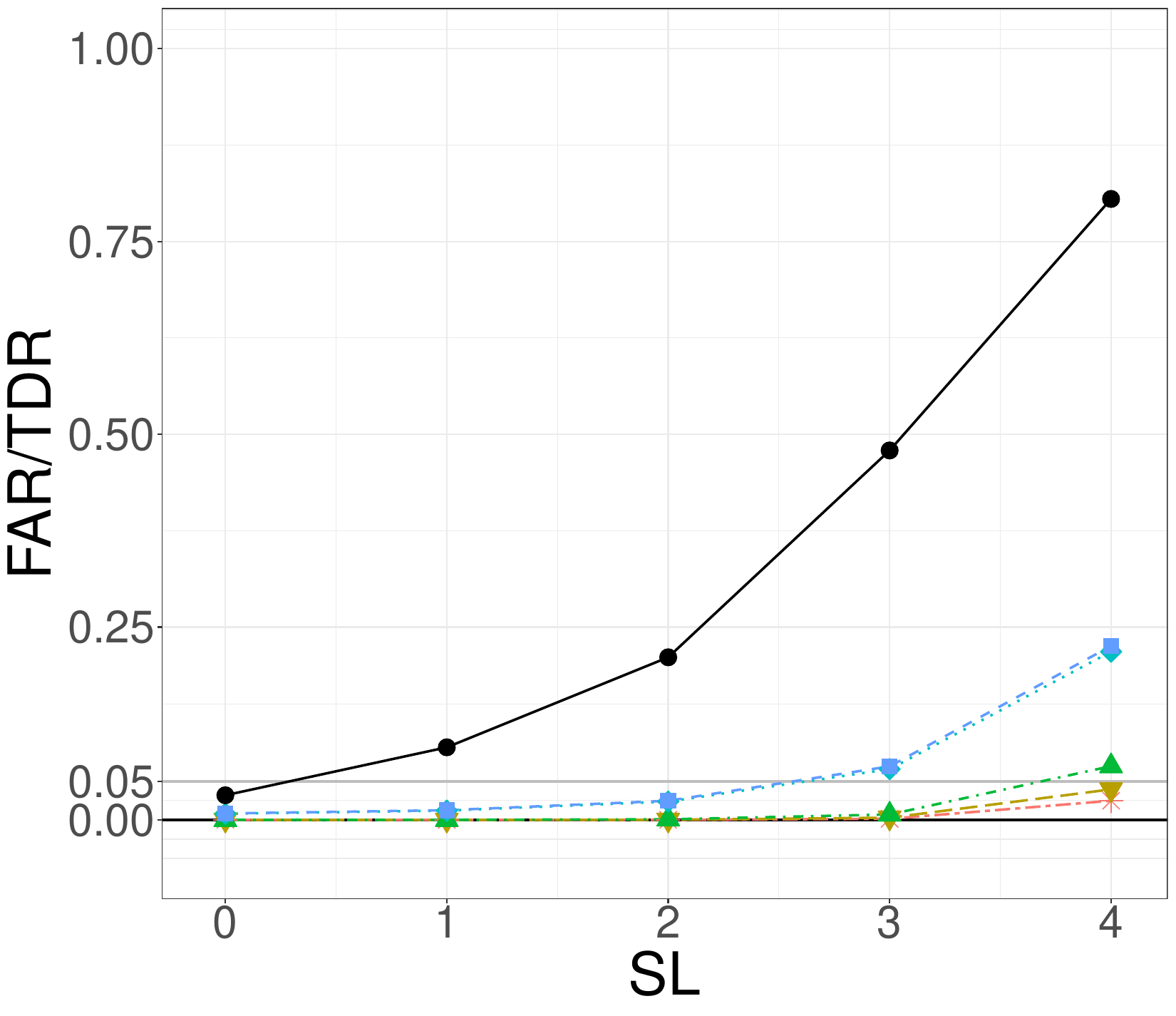}&\includegraphics[width=.25\textwidth]{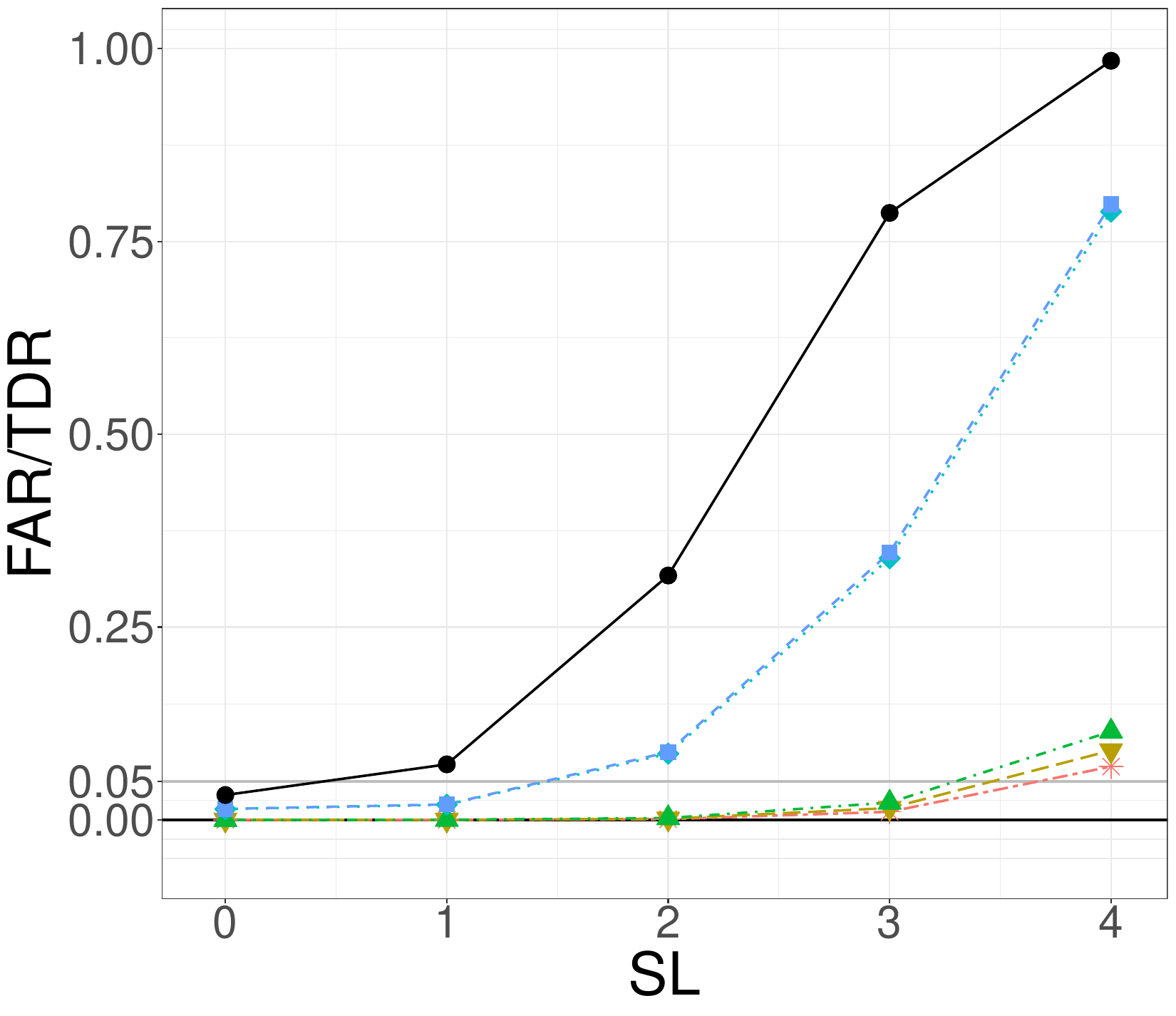}&\includegraphics[width=.25\textwidth]{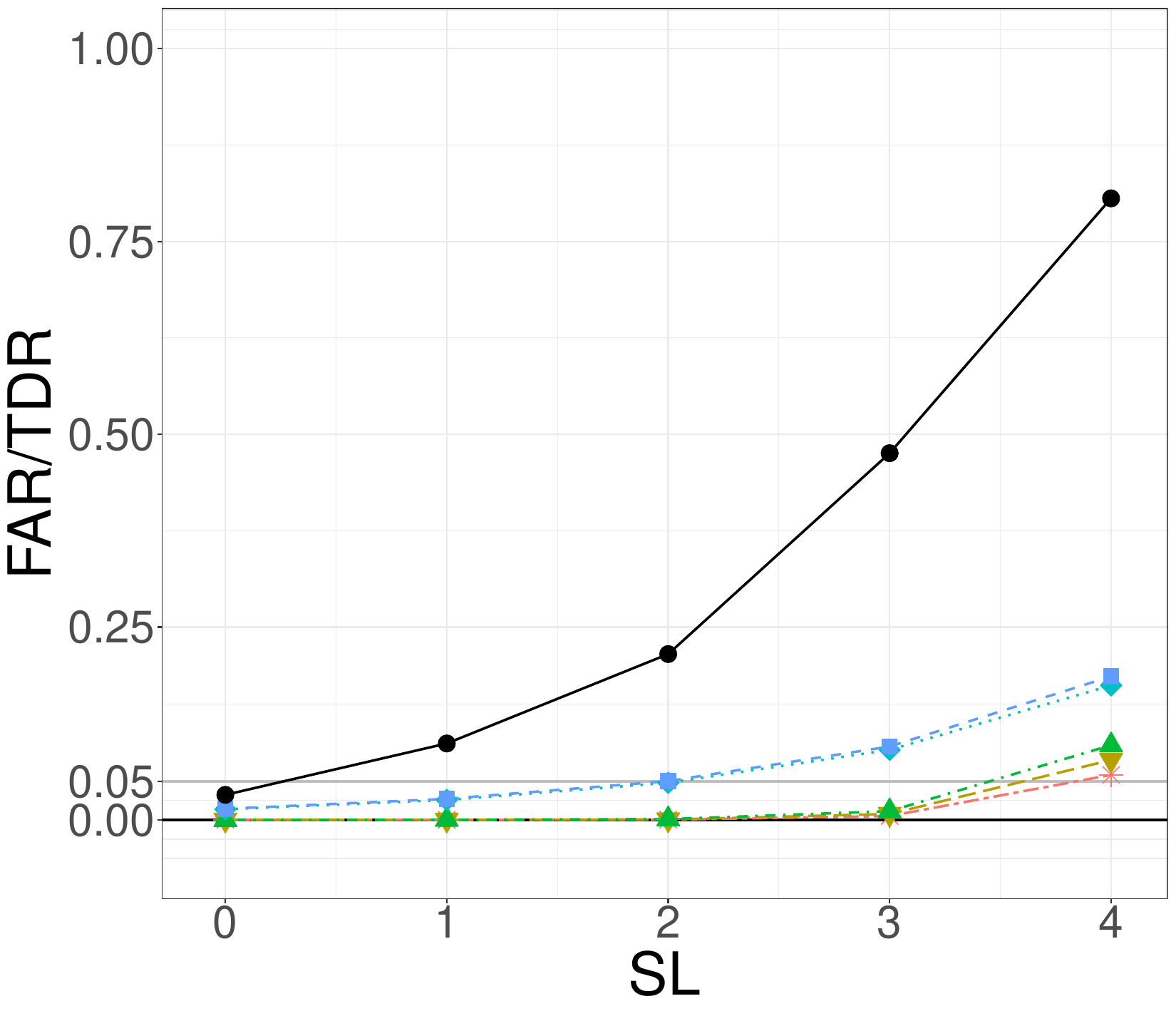}\\
	\end{tabular}
	\vspace{-.5cm}
\end{figure}

\begin{figure}[h]
	\caption{Mean FAR ($ SL=0 $) or TDR ($ SL\neq 0 $) achieved by MCC, iterMCC, RoMCC, MFCC, MFCCiter and RoMFCC for each contamination level (C1, C2 and C3), OC condition (OC-E and OC-P) as a function of the severity level $SL$ with contamination model Out-E and Out-P in Scenario 2 for $\tilde{p}=0.1$.}
	
	\label{fi_results_3_1}
	
	\centering
		\hspace{-2.5cm}
	\begin{tabular}{cM{0.24\textwidth}M{0.24\textwidth}M{0.24\textwidth}M{0.24\textwidth}}
			&\multicolumn{2}{c}{\hspace{0.12cm} \textbf{\large{Out-E}}}&	\multicolumn{2}{c}{\hspace{0.12cm} \textbf{\large{Out-P}}}\\
		&\hspace{0.6cm}\textbf{\footnotesize{OC-E}}&\hspace{0.6cm}\textbf{\footnotesize{OC-P}}&\hspace{0.5cm}\textbf{\footnotesize{OC-E}}&\hspace{0.5cm}\textbf{\footnotesize{OC-P}}\\
		\textbf{\footnotesize{C1}}&\includegraphics[width=.25\textwidth]{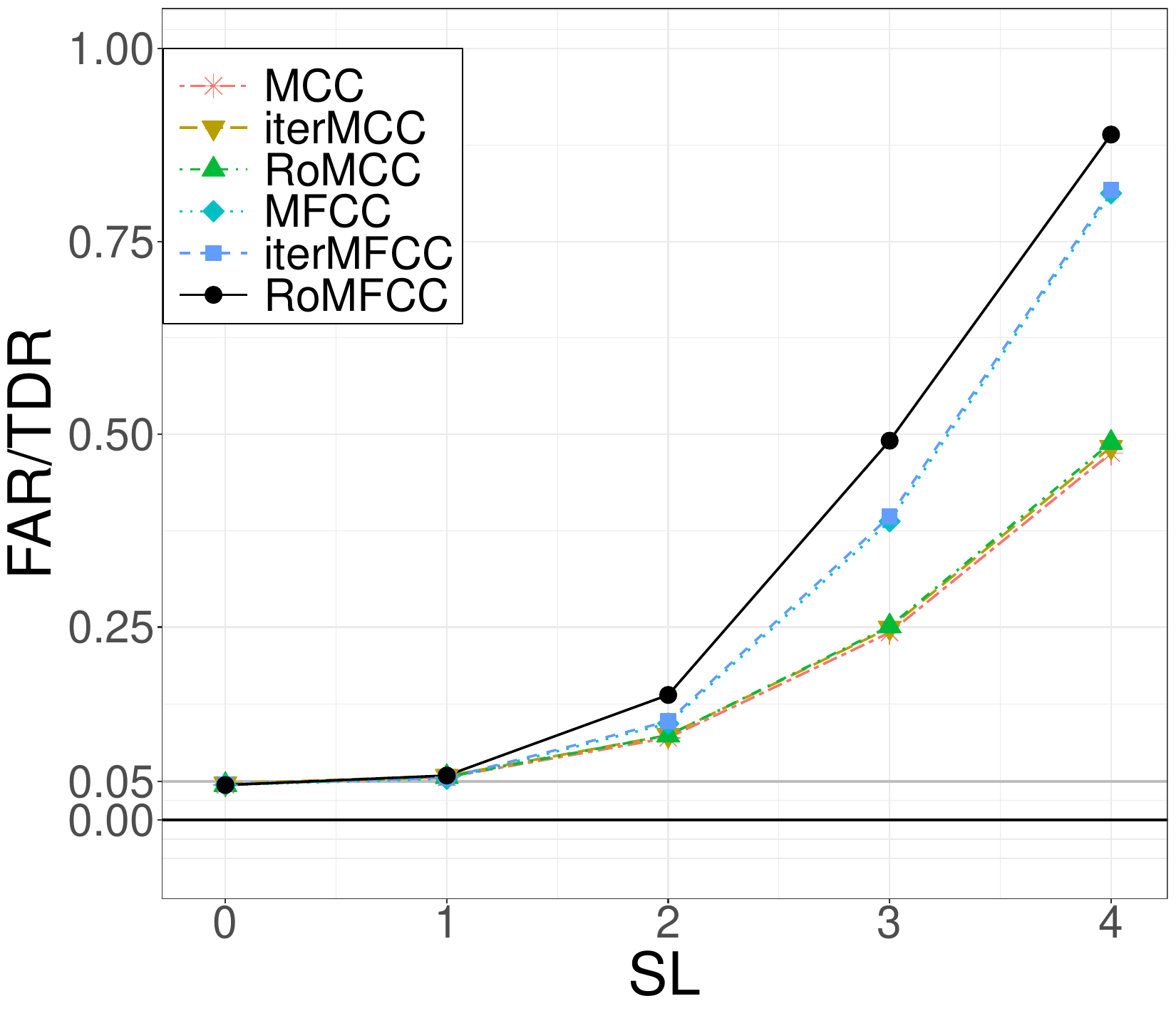}&\includegraphics[width=.25\textwidth]{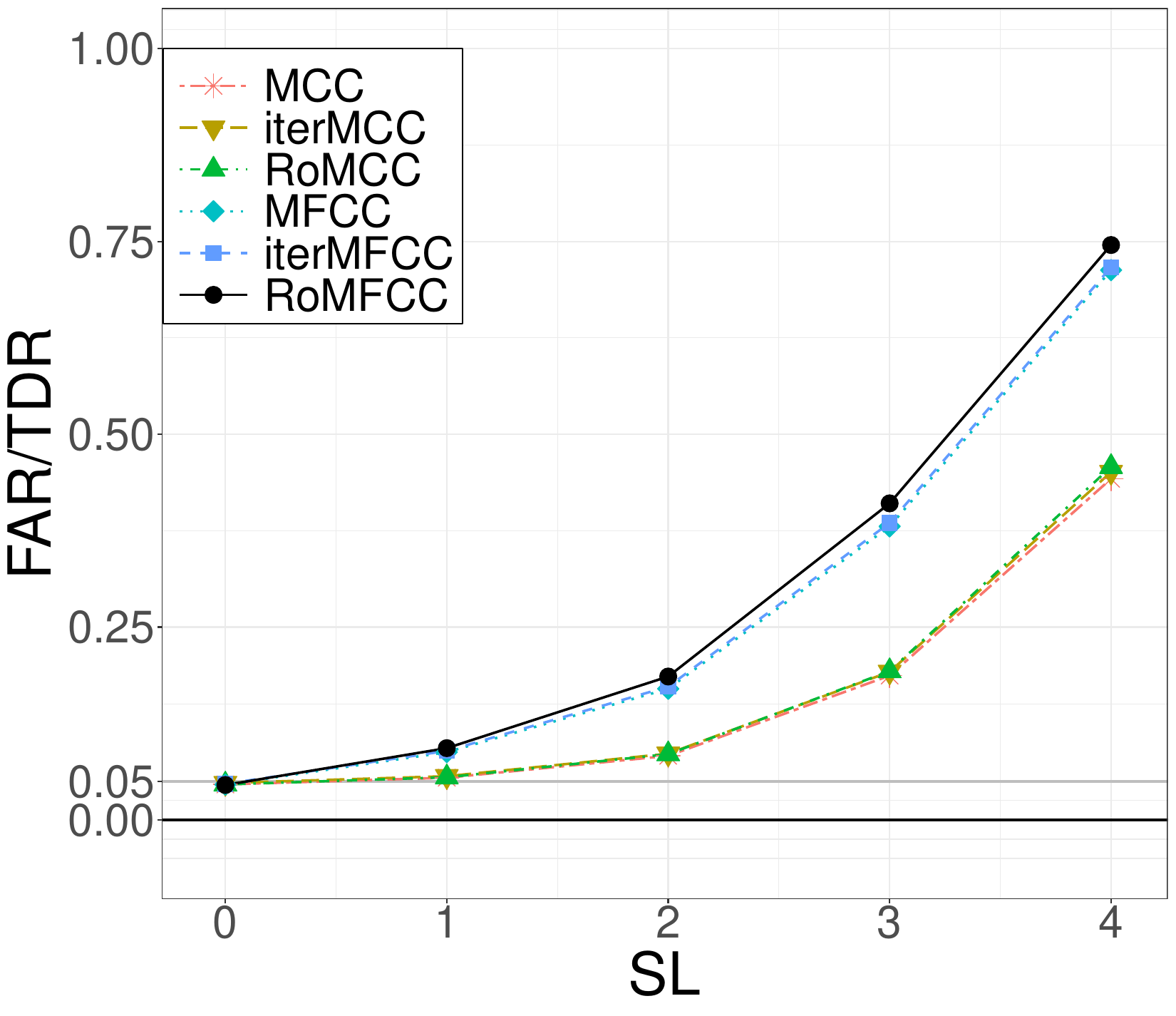}&\includegraphics[width=.25\textwidth]{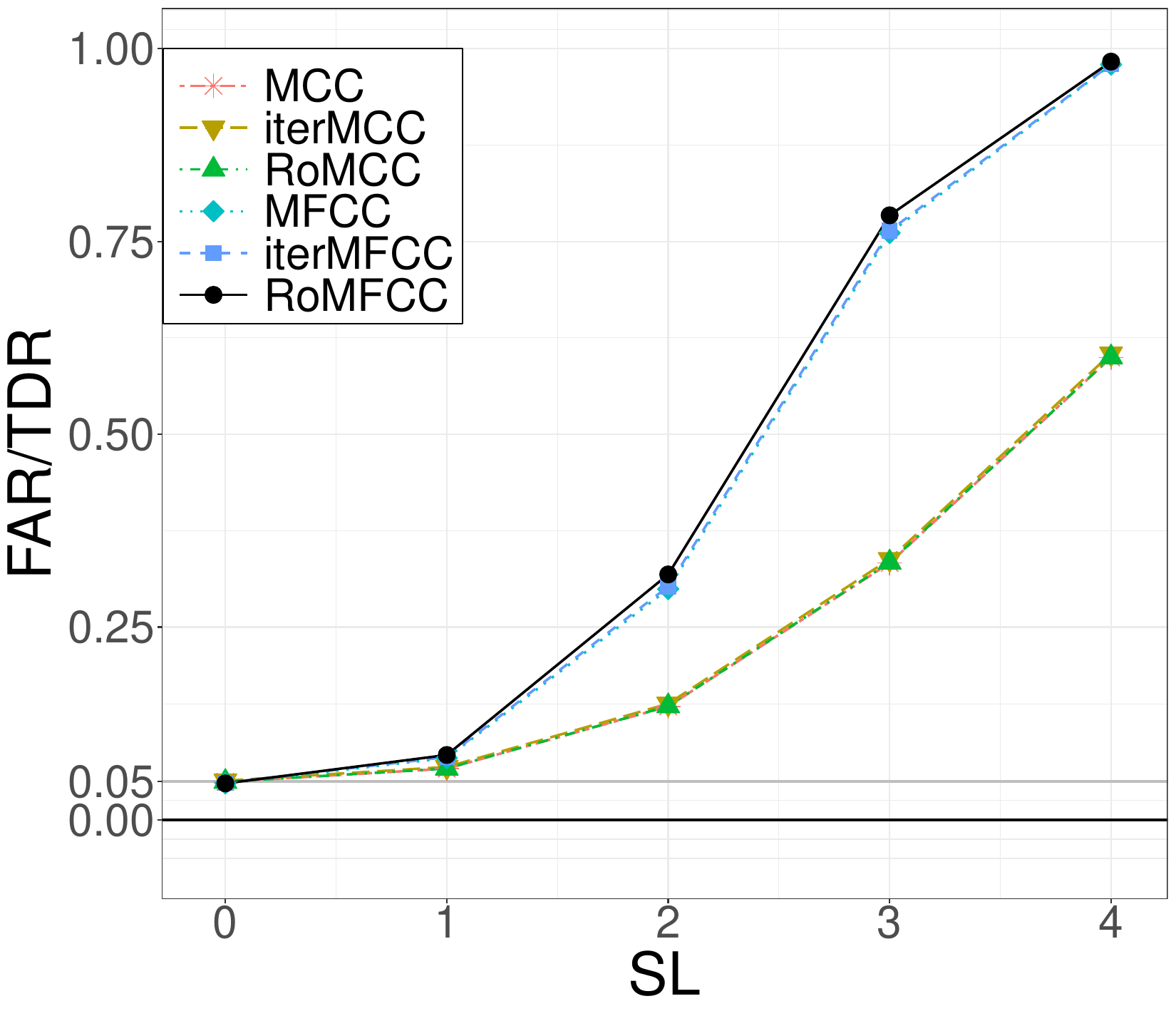}&\includegraphics[width=.25\textwidth]{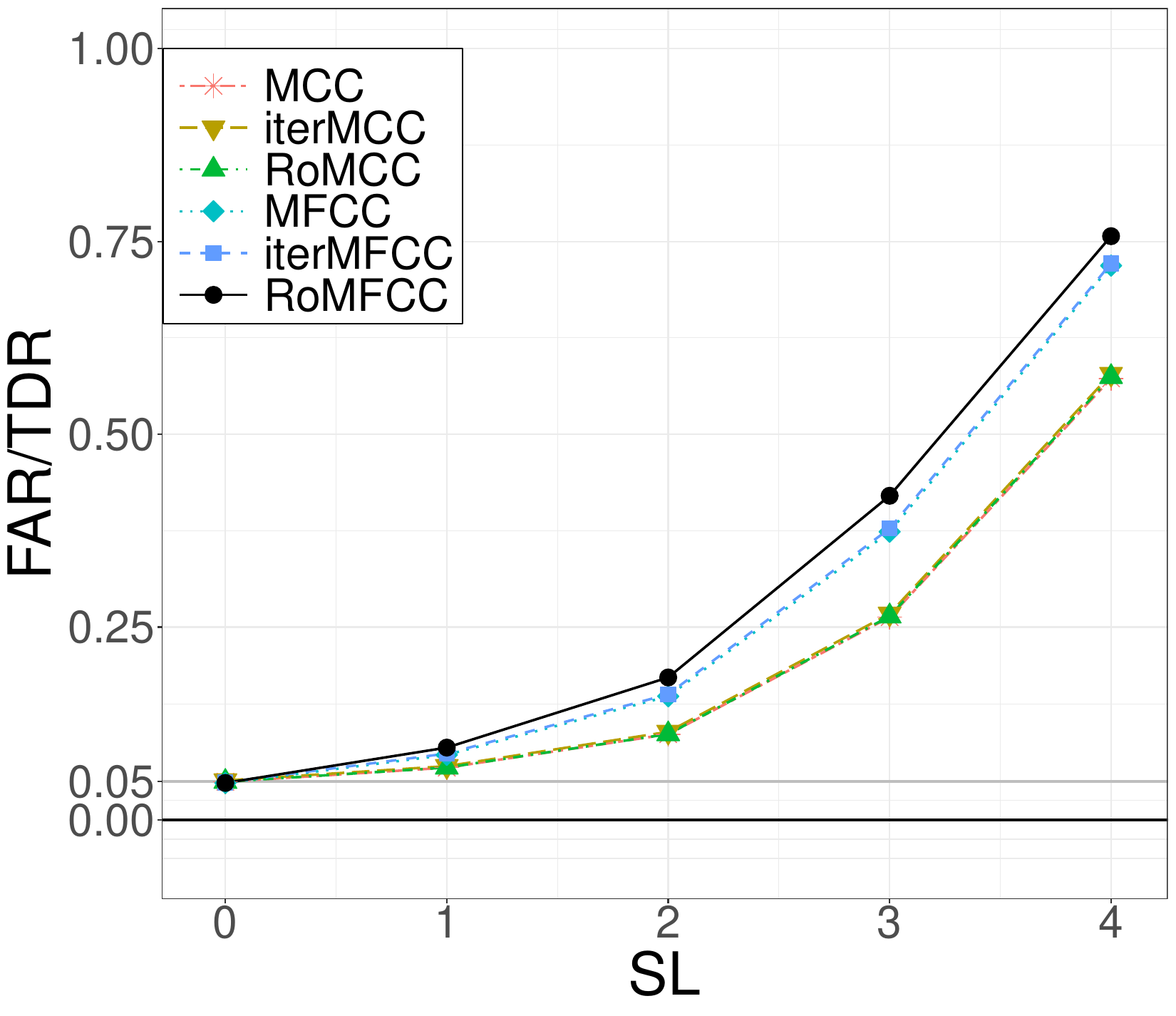}\\
		\textbf{\footnotesize{C2}}&\includegraphics[width=.25\textwidth]{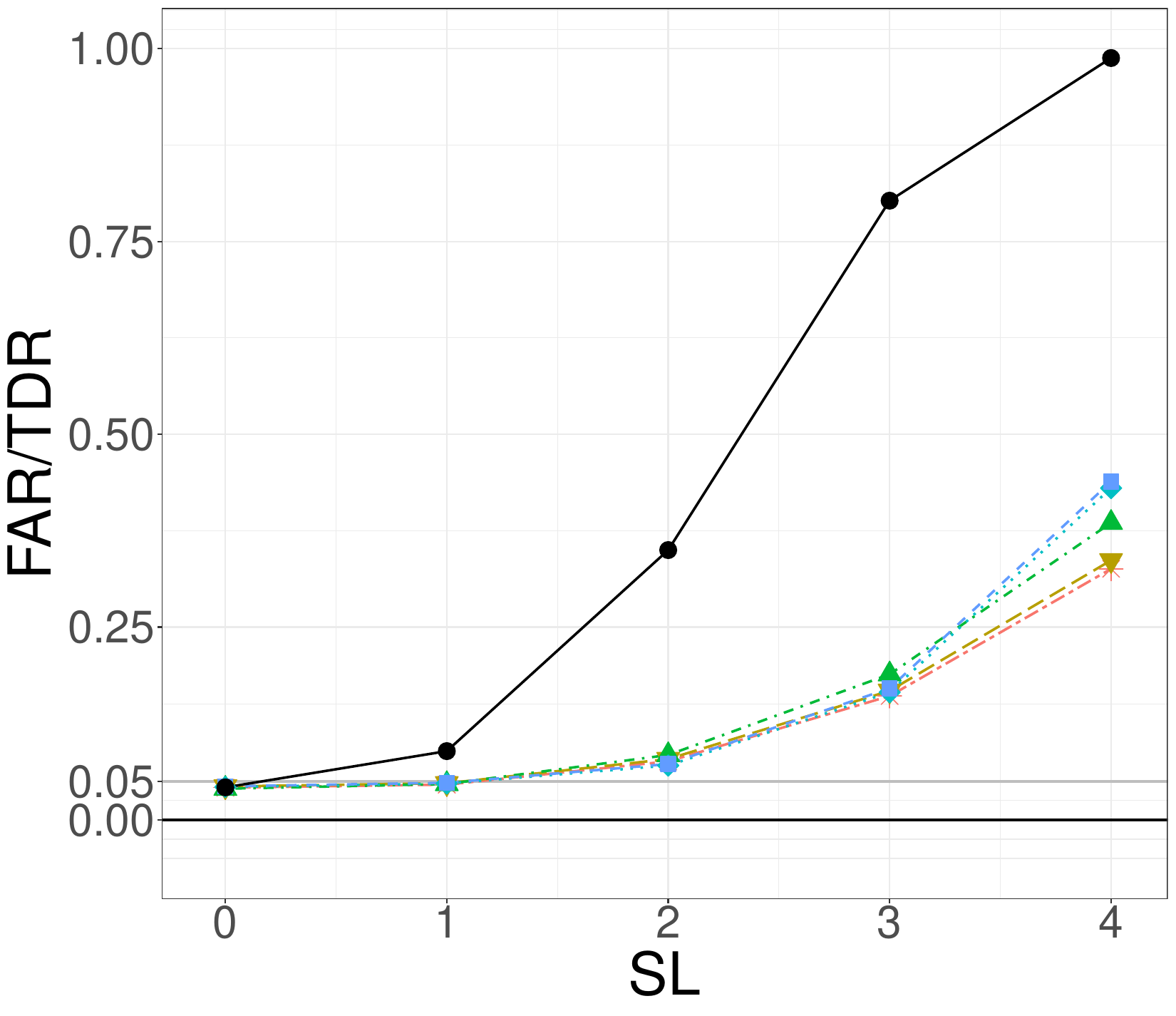}&\includegraphics[width=.25\textwidth]{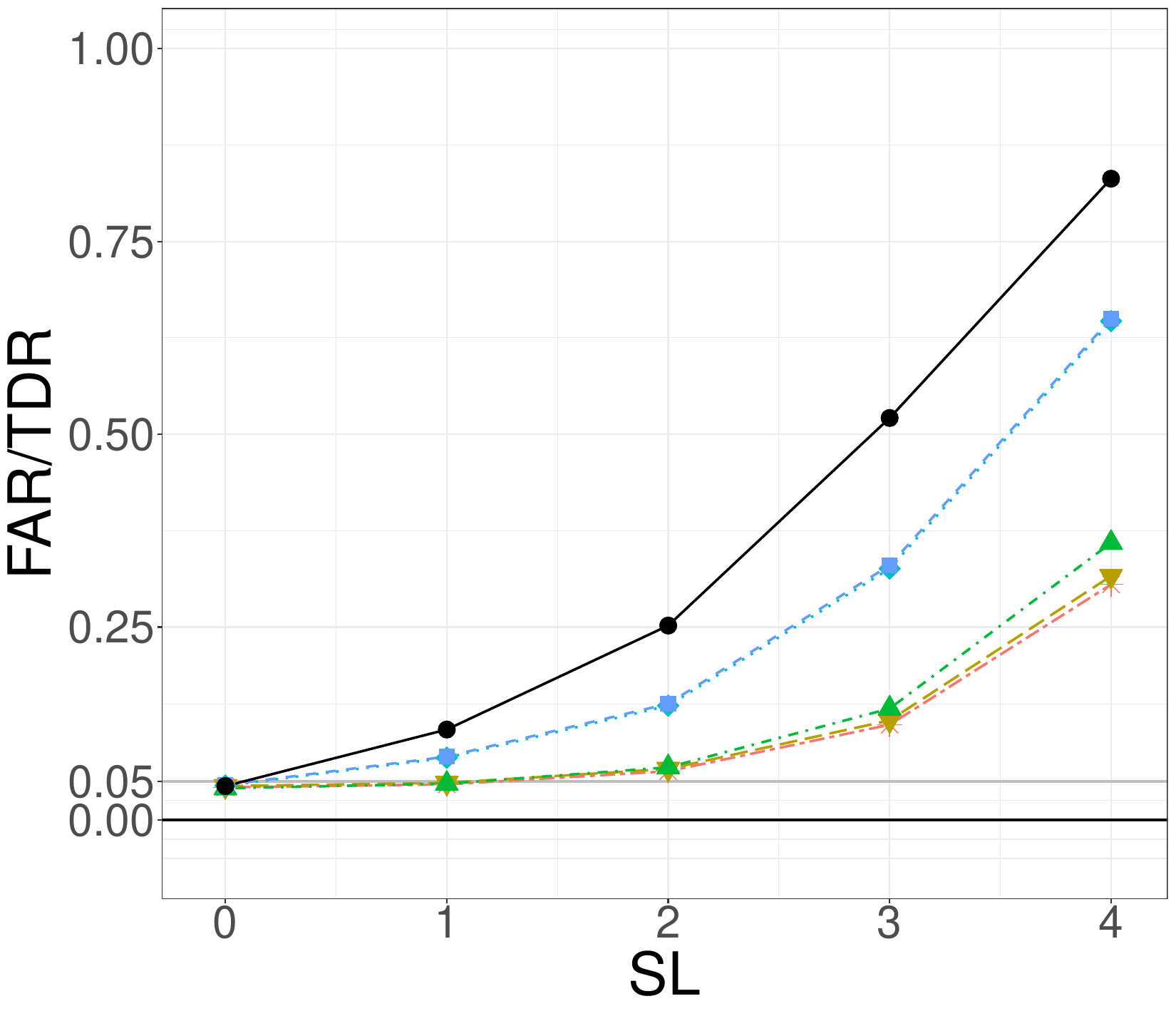}&\includegraphics[width=.25\textwidth]{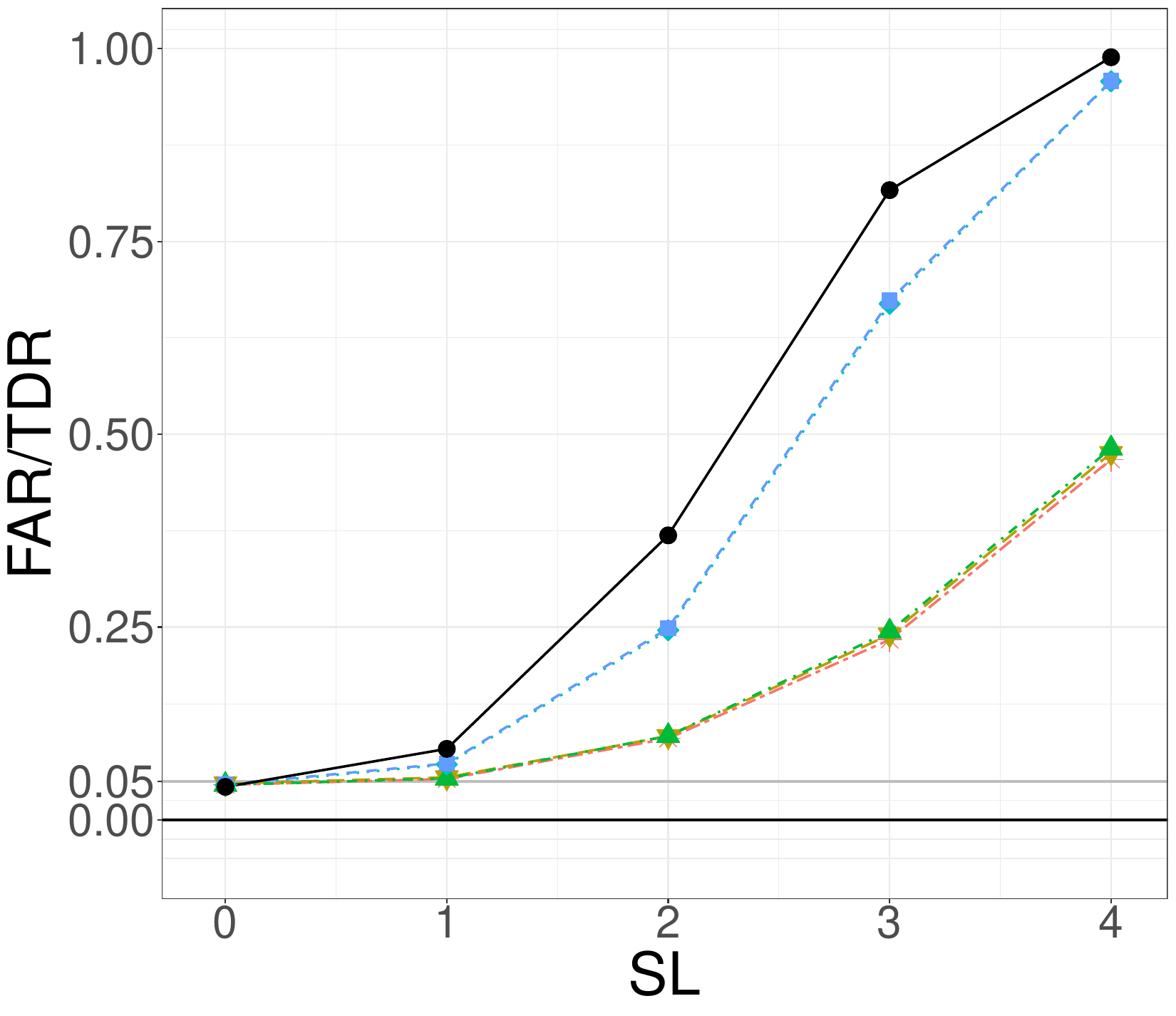}&\includegraphics[width=.25\textwidth]{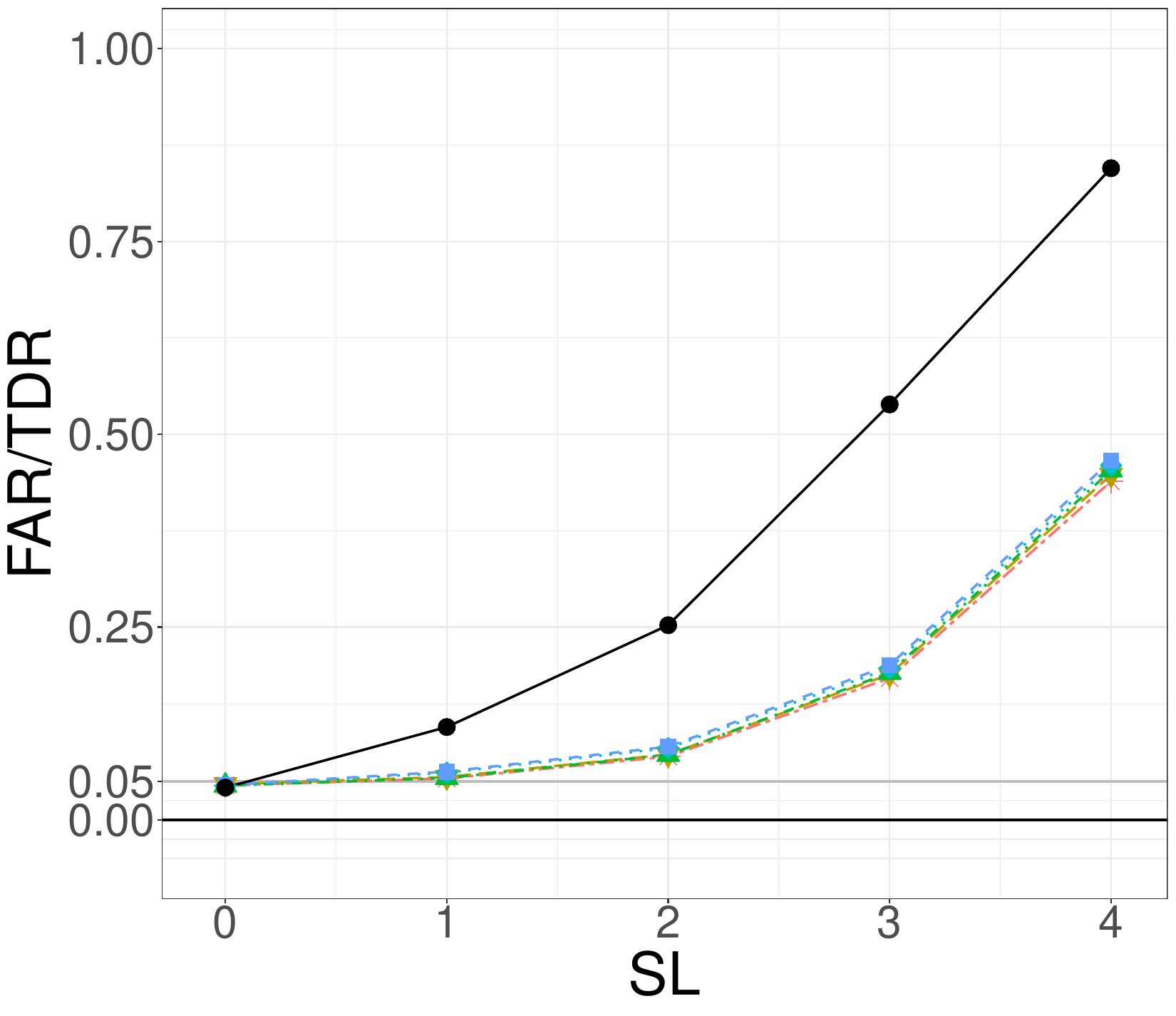}\\
		\textbf{\footnotesize{C3}}&\includegraphics[width=.25\textwidth]{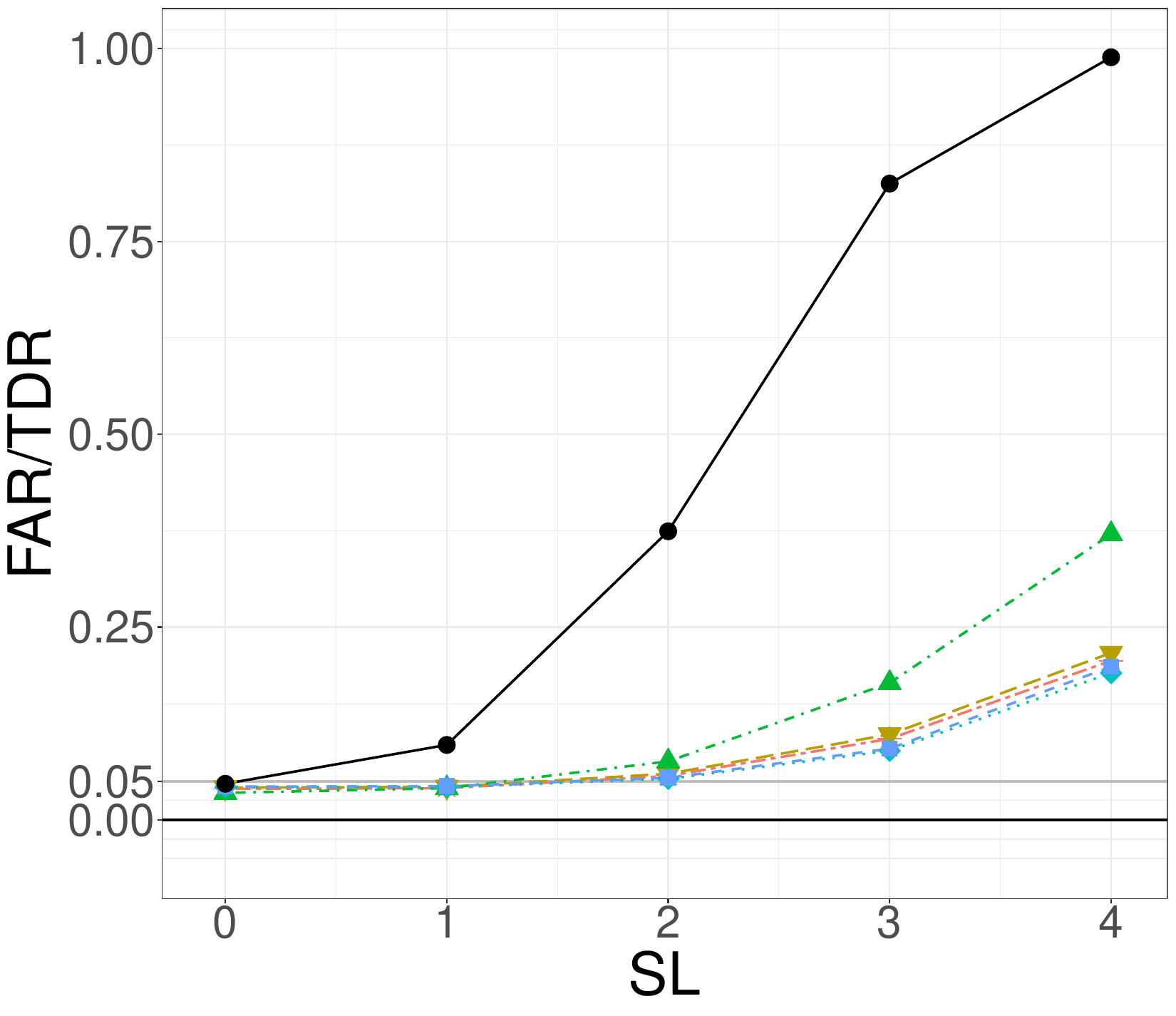}&\includegraphics[width=.25\textwidth]{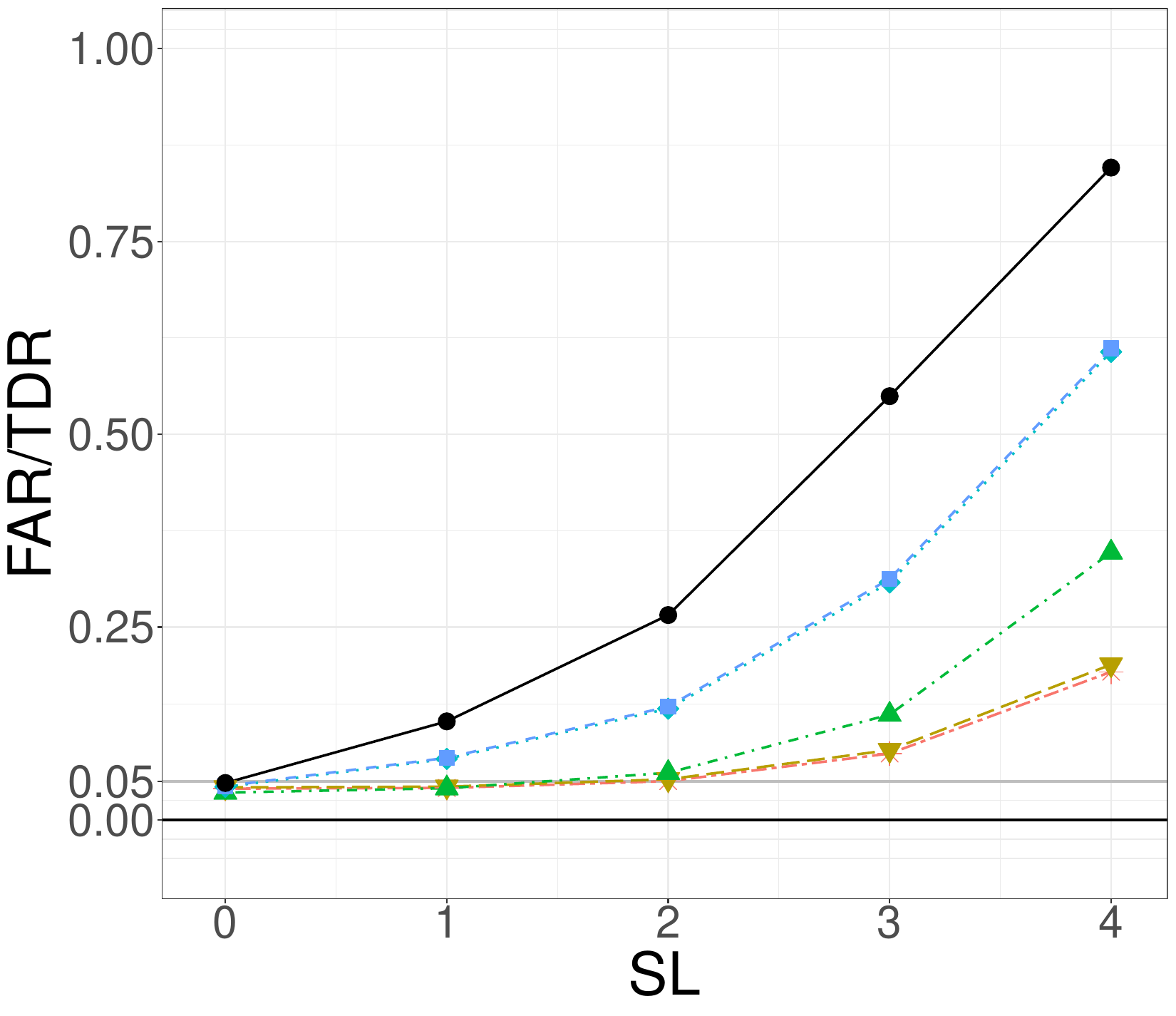}&\includegraphics[width=.25\textwidth]{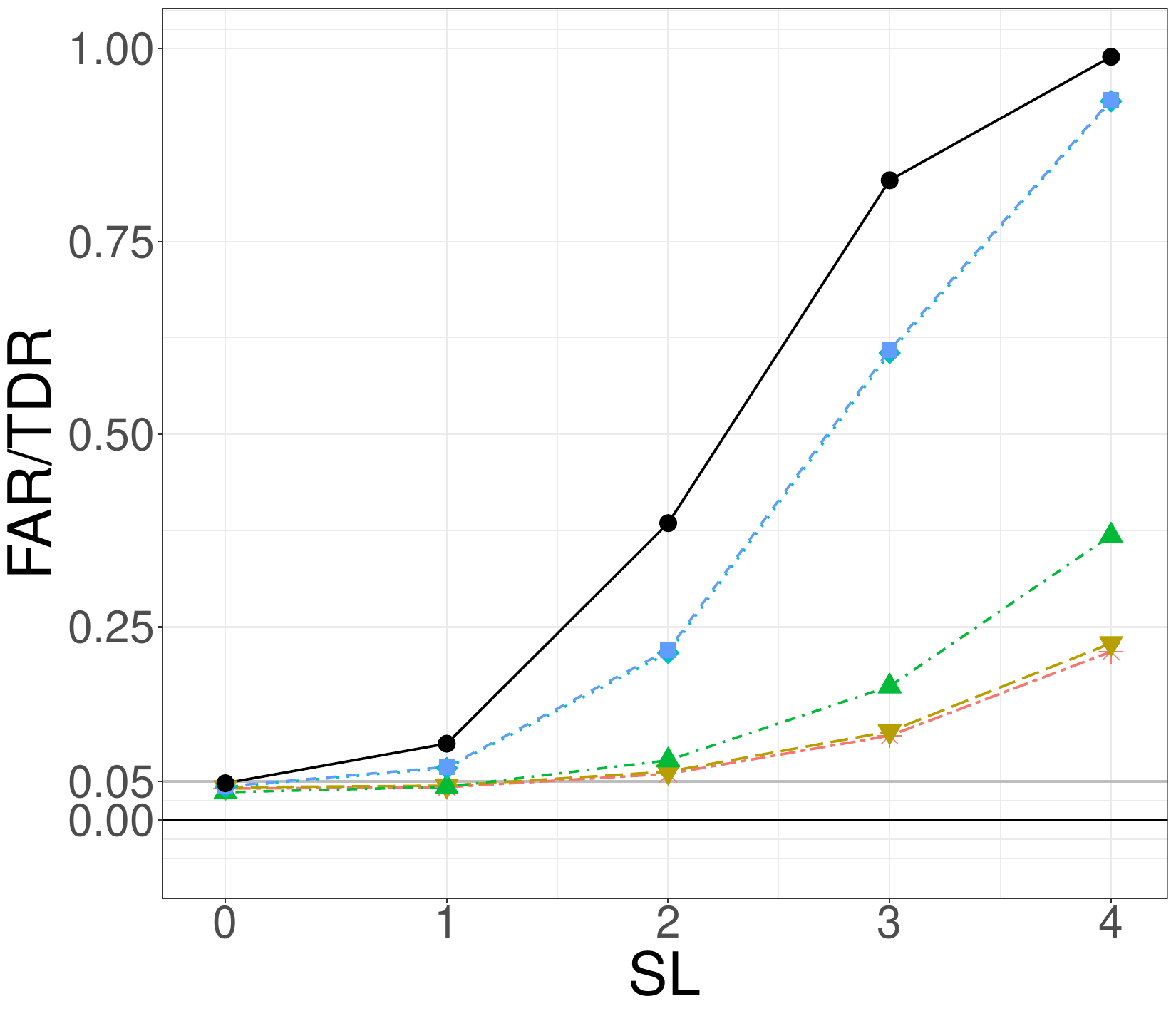}&\includegraphics[width=.25\textwidth]{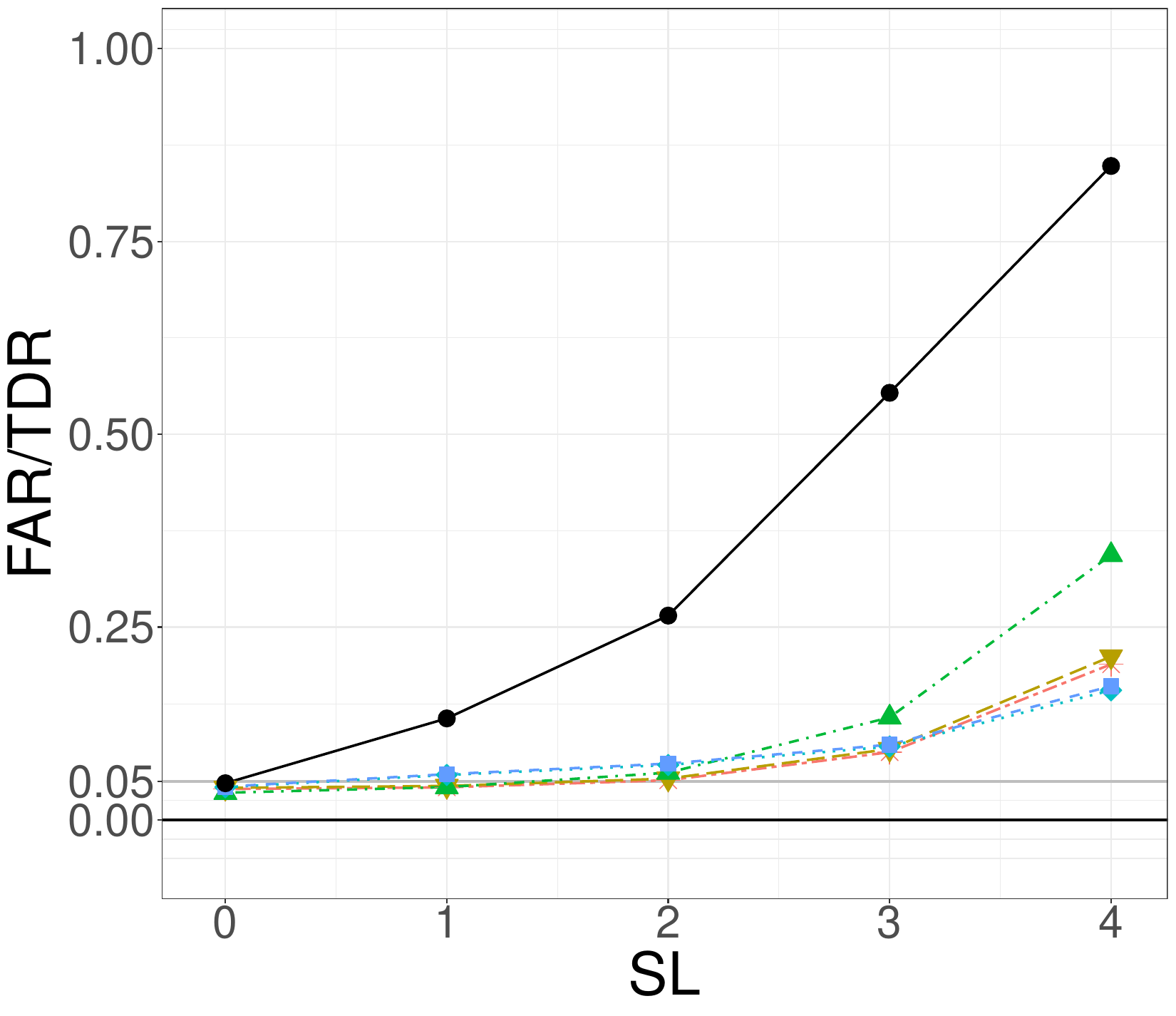}\\
	\end{tabular}
	\vspace{-.5cm}
\end{figure}

In this section, we present additional simulations to analyse the performance of the RoMFCC and the competing methods methods  when data are generated as described in Supplementary Material A with probability of contamination  $\tilde{p}$ equals to 0.1. The RoMFCC is implemented as described in Section 2.
Figure \ref{fi_results_2_1} and Figure \ref{fi_results_3_1} show for Scenario 1 and Scenario 2  the mean FAR ($ SL=0 $) or TDR ($ SL\neq 0 $) achieved by MCC, iterMCC, RoMCC, MFCC, MFCCiter and RoMFCC for each contamination level (C1, C2 and C3), OC condition (OC-E and OC-P) as a function of the severity level $SL$ with contamination model Out-E and Out-P.

By comparing Figure \ref{fi_results_2_1} and Figure \ref{fi_results_3_1} with Figure 4 and Figure 5, the RoMFCC  confirms itself as the best method in all the considered settings.
Specifically, the proposed method, differently from the competing methods, is almost insensible to the fact that a large fraction of the data in the Phase I sample is  now composed by either cellwise or case wise outliers. On the contrary, the competing methods are strongly affected by the different probability of contamination and show overall worse performance than in Section 3 and, thus, are totally inappropriate to monitor the multivariate functional quality characteristic in this setting.  This further confirms as shown in Section 3, i.e., in this simulation study  the RoMFCC performance is almost independent of the contamination of the Phase I sample.

As in Section 3, also in this case, performance differences between the proposed and competing methods are less pronounced in Scenario 2 due to the less severe contamination produced by functional casewise outliers.
However,  RoMFCC still outperforms all the cometing methods and, thus, stand out as the best method.

\bibliographystyle{chicago}
{\small
\bibliography{ref}}
%